\documentclass[11pt,a4paper]{article}
\usepackage{cite}
\usepackage{amsmath, amsthm,mathtools, amssymb,upgreek,slashed}
\usepackage{ifpdf}
\ifpdf
  \usepackage[pdftex]{graphicx}
  \usepackage{epstopdf}
\else
  \usepackage[dvips]{graphicx}
\fi
\parskip=\baselineskip
\def\Bbb{\mathbb}
\def\Tr{{\rm Tr}}
\def\16{{\bf 16}}
\def\1{{\bf 1}}
\def\2{{\bf 2}}
\def\3{{\bf 3}}
\def\4{{\bf 4}}
\def\sg{{\mathrm g}}
\def\i{{\mathrm i}}
\def\h{\widehat}
\def\u{u}
\def\sp{{\sigma}}
\def\E{{\mathcal E}}
\def\O{{\mathcal O}}
\def\PF{{\mathit{P}\negthinspace\mathit{F}}}
\def\tr{{\mathrm{tr}}}
\def\be{\begin{equation}}
\def\ee{\end{equation}}

 \def\ll{\langle\langle}
\def\rr{\rangle\rangle}
\def\la{\langle}
\def\ra{\rangle}
\def\T{{\mathcal T}}
\def\V{{\mathcal V}}
\def\bar{\overline}
\def\v{v}

\def\R{{\Bbb{R}}}\def\Z{{\Bbb{Z}}}

\def\B{{\mathcal B}}

\def\Pf{{\mathrm{Pf}}}
\def\D{{\mathcal D}}
\font\teneurm=eurm10 \font\seveneurm=eurm7 \font\fiveeurm=eurm5
\newfam\eurmfam
\textfont\eurmfam=\teneurm \scriptfont\eurmfam=\seveneurm
\scriptscriptfont\eurmfam=\fiveeurm

\font\teneusm=eusm10 \font\seveneusm=eusm7 \font\fiveeusm=eusm5
\newfam\eusmfam
\textfont\eusmfam=\teneusm \scriptfont\eusmfam=\seveneusm
\scriptscriptfont\eusmfam=\fiveeusm

\font\tencmmib=cmmib10 \skewchar\tencmmib='177
\font\sevencmmib=cmmib7 \skewchar\sevencmmib='177
\font\fivecmmib=cmmib5 \skewchar\fivecmmib='177
\newfam\cmmibfam
\textfont\cmmibfam=\tencmmib \scriptfont\cmmibfam=\sevencmmib
\scriptscriptfont\cmmibfam=\fivecmmib

\numberwithin{equation}{section}

\def\d{\mathrm d}
\def\C{{\Bbb C}}

\def\Ra{{\sf{R}}}
\def\sV{{\sf V}}
\def\Z{{\Bbb Z}}
\def\A{{\mathcal A}}
\def\S{{\mathcal S}}
\def\bar{\overline}
\def\b{{\vec b}}
\def\g{{\mathfrak g}}
\def\ga{\gamma}
\def\bg{\bar\ga}

\def\M{{\mathcal M}}
\def\bM{{\overline \M}}
\def\L{{\mathcal L}}
\def\sM{{\sf M}}
\def\gst{\mathrm{g}_{\mathrm{st}}}

\begin{document}
\begin{titlepage}
\begin{flushright}

\end{flushright}

\vskip 1.5in
\begin{center}
{\bf\Large{Developments In Topological Gravity}}
\vskip
0.5cm {Robbert Dijkgraaf and Edward Witten} \vskip 0.05in {\small{ \textit{Institute for Advanced Study}\vskip -.4cm
{\textit{Einstein Drive, Princeton, NJ 08540 USA}}}
}
\end{center}
\vskip 0.5in
\baselineskip 16pt
\begin{abstract}
This note aims to provide an entr\'ee to two developments in two-dimensional topological gravity -- that is,
intersection theory on the moduli space of Riemann surfaces -- that have not yet become well-known among physicists. 
A little over a decade ago, Mirzakhani discovered \cite{M1,M2} 
an elegant new proof of the formulas that result from the relationship between topological gravity and matrix models of two-dimensional
gravity. Here we will give a very partial introduction to that work, which hopefully will also serve as a modest tribute to the memory
of a brilliant mathematical pioneer.  More recently, Pandharipande, Solomon, and Tessler \cite{PST} (with further developments
in \cite{Tes,BT,STa}) generalized intersection theory on
moduli space to the case of Riemann surfaces with boundary, leading to generalizations of the familiar KdV and Virasoro formulas.  Though
the existence of such a generalization appears natural from the matrix model viewpoint -- it corresponds to adding vector degrees
of freedom to the matrix model --
constructing this generalization is not straightforward.   We will give some idea of the unexpected way that the difficulties
were resolved. \end{abstract}
\date{September, 2017}
\end{titlepage}
\def\Hom{\mathrm{Hom}}

\def\U{{\mathcal U}}

\tableofcontents
\section{Introduction}

There are at least two candidates for the simplest  model of quantum gravity in two spacetime dimensions.  Matrix models are certainly
one candidate, extensively studied since the 1980's.  These models were proposed in \cite{Wein,Kaz, David,Amb,KKM} and solved in \cite{BK,DS,GM};
for a comprehensive review with extensive references, see \cite{FGZ}.
A second candidate is provided by topological gravity, that is, intersection theory on the moduli space of Riemann surfaces.
It was conjectured some time ago that actually two-dimensional topological gravity is
 equivalent to the matrix model \cite{Wittenone,Witten}. 
 
 This equivalence led  to formulas expressing the intersection numbers of
certain natural cohomology classes on moduli space in terms of the partition function of the matrix model, which is governed by KdV equations
\cite{D}  or equivalently by Virasoro constraints \cite{DVV}.  
These formulas were first proved by Kontsevich \cite{K} by a direct calculation that expressed intersection
numbers on moduli space in terms of a new type of matrix model (which was again shown to be governed by the KdV and Virasoro constraints).

A little over a decade ago, Maryam Mirzakhani found a new proof of this relationship
as part of her Ph.D. thesis work \cite{M1,M2}.  (Several other proofs are known \cite{OP,KL}.)
She put the accent on understanding the Weil-Petersson volumes of  moduli spaces of hyperbolic
Riemann surfaces with boundary, showing that these volumes contain all the information in the intersection numbers.   A hyperbolic structure on a surface $\Sigma$ is determined by a flat $SL(2,\R)$ connection, so the moduli
space $\M$ of hyperbolic structures on $\Sigma$ can be understood as a moduli space of flat $SL(2,\R)$ connections.  Actually, the Weil-Petersson 
symplectic form
on $\M$ can be defined by the same formula that is used to define the symplectic form on the moduli space of flat connections on $\Sigma$ with
structure group a compact Lie group such as $SU(2)$.  For a compact Lie group, the volume of the moduli space can be computed by a direct
cut and paste method \cite{Wittengauge} that involves building $\Sigma$ out of simple building blocks (three-holed spheres).  Naively,
one might hope to do something similar for $SL(2,\R)$ and thus for the Weil-Petersson volumes.
But there is a crucial difference: in the case of $SL(2,\R)$, 
in order to define the moduli space whose volume one will calculate, one wants to divide by the action of the mapping class group on $\Sigma$.   
(Otherwise the volume is trivially infinite.)  But dividing by the mapping class group is not compatible with any
simple cut and paste method.   Maryam Mirzakhani overcame this
difficulty in a surprising and elegant way, of which we will give a glimpse in section \ref{curing}.

Matrix models of two-dimensional gravity  have a natural generalization in which vector degrees of freedom are added  \cite{Kostov,Minahan,J1,ZY,IT,J2}.  This generalization
is related, from a physical point of view, to two-dimensional gravity formulated on two-manifolds $\Sigma$ that carry a complex structure
but may have a boundary.  We will refer to such two-manifolds as open Riemann surfaces (if the boundary of $\Sigma$ is empty, we will call it a closed Riemann surface).  It is natural to hope that, by analogy with what
happens for closed Riemann surfaces, there would be an intersection theory on the moduli
space of open Riemann surfaces that would be related to matrix models with vector degrees of freedom.   In trying to construct such a theory,
one runs into immediate difficulties: the moduli space of open Riemann surfaces does not have a natural orientation and has a boundary; for
both reasons, it is not obvious how to define intersection theory on this space.   These difficulties were overcome  by Pandharipande,
Solomon, and Tessler in a rather unexpected way \cite{PST} whose full elucidation involves introducing spin structures in a problem in which at first sight they do not seem relevant \cite{Tes,BT,STa}.
 In section \ref{opentot}, we will explain some highlights of this story.  In section \ref{intmat}, we review matrix models with vector
 degrees of freedom, and show how they lead -- modulo a slightly surprising twist -- to precisely the same Virasoro constraints that
 have been found in intersection theory on the moduli space of open Riemann surfaces.
 
 The matrix models we consider are the direct extension of those studied in \cite{BK,DS,GM}.   The same problem has been treated in a rather different
 approach via Gaussian matrix models with an external source in \cite{BH} and in chapter 8 of \cite{BH2}. See also \cite{Alex} for another approach.  For an expository
 article on the relation of matrix models and intersection theory, see \cite{Ok}.

\section{Weil-Petersson Volumes And Two-Dimensional Topological Gravity}\label{curing}

\subsection{Background And Initial Steps}\label{initial}

Let $\Sigma$ be a closed Riemann surface of genus $g$
with  marked points\footnote{The marked points are labeled and are required to be always distinct.} 
$p_1,\dots,p_n$, and let $\L_i$ be the cotangent space to $p_i$ in $\Sigma$. As $\Sigma$ and the $p_i$
vary, $\L_i$ varies as the fiber of a complex line bundle -- which we also denote as $\L_i$ -- over $\M_{g,n}$, the moduli space of genus $g$ curves with $n$ punctures.
In fact, these line bundles extend naturally over $\bM_{g,n}$, the Deligne-Mumford compactification of $\M_{g,n}$.  We write $\psi_i$ for the first Chern class of $\L_i$; thus
$\psi_i=c_1(\L_i)$ is a two-dimensional cohomology class.  For a non-negative integer $d$, we set $\tau_{i,d}=\psi_i^d$, a cohomology class of dimension $2d$.
The usual correlation functions of 2d topological gravity are the intersection numbers
\be\label{zinc}\langle \tau_{d_1}\tau_{d_2}\dots \tau_{d_n}\rangle =\int_{\bM_{g,n}}\tau_{1,d_1}\tau_{2,d_2}\cdots \tau_{n,d_n}=\int_{\bM_{g,n}}\psi_1^{d_1}\psi_2^{d_2}\cdots \psi_n^{d_n},\ee
where $d_1,\dots,d_n$ is any $n$-plet of non-negative integers.  The right hand side of eqn. (\ref{zinc}) vanishes unless  $\sum_{i=1}^{n}d_i =3g-3+n$.  To be more
exact, what we have defined in eqn. (\ref{zinc}) is the genus $g$ contribution to the correlation function; the full correlation function is obtained by summing over $g\geq 0$.  (For a given set of $d_i$, there is at most one integer solution  $g$ of the condition $\sum_{i=1}^{n}d_i =
3g-3+n$, and this is the only value 
 that contributes to $\langle \tau_{d_1}\tau_{d_2}\dots \tau_{d_n}\rangle$.) 

Let us now explain how these correlation functions are related to the Weil-Petersson volume of $\M_{g}$.   In the special case $n=1$, we have just a single marked
point $p$ and a single line bundle $\L$ and cohomology class $\psi$.  We also have the forgetful map $\pi:\bM_{g,1}\to \bM_g$ that forgets the marked point.
We can construct a two-dimensional cohomology class $\kappa$ on $\bM_g$ by integrating the four-dimensional class $\tau_2=\psi^2$ over the fibers of this forgetful map:
\be\label{inc}\kappa= \pi_*(\tau_2). \ee
More generally, the Miller-Morita-Mumford (MMM) classes are defined by $\kappa_d=\pi_*(\tau_{d+1})$, so $\kappa$ is the same as the first MMM class $\kappa_1$.
$\kappa$ is cohomologous to a multiple of the Weil-Petersson symplectic form $\omega$ of the moduli space \cite{Wolpert1,Wolpert2}: 
\be\label{winc} \frac{\omega}{2\pi^2}=\kappa.\ee
Because of (\ref{inc}), it will be convenient to use $\kappa$, rather than $\omega$, to define a volume form.   With this choice, the
volume of $\M_g$ is 
\be\label{pinc} V_g=\int_{\bM_g}\frac{\kappa^{3g-3}}{(3g-3)!}=\int_{\bM_g}\exp(\kappa) .\ee

The relation between $\kappa$ and $\tau_2$ might make one hope that the volume $V_g$ would be one of the correlation functions of topological gravity:
\be\label{linc} V_g\overset{?}{=} \frac{1}{(3g-3)!}\big\langle \tau_2^{3g-3}\big\rangle . \ee
Such a simple formula is, however, not true, for the following reason.  To compute the right hand side of eqn. (\ref{linc}), we would have to introduce $3g-3$ marked points on $\Sigma$,
and insert $\tau_2$ (that is, $\psi_i^2$) at each of them.  It is true that for a single marked point, $\kappa$ 
can be obtained as the integral of $\tau_2$ over the fiber of the forgetful
map, as in eqn. (\ref{inc}).  However, when there is more than one marked point, we have to take into account that the Deligne-Mumford compactification of $\M_{g,n}$
is defined in such a way that the marked points are never allowed to collide.  Taking this into account leads  to corrections in
which, for instance, two copies of $\tau_2$ are replaced by a single copy of $\tau_3$.  The upshot is that $V_g$ can be expressed in terms of the correlation functions
of topological gravity, and thus can be computed using the KdV equations or the Virasoro constraints, but the necessary formula is more complicated.  See section \ref{volint} below.  For now, we just remark that this approach has been used  \cite{Z} to determine the large $g$ asymptotics
of $V_g$, but apparently does not easily lead to explicit formulas for $V_g$ in general.   Weil-Petersson volumes were originally studied and
their asymptotics estimated
by quite different methods \cite{Penner}. 

$\M_{g,n}$ likewise has a Weil-Petersson volume $V_{g,n}$ of its own, which likewise can be computed, in principle, using a knowledge of the intersection numbers on
$\bM_{g,n'}$ for $n'>n$.  Again this gives useful information but it is difficult to get explicit general formulas.

\begin{figure}
 \begin{center}
   \includegraphics[width=5.5in]{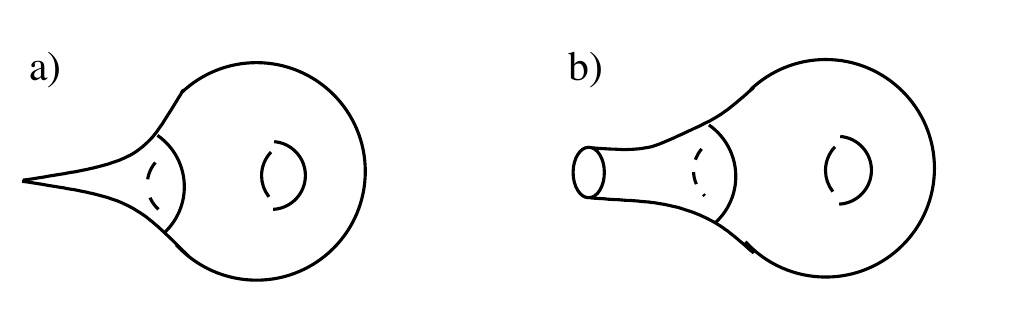}
 \end{center}
\caption{\small (a)  A marked point in a hyperbolic Riemann surface is treated as a cusp: it lies at infinity in the hyperbolic metric. (b) Instead of a cusp,
a hyperbolic Riemann surface might have a geodesic boundary, with  circumference any positive number $b$. \label{cusp}}
\end{figure}

Mirzakhani's procedure was different.  First of all, she worked in the hyperbolic world, so in the following
discussion $\Sigma$ is not
just a complex Riemann surface; it carries a hyperbolic  metric, by which we mean a Riemannian metric
 of constant scalar curvature $R=-1$.  We recall that  a complex Riemann surface admits a unique Kahler metric with
 $R=-1$.  We recall also that in studying hyperbolic Riemann surfaces, it is natural\footnote{This
is natural because the degenerations of the hyperbolic metric of $\Sigma$ that correspond to Deligne-Mumford  compactification of $\M_{g,n}$ generate cusps.  Since
the extra marked points that occur when $\Sigma$ degenerates (for example to two components that are glued together at new marked points) appear as cusps in the
hyperbolic metric, it is natural to treat all marked points that way.}
to think of a marked point as a cusp, which lies at infinity in the hyperbolic metric (fig. \ref{cusp}). 

 Instead of a marked point, we can consider a Riemann surface
with a boundary component.  In the hyperbolic world, one requires the boundary to be a geodesic in the hyperbolic metric.  Its circumference may be any positive number $b$.
Let us consider, rather than a closed Riemann surface $\Sigma$ of genus $g$ with $n$ labeled marked points, an open Riemann surface $\Sigma$ also of genus $g$,
but now with $n$ labeled boundaries.  In the hyperbolic world, it is natural to specify $n$ positive numbers $b_1,\dots,b_n$ and to require that $\Sigma$ carry
a hyperbolic metric such that the boundaries are geodesics of lengths $b_1,\dots,b_n$.   We denote the moduli space of such hyperbolic metrics as    
$\M_{g; b_1,b_2,\dots,b_n}$ or more briefly as $\M_{g,\b}$, where $\b$ is the $n$-plet $(b_1,b_2,\dots,b_n)$.  

As a topological space, $\M_{g,\b}$ is independent of $\b$. In fact, $\M_{g,\b}$ is an orbifold, and the topological type of an orbifold cannot
depend on continuously variable data such as $\b$.   In the limit that $b_1,\dots,b_n$ all go to zero, the boundaries turn into cusps
and $\M_{g,\b}$ turns into $\M_{g,n}$.  Thus topologically, $\M_{g,\b}$ is equivalent to $\M_{g,n}$ for any $\b$.  Very concretely, we
 can always convert a Riemann
surface with a boundary component to a Riemann surface with a marked point by gluing a disc, with a marked point at its center, to the given boundary component.
Thus we can turn a Riemann surface with boundaries into one with marked points without changing the parameters the Riemann surface can depend on,
and this leads to the topological equivalence of $\M_{g,\b}$ with $\M_{g,n}$.  If we allow the hyperbolic metric of $\Sigma$ to develop cusp singularities, we get
a compactification $\bM_{g,\b}$ of $\M_{g,b}$ which coincides with the Deligne-Mumford compactification $\bM_{g,n}$ of $\M_{g,n}$.  

$\M_{g,n}$ and $\M_{g,\b}$ have  natural Weil-Petersson symplectic forms that we will call $\omega$ and $\omega_{\vec b}$ (see \cite{Goldman}).  
Since $\M_{g,n}$ and $\M_{g,\b}$ are equivalent topologically, it makes
sense to ask if the symplectic form $\omega_\b$ of $\M_{g,\b}$ has the same cohomology class as the symplectic form $\omega$ of $\M_{g,n}$.  The answer
is that it does not.   Rather, one has (see \cite{M2}, Theorem 4.4)
\be\label{turf}\omega_\b=\omega+\frac{1}{2}\sum_{i=1}^n b_i^2\psi_i.\ee
(This is a relationship in cohomology, not an equation for differential forms.)
From this it follows that the Weil-Petersson volume of $\M_{g,\b}$ is\footnote{The reason for the factor of $1/(2\pi^2)^{3g-3+n}$ is just that we defined the
volumes in eqn. (\ref{pinc}) using $\kappa$ rather than $\omega$.}
\be\label{urf}V_{g,\b}=\frac{1}{(2\pi^2)^{3g-3+n}}\int_{\M_{g,\b}}\exp\left(\omega+\frac{1}{2}\sum_{i=1}^n b_i^2\psi_i\right) . \ee
Equivalently, since compactification by allowing cusps does not affect the volume integral,  and the compactification of $\M_{g,\b}$ is the same as $\bM_{g,n}$, one can write
this as an integral over the compactification:
\be\label{lurf}
V_{g,\b}=\frac{1}{(2\pi^2)^{3g-3+n}} \int_{\bM_{g,n}}\exp\left(\omega+\frac{1}{2}\sum_{i=1}^n b_i^2\psi_i\right).\ee

This last result tells us that at $\b=0$, $V_{g,\b}$ reduces to the volume $V_{g,n}=(1/2\pi^2)^{3g-3+n}\int_{\bM_{g,n}}e^\omega$ of $\M_{g,n}$.   Moreover, eqn. (\ref{lurf}) implies
that $V_{g,\b}$ is a polynomial in $\b^2=(b_1^2,\dots,b_n^2)$ of total degree $3g-3+n$.  In evaluating the term of top degree in $V_{g,\b}$, we can drop $\omega$ from the
exponent in eqn. (\ref{lurf}).  Then the expansion in powers of the $b_i$ tells us that this term of top degree is
\be\label{murf}\frac{1}{(2\pi^2)^{3g-3+n}} \sum_{d_1,\dots,d_n}\prod_{i=1}^n \frac{b_i^{2d_i}}{2^{d_i}d_i!} \left\langle \tau_{d_1}\tau_{d_2}\cdots \tau_{d_n}\right\rangle. \ee
(Only terms with $\sum_i d_i=3g-3+n$ make nonzero contributions in this sum.)
In other words, the correlation functions of two-dimensional topological gravity on a closed Riemann surface appear as coefficients in the expansion of $V_{g,\b}$.
Of course, $V_{g,\b}$ contains more information,\footnote{This additional information in principle is not really new.  Using facts that generalize the relationship
between $V_{g,n}$ and the correlation functions of topological gravity that we discussed at the outset, one can deduce also the subleading terms in $V_{g,\b}$ in terms of the correlation functions of
topological gravity.  However, it appears difficult to get useful formulas in this way.} since we can also consider the terms in $V_{g,\b}$ that are subleading in $\b$. 

Thus Mirzakhani's approach to topological gravity involved deducing the correlation functions of topological gravity from the volume polynomials $V_{g,\b}$.
We will give a few indications of how she computed these volume polynomials in section \ref{thecure}, after first recalling a much simpler problem.

\subsection{A Simpler Problem}\label{compact}

Before explaining how to compute the volume of $\M_{g,\b}$, we will describe how volumes can be computed in a simpler case.  In fact, the analogy was
noted in \cite{M2}.

Let $G$ be a compact Lie group, such as $SU(2)$, with Lie algebra $\g$, and let $\Sigma$ be a closed Riemann surface of genus $g$.  
Let $\sM_g$ be the moduli space of homomorphisms from
the fundamental group of $\Sigma$ to $G$. Equivalently, $\sM_g$ is the moduli space of flat $\g$-valued flat connections on $\Sigma$.   Then \cite{Goldman,ABott}
$\sM_g$ has a natural symplectic form that in many ways is analogous to the Weil-Petersson form on $\M_g$.  Writing $A$ for a flat connection on $\Sigma$ and $\delta A$
for its variation, the symplectic form of $\sM_g$ can be defined by the gauge theory formula
\be\label{prof}\upomega =\frac{1}{4\pi^2}\int_\Sigma \Tr\,\delta A\wedge \delta A,\ee
where (for $G=SU(2)$) we can take $\Tr $ to be the trace in the two-dimensional representation.  

Actually, the Weil-Petersson form of $\M_g$ can be defined by much the same formula.  The moduli space of hyperbolic metrics on $\Sigma$ is a component\footnote{The
moduli space of flat $SL(2,\R)$ connections on $\Sigma$ has various components labeled by the Euler class of a flat real vector bundle of rank 2 (transforming in the
2-dimensional representation of $SL(2,\R)$).  One of these components
parametrizes hyperbolic metrics on $\Sigma$ together with a choice of spin structure.  If we replace $SL(2,\R)$ by $PSL(2,\R)=SL(2,\R)/\Z_2$ (the symmetry
group of the hyperbolic plane), we forget the spin structure, so to be precise, $\M_g$ is a component of the moduli space of flat $PSL(2,\R)$ connections.
This refinement will not be important in what follows and we loosely speak of $SL(2,\R)$.   In terms of $PSL(2,\R)$, one can define $\Tr$ as $1/4$ of the trace
in the three-dimensional representation.}  of the moduli
space of flat $SL(2,\R)$ connections
 over $\Sigma$, divided by the mapping class group of $\Sigma$. Denoting the flat connection again as $A$ and taking $\Tr$ to be the trace in the two-dimensional representation
of $SL(2,\R)$, the right hand side of eqn. (\ref{prof}) becomes in this case a multiple of the Weil-Petersson symplectic form $\omega$ on $\M_g$.

There is also an analog for compact $G$ of the moduli spaces $\M_{g,\vec b}$ of hyperbolic Riemann surfaces with geodesic boundary.    For $\vec b=(b_1,\dots,b_n)$, 
$\M_{g,\vec b}$ can be interpreted as follows in the gauge theory language.
A point in $\M_{g,\vec b}$ corresponds,
in the gauge theory language, to a flat $SL(2,\R)$ connection on $\Sigma$ with the property that the holonomy around the $i^{th}$ boundary is conjugate in $SL(2,\R)$ to
the group element
$\mathrm{diag}(e^{b_i},e^{-b_i})$.  

In this language, it is clear how to imitate the definition of $\M_{g,\vec b}$ for a compact Lie group such as 
$SU(2)$.  For $k=1,\dots,n$, we choose a conjugacy class in $SU(2)$,
say the class that contains $U_k=\mathrm{diag}(e^{\i\alpha_k},e^{-\i\alpha_k})$, for 
some $\alpha_k$.   We write $\vec\alpha$ for the $n$-plet $(\alpha_1,\alpha_2,\dots,
\alpha_n)$, and we define $\sM_{g,\vec \alpha}$ to be the moduli space of 
flat connections on a genus $g$ surface $\Sigma$ with $n$ holes (or equivalently $n$
boundary components) with the property that the holonomy around the $k^{th}$ hole
is conjugate to $U_k$. 
With a little care,\footnote{On the gravity side, Mirzakhani's proof that the cohomology class of $\kappa_{\b}$ is linear in $\b^2$ did
not use eqn. (\ref{prof}) at all, but a different approach based on Fenchel-Nielsen coordinates.  
On the gauge theory side, in using eqn. (\ref{prof}), it can be convenient to consider a Riemann surface 
with punctures (i.e., marked points that have been deleted)
rather than boundaries.  This does not affect the moduli space of flat connections, 
because if $\Sigma$ is a Riemann surface with boundary, one can glue in to
each boundary component a once-punctured disc, thus replacing all boundaries by 
punctures, without changing the moduli space of flat connections.  
For brevity we will stick here with the language of Riemann surfaces with boundary.}
the right hand side of the formula (\ref{prof}) can be used in this situation to 
define the Weil-Petersson form $\kappa_{\vec b}$ of $\M_{g,\vec b}$,
and the analogous symplectic form $\upomega_{\vec \alpha}$ of $\sM_{g,\vec \alpha}$.  
Thus in particular, $\sM_{g,\vec\alpha}$ has a symplectic volume
$\sV_{g,\vec\alpha}$.    Moreover, $\sV_{g,\vec\alpha}$ is a polynomial in $\vec\alpha$, 
and the coefficients of this polynomial are the correlation functions of
a certain version of two-dimensional topological gauge theory -- they are the 
intersection numbers of certain natural cohomology classes on $\sM_{g,\vec\alpha}$.

These statements, which are analogs of what we described in the case of gravity in section \ref{initial}, 
were explained for gauge theory with a compact gauge
group in \cite{Wittengaugetwo}.    Moreover, for a compact gauge group, various relatively simple ways 
to compute the symplectic volume $\sV_{g,\vec\alpha}$
were described in \cite{Wittengauge}.  None of these methods carry over naturally to the gravitational case.  However, to 
appreciate Maryam Mirzakhani's
work on the gravitational case, it helps to have some idea how the analogous problem can be solved in the case of 
gauge theory with a compact gauge group.  
So we will make a few remarks.

\begin{figure}
 \begin{center}
   \includegraphics[width=6in]{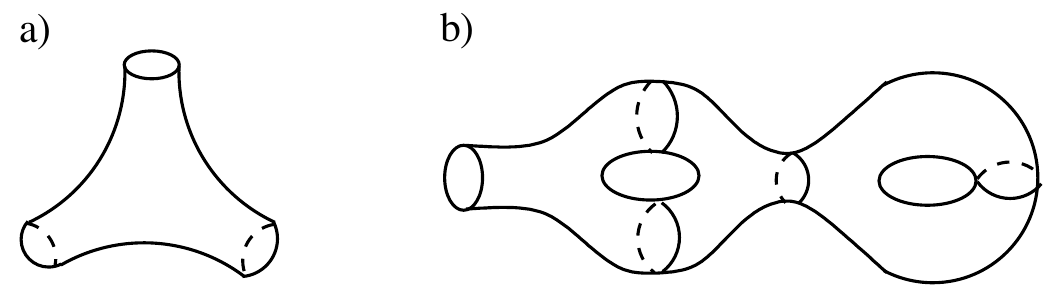}
 \end{center}  
\caption{\small (a) A three-holed sphere or ``pair of pants.'' (b) A Riemann surface $\Sigma$, possibly with boundaries, that is built by gluing three-holed spheres  along their boundaries.  Each boundary of one
of the three-holed spheres is either an external boundary -- a boundary of $\Sigma$ -- or an internal boundary, glued to a boundary of one of the three-holed
spheres (generically a different one).  The example shown has one external boundary and four internal ones.  \label{zurk}}
\end{figure}

First we consider the special case of a three-holed sphere (sometimes called a pair of pants; see fig. \ref{zurk}(a)).   
In the case of a three-holed sphere, for $G=SU(2)$,  $\sM_{0,\vec \alpha}$ 
is either a point, with volume 1, or an empty set, with volume 0, depending on $\vec\alpha$.  
The volumes of the three-holed sphere moduli spaces can also be computed (with a little more diffculty) for other compact $G$, but 
we will not explain the details as the case of $SU(2)$ will suffice
for illustration.

Now to generalize beyond the case of a three-holed sphere, we observe that any closed surface $\Sigma$ can be constructed by gluing together three-holed spheres along some
of their boundary components (fig. \ref{zurk}(b)).  If $\Sigma$
is built in this way, then the corresponding volume $\sM_{g,\vec \alpha}$ can be obtained by multiplying together
the volume functions of the individual three-holed spheres and integrating over the $\alpha$ parameters of the internal boundaries, where gluing occurs. (One also
has to integrate over some twist angles that enter in the gluing, but these give a trivial overall factor.)  Thus
for a compact group it is relatively straightforward to get  formulas 
 for the volumes $V_{g,\vec \alpha}$.  Moreover, these formulas turn out to be rather manageable.
 
 If we try to imitate this with $SU(2)$ replaced by $SL(2,\R)$, some of the steps work.  In particular, if $\Sigma$ is a three-holed sphere, then for any $\b$,
 the moduli space $\M_{0,\b}$ is a point and $V_{0,\b}=1$.  What really goes wrong for $SL(2,\R)$ is that, if $\Sigma$ is such that
 $\M_{0,\b}$ is not just a point, then the volume of the moduli space of flat $SL(2,\R)$
 connections on $\Sigma$ is infinite.    For $SU(2)$, the procedure mentioned in the last
 paragraph leads to an integral over the parameters $\vec \alpha$.  Those parameters are angular variables, valued in a compact set, and the integral over these
 parameters converges.  For $SL(2,\R)$ (in the particular case of the component of the moduli space of flat connections that is related to 
 hyperbolic metrics),
 we would want to replace the angular variables $\vec \alpha$ with the positive parameters $\vec b$.  The set of positive numbers is not compact and
 the integral over $\vec b$ is divergent.  
 
 This should not come as a surprise as it just reflects the fact that the group $SL(2,\R)$ is not compact.  The relation between  flat $SL(2,\R)$ connections and complex
 structures tells us what we have to do to get a sensible problem.  To go from (a component of) the moduli space of flat $SL(2,\R)$ connections to the moduli space of
 Riemann surfaces, we have to divide by the mapping class group of $\Sigma$ (the group of components of the group of diffeomorphisms of $\Sigma$).  It is the moduli space of Riemann surfaces that has a finite volume, not the moduli space of flat
 $SL(2,\R)$ connections. 
 
 But here is precisely where we run into difficulty with the cut and paste method to compute volumes.  Topologically, $\Sigma$ can be built by gluing three-holed
 spheres in many ways that are permuted by the action of the mapping class group.   Any one gluing procedure is not invariant under the mapping class group and in
 a calculation based  on any one gluing procedure, it is difficult to see how to divide by the mapping class group.
 
 Dealing with this problem, in a matter that we explain next, was the essence of Maryam Mirzakhani's approach to topological gravity.

\subsection{How Maryam Mirzakhani Cured Modular Invariance}\label{thecure}

\begin{figure}
 \begin{center}
   \includegraphics[width=6in]{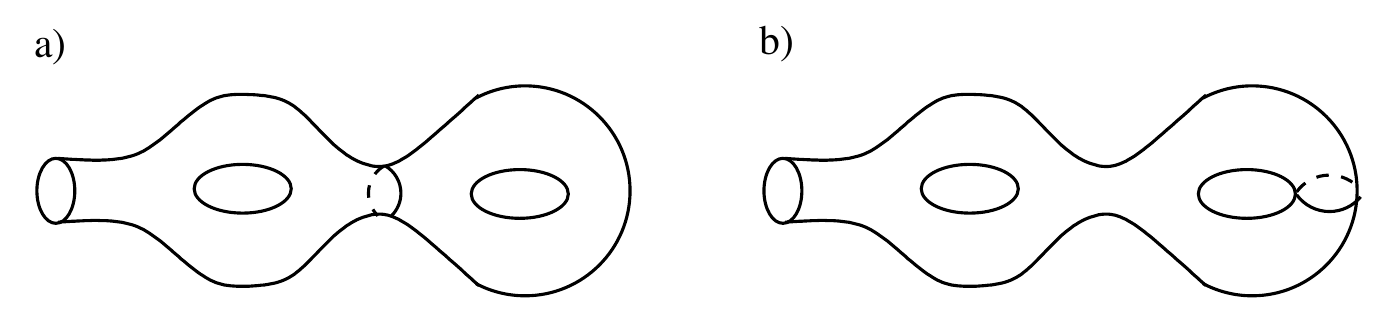}
 \end{center}  
\caption{\small A ``cut'' of a Riemann surface with boundary along an embedded circle may be separating as in (a) or non-separating as in (b).  \label{cut}}
\end{figure}
Let $\Sigma$ be a hyperbolic Riemann surface with geodesic boundary.  Ideally, to compute the volume of the corresponding
moduli space, we would ``cut'' $\Sigma$ on a simple closed geodesic $\ell$.   This cutting gives a way to build $\Sigma$ from
hyperbolic Riemann surfaces that are in some sense simpler than $\Sigma$.  If cutting along $\ell$ divides $\Sigma$ into two
disconnected components (fig. \ref{cut}(a)), then $\Sigma$ can be built by gluing along $\ell$ two hyperbolic Riemann surfaces
$\Sigma_1$ and $\Sigma_2$ of geodesic boundary.  If cutting along $\ell$ leaves $\Sigma$ connected (fig. \ref{cut}(b)), then $\Sigma$
is built by gluing together two boundary components of a surface $\Sigma'$.   We call these the separating and nonseparating cases.

In the separating case, we might naively hope to compute the volume function $V_{g,\vec b}$ for $\Sigma$ by multiplying together the corresponding
functions for $\Sigma_1$ and $\Sigma_2$ and integrating over the circumference $b$ of $\ell$.  Schematically,
\be\label{yonk}V_\Sigma \overset{?}{=} \int_0^\infty \d b \,V_{\Sigma_1,b}V_{\Sigma_2,b}, \ee
where we indicate that $\Sigma_1$ and $\Sigma_2$ each has one boundary component, of circumference $b$, that does not appear in
$\Sigma$.
In the nonseparating case, a similarly naive formula would be
\be\label{onk}V_\Sigma\overset{?}{=}\int_0^\infty \d b\, V_{\Sigma',b,b}, \ee
where we indicate that $\Sigma'$, relative to $\Sigma$,  has two extra boundary components each of circumference $b$.

The surfaces $\Sigma_1$, $\Sigma_2$, and $\Sigma'$ are in a precise sense ``simpler'' than $\Sigma$: their genus is less, or their
Euler characteristic is less negative.  So if we had something like (\ref{yonk}) or (\ref{onk}), a simple induction would lead to a general
formula for the volume functions.

The trouble with these formulas is that a hyperbolic Riemann surface actually has infinitely many simple 
closed geodesics $\ell_\alpha$, and there is
no natural (modular-invariant) way to pick one.  Suppose, however, that there were a function $F(b)$ of a positive real number $b$ with
the property that
\be\label{onz}\sum_\alpha F(b_\alpha)=1,\ee
where the sum runs over all simple closed geodesics $\ell_\alpha$ on a hyperbolic surface $\Sigma$, and $b_\alpha$ is the length of
$\ell_\alpha$.   In this case, by summing over all choices of embedded simple closed geodesic, and weighting each with a factor of $F(b)$, we would get a corrected version of the
above formulas.   In writing the formula, we have to remember that cutting along a given $\ell_\alpha$ either leaves $\Sigma$
connected or separates a genus $g$ surface
$\Sigma$ into surfaces $\Sigma_1,\Sigma_2$ of genera $g_1,g_2$ such that $g_1+g_2=g$.   In the separating case, the boundaries
of $\Sigma$ are partitioned  in some arbitrary way between $\Sigma_1$ and $\Sigma_2$ and each of $\Sigma_1,\Sigma_2$
has in addition one more boundary component whose circumference we will call $b'$.  So denoting as $\vec b$
the boundary lengths of $\Sigma$, the boundary lengths of $\Sigma_1$ and $\Sigma_2$ are respectively $\vec b_1,b'$
and $\vec b_2, b'$, where $\vec b=\vec b_1\sqcup \vec b_2$ (here $\vec b_1\sqcup\vec b_2$ denotes the disjoint union of two sets $\vec b_1$ and $\vec b_2$) and $\Sigma$
is built by gluing together $\Sigma_1$ and $\Sigma_2$ along their boundaries of length $b'$.  This is drawn in fig. \ref{cut}(a), but in the example shown, the set $\vec b$
consists of only one element.
In the nonseparating case of fig. \ref{cut}(b), $\Sigma$ is
made from gluing  a surface $\Sigma'$ of boundary lengths $\vec b,b',b'$ along its two boundaries of length $b'$.  
The genus $g'$ of $\Sigma'$ is $g'=g-1$.
Assuming the hypothetical sum rule
(\ref{onz}) involves a sum over all simple closed geodesics $\ell_\alpha$, regardless of topological properties,
 the resulting recursion relation for the volumes will also involve such a sum.
This recursion relation would be
\be\label{wonk}V_{g,\vec b}\overset{?}{=}\frac{1}{2}\sum_{g_1,g_2|g=g_1+g_2}\;\sum_{\vec b_1,\vec b_2|\vec b_1\sqcup\vec b_2=\vec b}
\int_0^\infty \d b' \,F(b')V_{g_1,\vec b_1,b'}V_{g_2,\vec b_2,b'}+\int_0^\infty \d b' \,F(b')V_{g-1,\vec b,b',b'}. \ee
In the first term, the sum runs over all topological choices in the gluing; the factor of $1/2$ reflects the possibility of exchanging
$\Sigma_1$ and $\Sigma_2$. The factors of $F(b')$ in the formula compensate for the fact that in deriving such a result, one has to sum over
cuts on all simple closed geodesics.   By induction (in the genus and the absolute value of the Euler characteristic of a surface), such a recursion relation would lead to explicit expressions for all $V_{g,\vec b}$.

\begin{figure}
 \begin{center}
   \includegraphics[width=4in]{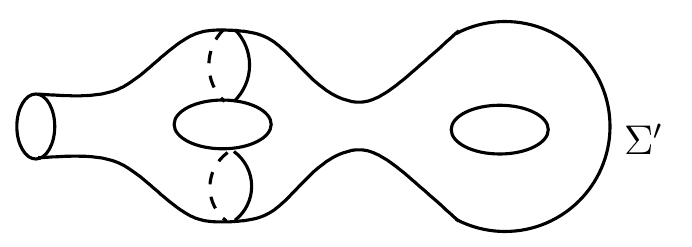}
 \end{center}  
\caption{\small Building a hyperbolic surface $\Sigma$  by gluing a hyperbolic pair of pants with geodesic boundary  onto a simpler hyperbolic surface $\Sigma'$. $\Sigma$ and $\Sigma'$ both have geodesic boundary.  (Shown here is the case that $\Sigma'$ is connected.)  \label{pair}}
\end{figure}

There is an important special case in which there actually is a sum rule \cite{McS}  precisely along the lines of eqn. (\ref{onz}) and therefore
there is an identity precisely along the lines of 
eqn. (\ref{wonk}).  This is the case that $\Sigma$ is a surface of genus 1 with one boundary component.  

The general case is more complicated.
In general, there is an  identity that involves pairs of simple closed geodesics in $\Sigma$ that have
the property that -- together with a specified boundary component of $\Sigma$ -- they bound a pair of pants (fig. \ref{pair}).  This identity was
proved for hyperbolic Riemann surfaces with punctures by McShane in \cite{McS} and generalized to surfaces with boundary by Mirzakhani
in \cite{M1}, Theorem 4.2.    

This generalized McShane identity leads to a recursive formula for Weil-Petersson volumes that is similar in spirit to eqn. (\ref{wonk}).
See Theorem 8.1 of Mirzakhani's paper \cite{M1} for the precise statement.   The main difference between the naive formula (\ref{wonk})
and the formula that actually works is the following.  In eqn. (\ref{wonk}), we imagine building $\Sigma$ from simpler building blocks
by a more or less arbitrary gluing.  In the correct formula -- Mirzakhani's Theorem 8.1 -- we more specifically build $\Sigma$ by gluing
a pair of pants onto something simpler, as in fig. \ref{pair}.  There is a function $F(b,b',b'')$, analogous to $F(b')$ in the above schematic
discussion, that enters in the generalized McShane
identity and therefore in the recursion relation.  It compensates for the fact that in deriving the recursion relation, one has to sum over infinitely many ways to cut off
a hyperbolic pair of pants from $\Sigma$.

In this manner, Mirzakhani arrived at a recursive formula for Weil-Petersson volumes that is similar to although somewhat more complicated
than eqn. (\ref{wonk}).  Part of the beauty of the subject is that this formula turned out to be surprisingly 
tractable.  In \cite{M1}, section 6, she used
the recursive formula to give a new proof -- independent of the relation to topological gravity that we reviewed in section \ref{initial} --
that the volume functions $V_{g,\vec b}$ are polynomials in $b_1^2,\dots,b_n^2$.  In \cite{M2}, she showed that these polynomials
satisfy the Virasoro constraints of two-dimensional gravity, as formulated for the matrix model in \cite{DVV}.  Thereby -- using
the relation between volumes and intersection numbers that we reviewed in section \ref{initial} and to which we will return in a moment -- she gave a new proof of the known
formulas  \cite{Witten,K} for intersection numbers on the moduli space of Riemann surfaces, or equivalently for correlation functions
of two-dimensional topological gravity.

\subsection{Volumes And Intersection Numbers}\label{volint}

We conclude this section by briefly describing the formula that relates  Weil-Petersson volumes to correlation functions of topological gravity.

Given a surface $\Sigma$ with $n+1$ marked points,
there is a forgetful map $\pi:\bM_{g,n+1}\to \bM_{g,n}$ that forgets one of the marked points $p$.  If we insert the class $\tau_{d+1}$ at $p$
and integrate over the fiber of $\pi$, we get the Miller-Morita-Mumford class $\kappa_d=\pi_*(\tau_{d+1})$, which is a class of degree $2d$ in the cohomology of $\bM_{g,n}$.  

As a first step in evaluating a correlation function $\langle \tau_{d+1}\prod_{j=1}^k \tau_{n_j}\rangle$, one might to try to integrate over the choice of the
point at which $\tau_{d+1}$ is inserted.   Integrating over the fiber of $\pi:\bM_{g,n+1}\to \bM_g$, one might hope to get a formula
\be\label{mortz}\left\langle \tau_{d+1}\prod_{j=1}^k \tau_{n_j}\right\rangle\overset{?}{=}\left\langle \kappa_d\prod_{j=1}^k \tau_{n_j}\right\rangle. \ee 
This is not true, however.  The right version of the formula has corrections that involve contact terms in which $\tau_{d+1}$ collides with $\tau_{n_j}$ for
some $j$.  Such a collision generates a correction that is an insertion of $\tau_{d+n_j}$.    For a fuller explanation, see \cite{Witten}.   Taking account of the contact
terms, one can express correlation functions of the $\tau$'s in terms of those of the $\kappa$'s, and vice-versa.

A special case is the computation of volumes.
As before, we write just $\kappa$ for $\kappa_1$, and
we  define the volume of $\bM_g$ as  $\int_{\bM_g} \kappa^{3g-3}/(3g-3)!$.   This can be expressed in terms of correlation functions of the $\tau$'s, but one has to take the contact terms into account.

As an example, we consider the case of a closed surface of genus 2.
 The volume of the compactified moduli space $\bM_2$ is
\be\label{mudd}V_2=\frac{1}{3!}\langle \kappa \kappa \kappa\rangle, \ee
and we want to compare this to topological gravity correlation functions such as
\be\label{udd}\frac{1}{3!}\langle \tau_2\tau_2\tau_2\rangle. \ee
By integrating over the position of one puncture, we can replace one copy of $\tau_2$ with $\kappa$, while also generating contact terms.  In such a contact term, $\tau_2$ collides with some $\tau_s$, $s\geq 0$, to generate
a contact term $\tau_{s+1}$.   Thus for example
\be\label{pudd}\langle \tau_2 \tau_2\tau_2\rangle =\langle \kappa \tau_2\tau_2\rangle +2\langle \tau_3\tau_2\rangle, \ee
where the factor of 2 reflects the fact that the first $\tau_2$ may collide with either of the two other $\tau_2$ insertions to generate a $\tau_3$.
The same process applies if factors of $\kappa$ are already present; they do not generate additional contact terms.
For example,
\be\label{zudd}\langle \kappa\tau_2\tau_2\rangle =\langle \kappa\kappa \tau_2\rangle+\langle \kappa \tau_3\rangle
=\langle \kappa\kappa\kappa\rangle +\langle \kappa\tau_3\rangle. \ee
Similarly
\be\label{pludd}\langle \tau_2\tau_3\rangle =\langle \kappa\tau_3\rangle +\langle \tau_4\rangle. \ee

Taking linear combinations of these formulas, we learn finally that 
\be\label{dudd}\langle \kappa\kappa\kappa\rangle = \langle \tau_2\tau_2\tau_2\rangle -3\langle \tau_2\tau_3\rangle +\langle \tau_4\rangle.\ee
This is equivalent to saying that $V_2$, which is  the term of order $\xi^3$ in
\be\label{wudd}\left\langle \exp(\xi\kappa)\right\rangle,\ee
is equally well the term of order $\xi^3$ in
\be\label{opludd}\left\langle \exp\left(\xi\tau_2 -\frac{\xi^2}{2!}\tau_3+\frac{\xi^3}{3!}\tau_4\right)\right\rangle. \ee

The generalization of this for higher genus is that
\be\label{gudd}\biggl\langle \exp\left(\xi\kappa\right)\biggr\rangle =\left\langle\exp\left (\sum_{k=2}^\infty \frac{(-1)^k\xi^{k-1}}{(k-1)!}\tau_k\right)
\right\rangle. \ee
The volume of $\bM_g$ is the coefficient of  $\xi^{3g-3}$ in the expansion of either of these formulas. 
To prove eqn. (\ref{gudd}), we write $W(\xi)$ for the right hand side, and we compute that
\be\label{ludd}\frac{\d}{\d \xi}W(\xi)=\left\langle \left(\tau_2+\sum_{r=3}^\infty(-1)^r\frac{\xi^{r-2}}{(r-2)!}\tau_r\right) \exp\left (\sum_{k=2}^\infty \frac{(-1)^k\xi^{k-1}}{(k-1)!}\tau_k\right)
\right\rangle. \ee
Next, one replaces the explicit $\tau_2$ term in the parentheses
on the right hand side with $\kappa$ plus a sum of contact terms between $\tau_2$ and the $\tau_k$'s that appear in the
exponential.  These contact terms cancel the $\tau_r$'s inside the parentheses, and one finds
\be\label{uudd} \frac{\d}{\d \xi}W(\xi) = \left\langle \kappa \,\exp\left (\sum_{k=2}^\infty \frac{(-1)^k\xi^{k-1}}{(k-1)!}\tau_k\right)
\right\rangle. \ee
Repeating this process gives for all $s\geq 0$
\be\label{zuudd} \frac{\d^s}{\d \xi^s}W(\xi) = \left\langle \kappa^s \,\exp\left (\sum_{k=2}^\infty \frac{(-1)^k\xi^{k-1}}{(k-1)!}\tau_k\right)
\right\rangle. \ee
Setting $\xi=0$, we get
\be\label{pluudd}\left.\frac{\d^s}{\d \xi^s}W(\xi)\right|_{\xi=0}=\left\langle \kappa^s \right\rangle, \ee
and the fact that this is true for all $s\geq 0$ is equivalent to eqn. (\ref{gudd}).

Eqn. (\ref{gudd}) has been deduced by comparing matrix model formulas to Mirzakhani's formulas for the volumes \cite{Eynard}. We will return to this when we discuss the spectral curve in section \ref{doublescaling}.
For algebro-geometric approaches and generalizations see \cite{KMZ,MZ}.  It is also possible to obtain similar formulas for the volume of $\bM_{g,\vec b}$.

\section{Open Topological Gravity}\label{opentot}

\subsection{Preliminaries}\label{prelims}

In this section, we provide an introduction to recent work \cite{PST,Tes,BT,STa} on topological gravity on open Riemann surfaces,
that is, on oriented two-manifolds with boundary.

From the point of view of matrix models of two-dimensional gravity, one should expect an interesting theory of this sort to exist because adding
vector degrees of freedom to a matrix model of two-dimensional gravity gives a potential model of two-manifolds with boundary.\footnote{Similarly,
by replacing the usual symmetry group $U(N)$ of the matrix model with $O(N)$ or $Sp(N)$, one can make a model associated to gravity on
unoriented (and possibly unorientable) two-manifolds.   It is not yet understood if this is related to some sort of topological field theory.}  We will discuss matrix models with vector degrees of freedom in section \ref{intmat}.   Here,
however, we discuss the topological field theory side of the story.

Let $\Sigma$ be a Riemann surface with boundary, and in general with marked points or punctures both in bulk and on the boundary.
Its complex structure gives $\Sigma$ a natural orientation, and this induces orientations on all boundary components.
  If
$p$ is a bulk puncture, then the cotangent space to $p$ in $\Sigma$ is a complex line bundle $\L$, and as we reviewed
in section \ref{initial}, one defines for every integer $d\geq 0$ the cohomology class $\tau_d=\psi^d$ of degree $2d$.  The operator $\tau_0=1$
is associated to a bulk puncture, and the $\tau_d$ with $d>0$ are called gravitational descendants.

A boundary puncture has no analogous gravitational descendants, because if $p$ is a boundary point
in $\Sigma$, the tangent bundle to $p$ in $\Sigma$  is naturally trivial.  It has a natural real
subbundle given by the tangent space to $p$ in $\partial\Sigma$,
and this subbundle is actually trivialized (up to homotopy) by the orientation of $\partial\Sigma$.
So  $c_1(\L)=0$ if $p$ is a boundary puncture.

Thus the list of observables in 2d topological gravity on a Riemann surface with boundary consists of the usual bulk observables $\tau_d$, $d\geq 0$,
and one more boundary observable $\sp$, corresponding to a boundary puncture.
Formally, the sort of thing one hopes to calculate for a Riemann surface $\Sigma$ with $n$ bulk punctures and $m$ boundary punctures is
\be\label{proft}\langle \tau_{d_1}\tau_{d_2}\cdots \tau_{d_n} \sp^m\rangle_\Sigma =\int_{\overline\M} \psi_1^{d_1}\psi_2^{d_2}\cdots \psi_n^{d_n},\ee
where $\overline\M$ is the (compactified) moduli space of conformal structures on $\Sigma$ with $n$ bulk punctures and $m$ boundary punctures.  The $d_i$
are arbitrary nonnegative integers, and we note that the cohomology class  $\prod_{i=1}^n\psi_i^{d_i}$ that is integrated over $\overline\M$ is generated only from data at the bulk
punctures (and in fact only from those bulk punctures with $d_i>0$).  The  boundary punctures (and those bulk punctures with $d_i=0$)
participate in the construction only because they enter the definition of $\overline\M$, the space on which the cohomology class in question
is supposed to be integrated.   Similarly to the case of a Riemann surface without boundary, to make the integral (\ref{proft}) nonzero,
$\Sigma$
must be chosen topologically so that the dimension of $\overline\M$ is the same as the degree of the cohomology class that we want to integrate:
\be\label{loft} \mathrm{dim}\,\overline\M=\sum_{i=1}^n 2d_i. \ee

Assuming that we can make sense of the definition in eqn. (\ref{proft}), 
the (unnormalized) correlation function $\langle \tau_{d_1}\tau_{d_2}\cdots \tau_{d_n} \sp^m\rangle$  of 2d gravity on Riemann surfaces
with boundary is then obtained by summing 
$\langle \tau_{d_1}\tau_{d_2}\cdots \tau_{d_n} \sp^m\rangle_\Sigma$ over all topological choices of $\Sigma$.  (If $\Sigma$ has more than one
boundary component, the sum over $\Sigma$ includes a sum over how the boundary punctures are distributed among those boundary components.)
It is also possible to slightly generalize the definition by weighting a surface $\Sigma$ in a way that depends on the number of its boundary
components.  For this, we introduce a parameter $w$, and weight a surface with $h$ boundary components with a factor of\footnote{Still
more generally, we could introduce a finite set $S$ of ``labels'' for the boundaries, so that each boundary is labeled by some
$s\in S$.  Then for each $s\in S$, one would have a boundary observable $\sigma_s$ corresponding to a puncture inserted on a boundary
with label $s$, and a corresponding parameter $v_s$ to count such punctures.  Eqn. (\ref{zoft}) below corresponds to the case that $w$
is the cardinality of the set $S$, and $v_s=v$ for all $s\in S$.  
  This generalization to include labels would correspond in eqn. (\ref{loboxx}) below to modifying the matrix integral with
a factor $\prod_{s\in S} \det(z_s-\Phi)$.  Similarly, one could replace $w^h$ by $\prod_{s\in S} w_s^{h_s}$, where $h_s$ is the number
of boundary components labeled by $s$ and there is a separate  parameter $w_s$ for each $s$.   This corresponds to including
in the matrix
integral a factor $\prod_{s\in S} (\det(z_s-\Phi))^{w_s}.$   Such generalizations have been treated in \cite{ABT}.}  $w^h$.

Introducing also the usual coupling parameters $t_i$ associated to the bulk observables $\tau_i$, and one more parameter $\v$ associated to $\sp$,
the partition function of 2d topological gravity on a Riemann surface with boundary is then formally
\be\label{zoft}Z(t_i;v,w)=\sum_{h=0}^\infty \sum_\Sigma w^h \left\langle \exp\left(\sum_{i=0}^\infty t_i \tau_i + \v \sp\right)\right\rangle_\Sigma. \ee
The sum over $\Sigma$ runs over all topological types of Riemann surface with $h$ boundary components, and specified bulk and boundary
punctures.  The exponential on the right hand side is expanded in a power series, and the monomials of an appropriate degree are then
evaluated via eqn. (\ref{proft}).  

There are two immediate difficulties with this formal definition:

(1) To integrate a cohomology class such as $\prod_i \psi_i^{d_i}$ over a manifold $M$, that manifold must be oriented.
But the moduli space of  Riemann surfaces with boundary is actually unorientable.

(2) For the integral of a cohomology class over an oriented manifold $M$ to be well-defined topologically, $M$ should have no
boundary, or the cohomology class in question should be trivialized along the boundary of $M$.  However, the compactified moduli space $\bar\M$ of
conformal structures on a Riemann surface with boundary is itself in general a manifold with boundary.

Dealing with these issues  requires some refinements
of the above formal definition  \cite{PST,Tes,BT,STa}.   The rest of this section is devoted to an introduction.
We begin with the unorientability of $\overline \M$.  Implications of the fact that $\overline \M$ is a manifold with boundary will be discussed in section
\ref{bd}.

\subsection{The Anomaly}\label{anomaly}

The problem in orienting the moduli space of Riemann surfaces with boundary can be seen most directly in the absence of boundary punctures.  Thus
we let $\Sigma$ be a Riemann surface of genus $g$ with $h$ holes or boundary components and no boundary punctures, but possibly containing
bulk punctures.

First of all, if $h=0$, then $\Sigma$ is an ordinary closed Riemann surface, possibly with punctures.  
The compactified moduli space $\bM$ of conformal structures on $\Sigma$ is
then a complex manifold (or more precisely an orbifold) and by consequence has a natural orientation.  This orientation is used in defining the usual intersection numbers on
$\bM$, that is, the correlation functions of 2d topological gravity on a Riemann surface without boundary.

This remains true if $\Sigma$ has punctures (which automatically are bulk punctures since so far $\Sigma$ has no boundary).  Now let us
replace some of the punctures of $\Sigma$ by holes.  Each time we replace a bulk puncture by a hole, we add one real modulus to the moduli
space.  If we view $\Sigma$ as a two-manifold with hyperbolic metric and geodesic boundaries, then the extra modulus is the circumference $b$
around the hole.

By adding $h>1$ holes, we add $h$ real moduli $b_1,b_2,\cdots,b_h$, which we write collectively as $\vec b$.
We denote the corresponding compactified moduli space as $\bM_{g,n,\vec b}$ (here $n$
is the number of punctures that have not been converted to holes).   A very important detail is the following. In defining Weil-Petersson volumes
in section \ref{curing}, we treated the $b_i$ as arbitrary constants; the ``volume'' was defined as the volume for fixed $\vec b$,
without trying to integrate over $\vec b$.  (Such an integral would have been divergent since the volume function  $V_{g,n,\vec b}$ 
is polynomial in $\vec b$. Moreover, what naturally enters Mirzakhani's recursion relation is the volume function defined for fixed $\vec b$.)
In defining two-dimensional gravity on a Riemann surface with boundary, the $b_i$ are treated as full-fledged moduli -- they are some of the
moduli that one integrates over in defining the intersection numbers.  Hopefully this change in viewpoint relative to section \ref{curing} will
not cause serious confusion.

If we suppress the $b_i$ by setting them all to 0, the holes turn back into punctures and $\bM_{g,n,\vec b}$ is replaced by $\bM_{g,n+h}$.
This is a complex manifold (or rather an orbifold) 
and in particular has a natural orientation.  Restoring the $b_i$, $\M_{g,n,\vec b}$ is a fiber bundle\footnote{This
assertion actually follows from  a fact that was exploited in section \ref{curing}:  for fixed $\vec b$, $\bM_{g,n,\vec b}$ is isomorphic
to $\bM_{g,n}$ as an orbifold (their symplectic structures are inequivalent, as we discussed in section \ref{curing}).   Given this equivalence
for fixed $\vec b$, it follows that upon letting $\vec b$ vary, $\bM_{g,n,\vec b}$ is a fiber bundle over $\bM_{g,n+h}$ with fiber
paramerized by $\vec b$.}  over
$\M_{g,n+h}$ with fiber a copy of $\R^h_+$ parametrized by $b_1,\dots,b_h$. (Here $\R_+$ is the space of positive real numbers and
$\R_+^h$ is the Cartesian product of $h$ copies of $\R_+$.)  Orienting $\M_{g,n,\vec b}$ is equivalent to orienting this copy of 
$\R^h_+$.  

If we were given an ordering of the holes in $\Sigma$ up to an even permutation, we would orient $\R^h_+$ by the differential form
\be\label{difff}\Omega=\d b_1\d b_2\cdots \d b_h.\ee   However, for $h>1$, in the absence of any information about how the holes should be ordered, $\R^h_+$ has no natural
orientation.  

Thus $\bM_{g,n,\vec b}$ has no natural orientation for $h>1$.  In fact it is unorientable.  This follows from the fact that a Riemann surface
$\Sigma$ with more than one hole has a diffeomorphism that exchanges two of the holes, leaving the others fixed.  (Moreover, this diffeomorphism can be
chosen to preserve the orientation of $\Sigma$.)  Dividing by this diffeomorphism
in constructing the moduli space $\bM_{g,n,\vec b}$ ensures that this moduli space is actually unorientable.

We can view this as a global anomaly in two-dimensional topological 
gravity on an oriented two-manifold  with boundary.   The moduli space is not oriented, or even orientable,
so there is no way to make sense of  the correlation functions that one wishes to define.

As usual, to cancel the anomaly, one can try to couple two-dimensional topological gravity to some matter system that carries a compensating anomaly.
In the context of two-dimensional topological gravity, the matter system in question should be a topological field theory.  To define a theory that
reduces to the usual topological gravity when the boundary of $\Sigma$ is empty, we need a topological field theory that on a Riemann surface
without boundary is essentially trivial, in 
a sense that we will see, and in particular is anomaly-free.   But the theory should 
become anomalous in the presence of boundaries.

These conditions may sound too strong, but there actually is a topological field theory with the right properties.  First of all, we endow $\Sigma$
with a spin structure.  (We will ultimately sum over spin structures to get a true topological field theory that does not depend on the choice of
an {\it a priori} spin structure on $\Sigma$.) 
When $\partial\Sigma=\varnothing$,  we can define a chiral Dirac operator on $\Sigma$ (a Dirac operator acting
on positive chirality spin 1/2 fields on $\Sigma$).   There is then a $\Z_2$-valued invariant
that we call $\zeta$, namely the mod 2 index of the chiral Dirac operator, in the sense of Atiyah and Singer \cite{AS,Aspin}.  $\zeta$ is defined as the number of zero-modes of the chiral
Dirac operator, reduced mod 2.  $\zeta$ is a topological invariant in that it does not depend on the choice of a conformal structure (or metric) on $\Sigma$.
A spin structure is said to be even or odd if the number of chiral zero-modes is even or odd (in other words if $\zeta=0$ or $\zeta=1$).
For an introduction to these matters, see \cite{WittenFP}, especially section 3.2. 

We define a topological field theory by summing over spin structures on $\Sigma$ with each spin structure weighted by a factor of $\frac{1}{2}(-1)^\zeta$.
The reason for the factor of $\frac{1}{2}$ is that a spin structure has a symmetry group that acts on fermions as $\pm 1$, with 2 elements.   As in Fadde'ev-Popov
gauge-fixing in gauge theory, to define a topological field theory, one needs to divide by the order of the unbroken symmetry group, which in this
case is the group $\Z_2$.  This accounts for the factor of $\frac{1}{2}$.  The more interesting factor, which will lead to a boundary anomaly, is $(-1)^\zeta$.
It may not be immediately apparent that we can define a topological field theory with this factor included.  We will describe two realizations of the theory in
question in section \ref{twor}, making it clear that there is such a topological field theory.   We will call it $\T$.

On a Riemann surface of genus $g$, there are $\frac{1}{2}(2^{2g}+2^g)$ even spin structures and $\frac{1}{2}(2^{2g}-2^g)$ odd ones.
The partition function of $\T$ in genus $g$ is thus
\be\label{onmo} Z_g=\frac{1}{2}\left(\left(\frac{1}{2}(2^{2g}+2^g)\right)-\left(\frac{1}{2}(2^{2g}-2^g)\right) \right)=2^{g-1}. \ee
This is not equal to 1, and thus the topological field theory $\T$  is nontrivial.  However, when we couple to topological gravity,
the genus $g$ amplitude has a factor $\sg_{\mathrm{st}}^{2g-2}$, where $\sg_{\mathrm{st}}$ is the string coupling constant.\footnote{In mathematical treatments, $\gst$ is often set to 1.  There is no essential loss of information, as the dependence on $\gst$ carries the
same information as the dependence on the parameter $t_1$ in the generating function.  This follows from the dilaton equation,
that is, the $L_0$ Virasoro constraint.}
The product of this with $Z_g$ is $(2\sg_{\mathrm {st}}^2)^{g-1}$. Thus,  as long as we are on a Riemann surface without boundary, coupling
topological gravity to $\T$ can be compensated\footnote{This point was actually made in 
\cite{Wittenveryold}, as
a special case of a more general discussion involving $r^{th}$ roots of the canonical bundle of $\Sigma$ for arbitrary $r\geq 2$.} by absorbing a factor of $\sqrt 2$ in the definition of $\sg_{\mathrm{st}}$.   In that sense, coupling of topological gravity to $\T$ has no effect, as long as we consider
only closed Riemann surfaces.

Matters are different if $\Sigma$ has a boundary.   On a Riemann surface with boundary, it is not possible
to define  a local boundary condition for the chiral Dirac operator that is complex linear and sensible (elliptic), and there is no topological invariant corresponding to $\zeta$.
Thus  theory $\T$ cannot be defined as a topological field theory on a manifold with boundary.

It is possible to define theory $\T$ on a manifold with boundary as a sort of anomalous topological field theory, with an anomaly that will help compensate
for the problem that we found above with the orientation of the moduli space.  To explain this, we will first describe some more physical constructions
of theory $\T$.  First we discuss how  theory $\T$ is related to contemporary topics in condensed matter physics.

\subsection{Relation To Condensed Matter Physics}\label{cm}

Theory $\T$ has a close cousin that is familiar in condensed matter physics.   One considers a chain of fermions in $1+1$ dimensions with the property
that in the bulk of the chain there is an energy gap to the first excited state above the ground state, and the further requirement that the chain is  in an ``invertible'' phase, meaning that the tensor product of a suitable number of
 identical chains would be completely
trivial.\footnote{Triviality here means that by deforming the Hamiltonian without losing the gap in the bulk spectrum, one can reach a Hamiltonian whose
ground state is the tensor product of local wavefunctions, one on each lattice site.}  There are two such phases, just
one of which is nontrivial.  The nontrivial phase is called the Kitaev spin chain \cite{Kitaev}. It is characterized by the fact that at the end of an extremely long chain, there is an unpaired Majorana fermion
mode, usually called a zero-mode because in the limit of a long chain, it commutes with 
the  Hamiltonian.\footnote{A long but finite chain has a pair of such Majorana modes, one at each end.  Upon quantization, 
they generate a rank two Clifford
algebra, whose irreducible representation is two-dimensional.  As a result, a long chain is exponentially close to having a two-fold 
degenerate ground state.  In condensed matter physics, this degeneracy is broken by tunneling effects in which a fermion propagates between
the two ends of the chain. In the idealized model considered below, the degeneracy is exact.} 

The Kitaev spin chain is naturally studied in condensed matter physics from a Hamiltonian point of view, which in fact we adopted in the last paragraph.
From a relativistic point of view, the Kitaev spin chain corresponds to a topological field theory theory that is defined on an oriented two-dimensional spin manifold $\Sigma$,
and whose partition function if $\Sigma$ has no boundary is $(-1)^\zeta$.  We will see below how this statement relates to more standard characterizations of the Kitaev
spin chain.  Our theory $\T$ differs from the Kitaev spin chain simply in that we sum over spin structures in defining it, while the Kitaev model is a theory of fermions
and is defined on a two-manifold with a particular spin structure.   Moreover, as we discuss in detail below, when $\Sigma$ has a boundary, the appropriate boundary
conditions in the context of condensed matter physics are different from what they are in our application to two-dimensional gravity.   Despite these differences,
the comparison between the two problems will be illuminating.

Because we are here studying two-dimensional gravity on an oriented two-manifold $\Sigma$, time-reversal symmetry, which corresponds to a diffeomorphism that
reverses the orientation of $\Sigma$, will not play any role.  The Kitaev spin chain has an interesting time-reversal symmetric refinement, but this will not be relevant.
 
Theory $\T$ has another interesting relation to condensed matter physics: it is associated to the high temperature phase of the two-dimensional Ising model.   In this interpretation \cite{KS}, the triviality of theory $\T$ corresponds to the fact that the Ising model
in its high temperature phase has only one equilibrium state, which moreover is gapped.

\subsection{Two Realizations Of  Theory $\T$}\label{twor}

We will describe two realizations of theory $\T$, one in the spirit of condensed matter physics, where we get a topological field theory as the
low energy limit of a physical gapped system, and one in the spirit of topological sigma 
models \cite{Wittentopsigma}, where a supersymmetric theory is twisted to
get a topological field theory.   

First we consider a massive Majorana fermion in two spacetime dimensions. It is convenient to work in Euclidean signature.
  One can choose the Dirac operator to be
\be\label{chd}\D_m=\ga^1 D_1+\ga^2 D_2+m \bg, \ee
where one can choose the gamma matrices to be real and symmetric, for instance
\be\label{hd}\ga^1=\begin{pmatrix} 0&1\cr 1&0 \end{pmatrix},~~~ \ga^2=\begin{pmatrix}1&0 \cr 0&-1\end{pmatrix}. \ee
This ensures that $\bg=\ga^1\ga^2$ is real and antisymmetric:
\be\label{pd}\bg=\begin{pmatrix}0&-1\cr 1&0\end{pmatrix}. \ee
These choices  ensure that the Dirac operator $\D_m$ is real and antisymmetric.  We call $m$ the mass parameter; the mass
of the fermion is actually $|m|$.

Formally, the path integral for a single Majorana fermion is $\Pf(\D_m)$, the Pfaffian of the real antisymmetric operator $\D_m$.  The Pfaffian of
a real antisymmetric operator is real, and its square is the determinant; in the present context, the determinant $\det\,\D_m=(\Pf(\D_m))^2$ can
be defined satisfactorily by, for example, zeta-function regularization.  However, the sign of the Pfaffian is subtle.  For a finite-dimensional
real antisymmetric matrix $M$, the sign of the Pfaffian depends on an orientation of the space that $M$ acts on.    In the case of the infinite-dimensional
matrix $\D_m$, no such natural orientation presents itself and therefore, for a single Majorana fermion, there is no natural choice of the 
sign of $\Pf(\D_m)$.

Suppose, however, that we consider a {\it pair} of Majorana fermions with the same (nonzero) mass parameter 
$m$.  Then the path integral is $\Pf(\D_m)^2$ and
(since $\Pf(\D_m)$ is naturally real) this is real and positive.  This actually ensures that the topological field theory obtained from the low energy
limit of a pair of massive Majorana fermions of the same mass parameter is completely trivial.  Without losing the mass gap, we can take $|m|\to\infty$,
and in that limit, $\Pf(\D_m)^2$ produces no physical effects at all except for a renormalization of some of the parameters in the effective action.\footnote{In two
dimensions, when we integrate out a massive neutral field, 
the only parameters that have to be renormalized are the vacuum energy, which corresponds to a term in the effective action proportional to the 
volume $\int_\Sigma\d^2x\sqrt g$ of $\Sigma$, and the coefficient of another term $\int_\Sigma\d^2x\sqrt g R$ proportional to the Euler characteristic of $\Sigma$ (here $R$ is
the Ricci scalar of $\Sigma$).}

To get theory $\T$, we consider instead a pair of Majorana fermions, one of positive mass parameter and one of negative mass parameter.  
Just varying mass parameters,
to interpolate between this theory and the equal mass parameter case, we would have to let a mass parameter pass through 0, losing the mass gap.

This suggests that a theory with opposite sign mass parameters might be in an essentially different phase from the trivial case of equal mass parameters.   
To establish
this and show the relation to theory $\T$, we will analyze what happens to the partition function of the theory when the mass parameter of a single
Majorana fermion is varied between positive and negative values.

The absolute value of $\Pf(\D_m)$ does not depend on the sign of $m$.  This follows from the fact that the operator $\D_m^2$ is invariant
under $m\to -m$.  (The determinant of $-\D_m^2$ is a power of $\Pf(\D_m)$, and the fact that it is invariant under $m\to -m$ implies
that $\Pf(\D_m)$ is independent of $\mathrm{sign}(m)$ up to sign.)   Therefore the partition functions of the two theories with opposite masses or with equal masses have the same absolute
value.    They can differ only in sign, and this sign is what we want to understand.

To determine the sign, we ask what happens to $\Pf(\D_m)$ when $m$ is varied from large positive values to large negative ones. To change
sign, $\Pf(\D_m)$ has to vanish, and it   vanishes only when $\D_m$ has a zero-mode.   This can only happen at $m=0$.   
So the question is just to determine
what happens to the sign of $\Pf(\D_m)$ when $m$ passes through 0. 

Zero-modes of $\D_m$ at $m=0$ are simply zero-modes of the massless Dirac operator $\D=\ga^1 D_1+\ga^2 D_2$.  Such modes appear
in pairs of equal and opposite chirality.  To be more precise, let $\h\ga=\i\bg$ be the chirality operator, with eigenvalues $1$ and $-1$ for fermions of
positive or negative chirality.  What we called
the chiral Dirac operator when we defined the mod 2 index in section \ref{anomaly} is the operator $\D$ restricted to act on states of $\h\ga=+1$.
Since $\h\ga$ is imaginary and $\D$ is real, 
complex conjugation of an eigenfunction reverses its $\h\gamma$ eigenvalue while commuting with $\D$; thus zero-modes of $\D$
occur in pairs with equal and opposite chirality.

Restricted to such a pair of zero-modes of $\D$, the antisymmetric operator $\D_m$ looks like
\be\label{nyt}\begin{pmatrix} 0 & -m \cr m & 0 \end{pmatrix},\ee
and its Pfaffian is $m$, up to an $m$-independent sign that depends on a choice of orientation in the two-dimensional space.  The important
point is that this Pfaffian changes sign when $m$ changes sign.  If there are $s$ pairs of zero-modes, the Pfaffian in the zero-mode space is
$m^s$, and so changes in sign by $(-1)^s$ when $m$ changes sign.  But $s$ is precisely the number of zero-modes of positive chirality, and $(-1)^s$
is the same as $(-1)^\zeta$, where $\zeta$ is the mod 2 index of the Dirac operator.  

Now we can answer the original question.  Since the partition function for the theory with two equal mass parameters is trivial (up to a renormalization
of some of the low energy parameters), the partition function of the theory with one mass of each sign is $(-1)^\zeta$ (up to such a renormalization).
Thus we have found a physical realization of theory $\T$.

The result we have found can be interpreted in terms of a discrete chiral anomaly.     At the classical level, for $m=0$, the Majorana fermion has
a $\Z_2$ chiral symmetry\footnote{With our conventions, the operator $\h\ga$ is imaginary in Euclidean signature and one might
wonder if this symmetry makes sense for a Majorana fermion.  However, after Wick rotation to Lorentz signature (in which $\ga^0$ acquires
a factor of $\i$), $\h\ga$ becomes real, and it is always in Lorentz
signature that reality conditions should be imposed on fermion fields and their symmetries.  Thus actually $\psi\to \h\ga\psi$ is a physically
meaningful symmetry and $\psi\to\bg\psi$ (which may look more natural in Euclidean signature) is not.   Under the latter transformation,
the massless Dirac action actually changes sign, so it is indeed $\psi\to\h\ga\psi$ and not $\psi\to\bg\psi$ that is a symmetry at the classical
level.}   $\psi\to \h\ga\psi$.   The mass parameter is odd under this symmetry, so classically the
theories with positive or negative $m$ are equivalent.  Quantum mechanically, one has to ask whether the fermion measure is invariant under
the discrete chiral symmetry.  As usual, the nonzero modes of the Dirac operator are paired up in such a way that the measure for those
modes is invariant; thus one only has to test the zero-modes.  Since $\psi\to \h\ga\psi$ leaves invariant the positive chirality zero-modes
and multiplies each negative chirality zero-mode by $-1$, this operation transforms the measure by a factor $(-1)^s=(-1)^\zeta$, where
$s$ is the number of negative (or positive) chirality zero-modes, and $\zeta$ is the mod 2 index. 

Finally, we will describe another though closely related way to construct the same topological field theory.  The extra machinery required will
be useful  later.  We consider in two dimensions a theory with $(2,2)$ supersymmetry and a single complex chiral superfield $\Phi$.
We work in flat spacetime to begin with and
assume a familiarity with the usual superspace formalism of $(2,2)$ supersymmetry and its twisting to make a topological field
theory.    $\Phi$ can be expanded
\be\label{zelp} \Phi=\phi+\theta_-\psi_+ +\theta_+\psi_-+\theta_+\theta_- F. \ee
Here $\phi$ is a complex scalar field; $\psi_+$ and $\psi_-$ are the chiral components of a Dirac fermion field; and $F$ is a complex auxiliary field.
We consider the action
\be\label{belp}S=\int \d^2x \d^4\theta \bar\Phi\Phi+\frac{\i}{2} \int \d^2x \d^2\theta \,m\Phi^2-\frac{\i}{2}\int\d^2x\d^2\bar\theta \,\bar m\bar\Phi^2. \ee
Thus the superpotential is $W(\Phi)={\i m}\Phi^2/2$.
In general, here $m$ is a complex mass parameter, but for our purposes  we can assume that $m>0$.  After integrating over the $\theta$'s
and integrating out the auxiliary field $F$, the action becomes\footnote{Here $\epsilon^{\alpha\beta}$ is the Levi-Civita antisymmetric
tensor in the two-dimensional space spanned by $\psi_+,\psi_-$.}
\be\label{nelp}S=\int\d^2x\left(\partial_\mu\bar\phi\partial^\mu\phi +m^2|\phi|^2+\bar\psi\gamma^\mu\partial_\mu\psi +\i m\epsilon^{\alpha\beta}
(\psi_\alpha\psi_\beta -\bar\psi_\alpha\bar\psi_\beta)\right). \ee 

If we expand the Dirac fermion $\psi$ in terms of a pair of Majorana fermions $\chi_1,\chi_2$ by $\psi=(\chi_1+\i\chi_2)/\sqrt 2$, we find that
$\chi_1$ and $\chi_2$ are massive Majorana fermions with  a mass matrix that has one positive and one negative eigenvalue, as in our previous
construction of theory $\T$.  The massive field $\phi$ does not
play an important role at low energies: its path integral is positive definite, and in the large $m$ or low energy limit, just contributes renormalization
effects.  So at low energies the supersymmetric theory considered here gives another realization of theory $\T$.  However, the supersymmetric
machinery gives a way to obtain theory $\T$ without taking a low energy limit, and this will be useful later.  Because the superpotential 
$W=\i m\Phi^2/2$ is homogeneous in $\Phi$, the theory has a $U(1)$ $R$-symmetry that acts on the superspace coordinates
as $\theta_\pm \to e^{\i\alpha}\theta_\pm$.   Because $W$ is quadratic in $\Phi$, one has to define this symmetry to leave $\psi$ invariant
and to transform $\phi$ by $\phi\to e^{\i\alpha}\phi$.  When one ``twists'' to make a topological field theory, the spin of a field is shifted by one-half
of its $R$-charge.  In the present case, as $\psi$ is invariant under the $R$-symmetry, it remains a Dirac fermion after twisting, but $\phi$ acquires
spin $+1/2$ (it transforms under rotations like the positive chirality part of a Dirac fermion).  

The twisted theory can be formulated as a topological field theory on any Riemann surface $\Sigma$ with any metric tensor.  We use the phrase
``topological field theory'' loosely since the twisted theory, as it has fields of spin 1/2, requires a choice of spin structure.  To get a true topological
field theory, one has to sum over the choice of spin structure.  The supersymmetry of the twisted theory ensures that the path integral
over $\phi$ cancels the absolute value of the path integral over $\psi$, leaving only the sign $(-1)^\zeta$.  Thus the twisted theory is precisely
equivalent to theory $\T$, without taking any low energy limit.

In \cite{Wittenveryold}, this last statement is deduced in another way
as a special case of an analysis of a theory with $\Phi^r$ superpotential for any $r\geq 2$.

For our later application, it will be useful to know that the condition for a configuration of the $\phi$ field to preserve the supersymmetry
of the twisted theory is
\be\label{milk}\bar\partial\phi+\i m\bar\phi=0. \ee
The generalization of this equation for arbitrary superpotential is \be\label{ggf}\bar\partial \phi+\frac{\partial \bar W}{\partial\bar\phi}=0,\ee  which has been called
the $\zeta$-instanton equation  \cite{GMW}.
 A small calculation shows that if we set $\phi=\phi_1+\i\phi_2$ with real $\phi_1,\phi_2$, and set $\h\phi=\begin{pmatrix}\phi_1\cr -\phi_2\end{pmatrix}$,
then eqn. (\ref{milk}) is equivalent to
\be\label{zilk}\D_m \h\phi= 0, \ee
with $\D_m$ the massive Dirac operator (\ref{chd}).

\subsection{Boundary Conditions In Theory $\T$}\label{bound}

Our next task is to consider theory $\T$ on a manifold with boundary.   Here of course we must begin by discussing possible boundary conditions.
In this section, we will use the realization of theory $\T$ in terms of a pair of Majorana fermions with opposite masses.  

The main requirement for a boundary condition is that it should preserve the antisymmetry of the operator $\D_m$.  If $\tr$ denotes the transpose,
then the antisymmetry means concretely that
\begin{equation}\label{zeff}\int\left( \chi^\tr \D_m\psi +(\D_m\chi)^\tr \psi\right)=0. \ee
In verifying this, one has to integrate by parts, and one encounters a surface term, which is the boundary integral of $\chi^\tr \gamma_\perp \psi$,
where $\gamma_\perp$ is the gamma matrix normal to the bondary.   This will vanish if we impose the boundary condition 
\be\label{bc}\left.\gamma_\parallel\psi\right|= \eta \psi,\ee 
where $\eta=+1$ or $-1$, 
 $\gamma_\parallel$ is the gamma matrix tangent to the boundary, and $\vert$ represents restriction to the boundary.  
Just to ensure the antisymmetry of the operator $\D_m$, either
choice of sign will do.   With either choice of sign, $\D_m$ is a real operator, so its Pfaffian $\Pf(\D_m)$ remains real.

The boundary conditions $\left.\gamma_\parallel\psi\right|=\pm\psi$ have a simple interpretation.  Tangent to the boundary, there is a single gamma matrix
$\gamma_\parallel$.  It generates a rank 1 Clifford algebra, satisfying $\gamma_\parallel^2=1$.  In an irreducible representation, it satisfies
$\gamma_\parallel=1$ or $\gamma_\parallel=-1$.    Thus the spin bundle of $\Sigma$, which is a real vector bundle of rank 2, decomposes along
$\partial\Sigma$ as the direct sum of two spin bundles of $\partial\Sigma$, namely the subbundles defined respectively by
$\gamma_\parallel\psi=\psi$ and by $\gamma_\parallel\psi=-\psi$.     These two spin bundles of $\partial\Sigma$ are isomorphic, since they
are exchanged by multipication by $\gamma_\perp,$ which is globally-defined along $\partial\Sigma$.  Thus the spin bundle 
of $\Sigma$ decomposes
along $\partial\Sigma$ in a natural way as the direct sum of two copies of the spin bundle of $\partial\Sigma$, and the boundary condition
says that along the boundary, $\psi$ takes values in one of these bundles.   We will write $\S$ for the spin bundle of $\Sigma$ and
$\E$ for the spin bundle of $\partial\Sigma$.

Now let us discuss the behavior near the boundary of a Majorana fermion that satisfies one of these boundary conditions.
We work on a half-space in $\R^2$,
say  the half-space $x_1\geq 0$ in the 
$x_1x_2$ plane.  For a mode that is independent of $x_2$, the Dirac equation $\D_m\psi=0$ becomes 
\be\label{polf}\left(\frac{\d}{\d x_1}+m\gamma_2\right)\psi=0,\ee
with solution
\be\label{nolf} \psi=\exp(-mx_1\gamma_2)\psi_0,\ee
for some $\psi_0$.   In this geometry, $\ga_2$ is the same as $\ga_\parallel$.
We see that if $\psi$ satisfies the boundary condition $\left.\gamma_{\parallel}\psi\right| =\eta\psi$, then this mode
is normalizable if and only if
\be\label{pollf} m\eta >0.  \ee

If $m\eta<0$, the theory remains gapped, with a gap of order $m$, even along the boundary.  But if $m\eta>0$, the mode that we have just
found propagates along the boundary as a $0+1$-dimensional massless Majorana fermion.

We will now use these results to study the boundary anomaly of theory $\T$, with several possible boundary conditions.

\subsection{Boundary Anomaly Of Theory $\T$}\label{boundanom}

Let us first recall that for a real fermion field with a real antisymmetric Dirac operator such as $\D_m$, in general there is an anomaly
in the sign of the path integral $\Pf(\D_m)$.  The anomaly is naturally described mathematically by saying that there is a real Pfaffian
line $\PF$ associated to the Dirac operator, and the fermion Pfaffian $\Pf(\D_m)$ is well-defined as a section of $\PF$.  

In our problem, there are two Majorana fermions, say $\psi_1$ and $\psi_2$, with possibly different masses and possibly different boundary
conditions.  Correspondingly there are two Pfaffian lines, say $\PF_1$ and $\PF_2$, and the overall Pfaffian line is the tensor product\footnote{There is a potential
subtlety here.  If a fermion field
has an odd number of zero-modes, its Pfaffian line should be considered odd or fermionic.  Accordingly, if $\psi_1$ and $\psi_2$ each have an odd number of zero-modes,
then $\PF_1$ and $\PF_2$ are both odd and the correct statement is that $\PF=\PF_1\h\otimes \PF_2$, where $\h\otimes$ is a $\Z_2$-graded tensor product (this notion
is described in section \ref{cndmat}).  We will not encounter this subtlety, because always at least one of $\psi_1$ and $\psi_2$ will satisfy one of the boundary conditions
(\ref{bc}).  A fermion field obeying one of those boundary conditions has an even number of zero-modes, since there are none at all if $m\eta<0$ and the number is independent of $m$ mod 2.   Note that on a Riemann surface with boundary, there is no notion of the
chirality of a zero-mode and we simply count all zero-modes. By contrast, the mod 2 index that is used in defining theory $\T$ on a surface without boundary is defined by counting positive chirality zero-modes only.}
$\PF=\PF_1\otimes \PF_2$.

In general, the Pfaffian line of a Dirac operator does not depend on a fermion mass, but it may depend on the boundary conditions.  Indeed,
as we will see, there is such a dependence in our problem and it will play an essential role. 

We will now consider the boundary path integral and boundary anomaly in our problem for several choices of boundary condition.  

\subsubsection{The Trivial Case}\label{triv}

The most trivial case is that the two masses and also the two boundary conditions are the same.  Moreover, we choose the masses and the signs
so that $m\eta<0$.

Since the two boundary conditions are the same,
$\PF_1$ is canonically isomorphic to $\PF_2$, and therefore $\PF=\PF_1\otimes \PF_2$ is canonically trivial.

Since the two Majorana fermions have the same mass and boundary condition, the combined Dirac operator $\D$ of the two modes is just the direct sum of
two copies of the same Dirac operator $\D_m$.  Thus the fermion path integral $\Pf(\D)$ satisfies $\Pf(\D)=\Pf(\D_m)^2$, and in particular $\Pf(\D)$ is naturally positive (relative to the trivialization of $\PF$
that reflects the isomorphism $\PF_1\cong \PF_2$).   

Since $m\eta<0$, there are no low-lying modes near the boundary and the theory has a uniform mass gap of order $m$ along the boundary as well as in the bulk.
Therefore, after renormalizing a few constants in the low energy effective action, the path integral $\Pf(\D)$ is just 1.

In other words, with equal boundary conditions for the two modes, the trivial theory with equal masses remains trivial along the boundary.

Assuming we allow ourselves to make a generic relevant deformation of the theory (as we would certainly do in condensed matter physics, for example),
this is still true  if we pick the boundary conditions for the two Majorana fermions to be equal but such that $m\eta>0$.  Then we generate
two $0+1$-dimensional massless Majorana fermions, say $\chi_1,\chi_2$.  But given any such pair of Majorana modes in $0+1$ dimensions, 
one can add a mass term $\i\mu\chi_1\chi_2$ to the Hamiltonian (or the Lagrangian), with some constant $\mu$, removing them from the low energy theory.  The theory becomes gapped
and the renormalized partition function is again 1.

Fermi statistics do not allow the addition of a mass term for a {\it single} massless 1d Majorana fermion.   Hence the number of 1d Majorana modes along the 
boundary is a topological invariant mod 2.   We will discuss next the case that this invariant is nonzero.

\subsubsection{Boundary Condition In Condensed Matter Physics}\label{cndmat}

 For theory $\T$, or for the Kitaev spin chain, we consider two Majorana fermions, with opposite signs of $m$.   In the context of condensed matter physics,
 to study the theory on a manifold with boundary,
  we want a boundary condition that makes the theory fully anomaly-free.  In other words, we want to ensure that the Pfaffian line bundle $\PF=\PF_1\otimes \PF_2$
 remains canonically trivial.  This is straightforward: since $\PF$ is in general independent of the masses, we simply use the same boundary condition as
 in section \ref{triv} -- namely the same sign of $\eta$ for both Majorana fermions -- and then $\PF$ remains canonically trivial, regardless of the masses.  
 
However, since the two Majorana fermions have opposite signs of $m$, we see now that regardless of the common choice of $\eta$, 
precisely one of them has a normalizable zero-mode (\ref{nolf}) along the boundary.
This means that the mass gap of the theory breaks down along the boundary.
Although it is gapped in bulk, there is a single $0+1$-dimensional massless Majorana fermion propagating along the boundary.   
As we noted in section \ref{cm}, this is regarded in condensed matter physics as the defining property of the Kitaev spin chain.

Now let us discuss a consequence of this construction that has been important in mathematical work \cite{PST,Tes,BT,STa} on 2d gravity on
a manifold with boundary.  We will see later the reason for its importance.
 In general, suppose that $\Sigma$ has $h$ boundary components $\partial_1\Sigma,\partial_2\Sigma,\cdots, \partial_h\Sigma$.
On each boundary component, one makes a choice of sign in the boundary condition, and this determines a real spin bundle $\E_i$
of $\partial_i\Sigma$.  Along each $\partial_i\Sigma$, there propagates a massless 1d Majorana fermion $\chi_i$.  In propagating around $\partial_i\Sigma$,
$\chi_i$ may obey either periodic or antiperiodic boundary conditions.  Indeed, on the circle $\partial_i\Sigma$, there are two possible
spin structures, which in string theory are usually called the Neveu-Schwarz or NS (antiperiodic) spin structure and the Ramond (periodic) spin structure.   The NS spin structure is bounding and the R spin structure is unbounding.
The underlying spin bundle $\S$ of $\Sigma$ determines whether $\E_i$ is of NS or Ramond type.  For general $\S$, the only general constraint on the $\E_i$ is that the number 
$\Ra$ of boundary components
with Ramond spin structure is even.  

The field $\chi_i$, in propagating around the circle $\partial_i\Sigma$, has a zero-mode if and only if $\E_i$ is of Ramond type.
This is not an exact zero-mode, but it is exponentially close to being one if $m$ is large
(compared to the inverse of the characteristic length scale of $\Sigma$).  Let us write $\nu_i,$ $i=1,\dots,\Ra$, for these modes.  The $\nu_i$ have much smaller eigenvalues of $\D_m$ than any
other modes of $\psi_1$ and $\psi_2$, so there is a consistent procedure in which we integrate out all other modes and leave
an effective theory of the $\nu_i$ only.   

Since the underlying theory was chosen to be anomaly-free, it must determine a well-defined measure
for the $\nu_i$.   This condition is not as innocent as it may sound.  A measure on the space parametrized by the $\nu_i$ is something like
\be\label{molt}\d \nu_1\d \nu_2\cdots \d \nu_\Ra. \ee
However, {\it a priori}, this expression does not have a well-defined sign.  First of all, its sign is obviously changed if we make an odd permutation of
the $\nu_i$, that is of the Ramond boundary components.   But in addition, we should worry about the signs of the individual $\nu_i$.  Since the $\nu_i$
are real, we can fix their normalization up to sign by asking them to have, for example, unit $\mathrm{L}^2$ norm.  But there is no natural way to choose
the signs of the $\nu_i$, and obviously, flipping an odd number of the signs will reverse the sign of the measure (\ref{molt}).

There is no natural way to pick the signs of the $\nu_i$ up to an even number of sign flips, and likewise, there is no natural way to pick
an ordering of the $\nu_i$ up to even permutations.  However, the fact that there is actually a well-defined measure on the space
spanned by $\nu_1,\cdots,\nu_\Ra$ means that one of these choices determines the other.  This fact (originally proved in a very different way)
is an important lemma in \cite{PST,Tes,BT,STa}.

The existence of a natural measure on the space spanned by the $\nu_i$ can be expressed in the following mathematical language.  
For $i=1,\dots, \Ra$, let $\varepsilon_i$
be the 1-dimensional real vector space generated by $\nu_i$.  The $\Z_2$-graded tensor product\footnote{\label{ztwo} Since this notion may be unfamiliar,
we give an example, following P. Deligne.  Let $S_i$, $i=1,\dots,t$ be a family of circles, and let $T$ be the torus $\prod_{i=1}^t S_i$.  Then $\varepsilon_i
=H^1(S_i,\R)$ is a 1-dimensional vector space, as is $\alpha=H^n(T,\R)$.  There can be no natural isomorphism between $\alpha$ and the
ordinary tensor product $\otimes_i \varepsilon_i$, since the exchange of two of the circles acts trivially on $\otimes_i\varepsilon_i$, while
acting on $\alpha$ as $-1$.  But there is a canonical isomorphism $\alpha\cong \h\otimes_i\varepsilon_i$.} of the $\varepsilon_i$, denoted
$\h\otimes_i\varepsilon_i$, is equivalent to the ordinary tensor product once an ordering of the $\varepsilon_i$ is picked, by an isomorphism
that reverses sign if two of the $\varepsilon_i$ are exchanged.  The lemma that we have been describing is equivalent to the statement that the
$\Z_2$-graded tensor product of the $\varepsilon_i$ is canonically trivial:
\be\label{mifo}\h\otimes_{i=1}^\Ra\varepsilon_i\cong \R. \ee

\subsubsection{Boundary Condition In Two-Dimensional Gravity}\label{btg}

For the application of theory $\T$ to two-dimensional gravity -- or at least to the theory studied in \cite{PST,Tes,BT,STa} -- we need a different boundary
condition.   In this application, we want theory $\T$ to remain gapped along the boundary as well as in bulk.  But it will have an anomaly that will help in canceling
the gravitational anomaly.

Thus, the two Majorana fermions must remain gapped along the boundary, even though they have opposite masses.  To achieve this, we must give
the two Majorana fermions opposite boundary conditions, so that $m\eta<0$ for each of the two modes. 

Given that the theory has a uniform mass gap of order $m$ even near the boundary, its path integral, after renormalizing a few parameters in the effective
action, is of modulus 1. Moreover, this path integral is naturally real.
 Thus it is fairly natural to write the path integral as $(-1)^\zeta$, just as we did in the absence of a boundary.\footnote{This is a path integral for a particular
spin structure.  As usual, to make the partition function of theory $\T$, we sum over spin structures and divide by 2.}  However, $(-1)^\zeta$ is no longer a number 
$\pm 1$; it now takes values in the real line bundle $\PF$.  In fact, since it is everywhere nonzero, $(-1)^\zeta$ is a trivialization of $\PF$.

This theory actually challenges some of the standard terminology about anomalies.  The line bundle $\PF$ is clearly trivial, because the renormalized partition function $(-1)^\zeta$
provides a trivialization.  However, because this trivialization is provided by the path integral itself, rather than by more local or more elementary considerations, it is not natural to call
the theory anomaly-free. When we say that a theory is anomaly-free, we usually mean that its path integral can be defined as a number, rather than as a section of a line
bundle; that is not the case here.

In our problem, $\PF$ cannot be trivialized by local considerations.  Rather, local considerations will give an isomorphism
\be\label{inz}\PF\cong \h\otimes_{i=1}^\Ra\varepsilon_i,\ee
where the product is over all boundary components with Ramond spin structure.
This claim is consistent with the claim that $\PF$ is trivial, because we have shown in eqn. (\ref{mifo}) that $ \h\otimes_{i=1}^\Ra\varepsilon_i$ is trivial.

To explain what we mean in saying that (\ref{inz}) can be established by local considerations, first set 
\be\label{belvo} V=\oplus_{i=1}^\Ra \varepsilon_i.\ee
  Then the statement (\ref{inz}) is equivalent to
\be\label{pinz} \PF\cong \det V, \ee
where for a vector space $V$, $\det V$ is its top exterior power.   (Note that exchanging two summands $\varepsilon_i$ and $\varepsilon_j$ in $V$ acts as $-1$
on $\det V$, and likewise acts as $-1$ on the $\Z_2$-graded tensor product in (\ref{inz}).)

We will use the following fact about Pfaffian line bundles.  Consider a family of real Dirac operators  parametrized by some space $W$ (in our case,
$W$ represents the choice of metric on $\Sigma$).
As long as the space of zero-modes of the Dirac operator has a fixed dimension, it furnishes the fiber of a vector bundle $V\to W$.  The Pfaffian line bundle $\PF\to W$
is then $\det V$, the top exterior power of $V$. 

More generally, instead of considering zero-modes, we can consider any positive number $a$ that (in a given portion of $W$) is not an eigenvalue of $\i\D_m$,
and let $V$ be the space spanned by eigenvectors of the Dirac operator
with eigenvalue less than $a$ in absolute value.  One still has an isomorphism $\PF\cong \det V$.

Furthermore, the Pfaffian line bundle $\PF$ is independent of fermion masses.  This means that to compute $\PF$ in our problem, instead of considering the case
that the masses are opposite and the signs in the boundary conditions are also opposite, we can take the masses to be the same while the boundary conditions remain 
opposite.

In this situation, one of the fields $\psi_1,\psi_2$ has positive $m\eta$ and one has negative $m\eta$.  So although the interpretation is different,
we are back in the situation considered in section \ref{cndmat}: one fermion has a mass gap $m$ that persists even along the boundary, and the other has a single low-lying made 
along each Ramond boundary component.  The space of low-lying fermion modes is thus $V=\oplus_{i=1}^\Ra \varepsilon_i$, and this leads to eqn. (\ref{inz}).

Eqn. (\ref{inz}) will suffice for our purposes, but it is perhaps worth pointing out that it has the following generalization, which is analogous to Theorem B in \cite{Freed}.
Instead of flipping the boundary condition simultaneously along all boundary components of $\Sigma$, it makes sense to flip the boundary condition along one boundary
component at a time.  Let $S$ be a particular boundary component of $\Sigma$ and let $\PF$ and $\PF'$ be the Pfaffian line bundles before and after flipping the
boundary condition of one fermion along $S$.  If the spin structure along $S$ is of NS type, then
\be\label{pinl}\PF'\cong \PF, \ee
that is, changing the boundary condition has no effect.  But if it is of Ramond type, then\footnote{This formula shows that if we flip the boundary condition for one of the
Majorana fermions along just one of the Ramond boundaries or more generally along some but not all of them, then the Pfaffian line becomes nontrivial and the theory becomes
genuinely anomalous.} 
\be\label{inl}\PF'\cong \varepsilon\h\otimes \PF,\ee
where $\varepsilon$ is the space of fermion zero-modes along $S$.  Repeated application of these rules, starting with the fact that $\PF$ is trivial if $\psi_1$ and $\psi_2$
have the same sign of $\eta$, leads to eqn. (\ref{inz}) for the case that they have opposite signs of $\eta$.

To justify the statements (\ref{pinl}) and (\ref{inl}), we use the fact that by the excision property of index theory, the change in the Pfaffian line when we flip the boundary condition
 along $S$ depends only on the geometry along $S$ and not on the rest of $\Sigma$.  Thus we can embed $S$ in any convenient $\Sigma$ of our choice.
It is convenient to take $\Sigma$ to be the annulus $S\times I$, where $I=[0,1]$ is a unit interval, and we consider $S$ to be embedded  in $S\times I$ as the left boundary
$S\times \{0\}$.  
 We want to compute the effect of flipping the boundary condition at $S\times \{0\}$, keeping it fixed at $S\times \{1\}$.  We can take the fermion mass to be 0, so
 the Dirac operator becomes conformally invariant and we can take the metric on the annulus to be flat.  A fermion zero-mode is then simply a constant mode that
 satisfies the boundary conditions.  For the case of an NS spin structure, the fermions are antiperiodic in the $S$ direction and so have no zero-modes.  Thus the space
 of zero-modes is $V=0$, so that $\det V=\R$.  This justifies (\ref{pinl}) in the NS case.  In the R case, flipping the boundary condition at one end adds or removes
 a zero-mode (depending on the boundary condition at the other end).  The relevant space of zero-modes is $V\cong \varepsilon$, so that $\det V\cong \varepsilon$, leading to (\ref{inl}).

\subsection{Anomaly Cancellation}\label{anomcan}

We are finally ready to explain how the anomaly that we described in section \ref{anomaly} has been canceled in \cite{PST,Tes,BT,STa}.
We consider first the case that all boundaries of $\Sigma$ are of Ramond type, and to start with, we omit boundary punctures.
We denote the circumference of the $i^{th}$ boundary as $b_i$.  We recall that the reason for the anomaly is that there is no natural sign
of the differential form $\Omega=\d b_1\d b_2\cdots \d b_\Ra$ (eqn. (\ref{difff})).  However, after coupling to theory $\T$, 
what needs to have a natural
sign is the product of this with $(-1)^\zeta$, the path integral of theory $\T$:
\be\label{modo}\h\Omega= \d b_1\d b_2\cdots \d b_\Ra \, (-1)^\zeta. \ee
We recall, in addition, that $(-1)^\zeta$ takes values in $\h\otimes_i \varepsilon_i$, where $\varepsilon_i $ is a 1-dimensional 
vector space of zero-modes along  the $i^{th}$ Ramond boundary.   

Here $\h\otimes_i$ is a $\Z_2$-graded tensor product, meaning that $\h\otimes_i \varepsilon_i$ changes sign 
if any two of the boundary components are
exchanged.  But the original anomaly was that $\d b_1\d b_2\cdots \d b_\Ra$ likewise changes sign if any two 
boundary components are exchanged.  The upshot
then is that the product $\h\Omega$ does {\it not} change sign under permutations of boundary components.  
It naturally takes values in the {\it ordinary} tensor product of
the $\varepsilon_i$:
\be\label{odo}\h\Omega \in \otimes_{i=1}^\Ra \varepsilon_i . \ee

What have we gained?  The anomaly has not disappeared, but it has become local: it has turned into an 
ordinary tensor product of factors associated to individual
boundary components; because it is an ordinary tensor product, it can be canceled by a local choice made
independently on each boundary component.  

The last step in canceling the anomaly is to say that a boundary of $\Sigma$ is not just a ``bare'' boundary: it comes with additional structure.
Let $S_i$ be the $i^{th}$ boundary component of $\Sigma$, and let $\E_i$ be its spin structure.   In the theory developed in \cite{PST,Tes,BT,STa}
(but for the moment still ignoring boundary punctures) each $S_i$ is endowed with a trivialization of $\E_i$, up to 
homotopy. For the moment we consider Ramond boundaries only. Since $\E_i$ is a real
line bundle, and is trivial on a Ramond boundary, it has  two homotopy classes of trivialization over each Ramond boundary.  
In addition to summing over spin structures on $\Sigma$ and 
integrating over its moduli, one is supposed
to sum over (homotopy classes of) trivializations of $\E_i$ for each Ramond boundary $S_i$.  

A fermion zero-mode on $S_i$ is a ``constant'' mode that is everywhere non-vanishing, so the choice of such a zero-mode 
gives a trivialization of $\E_i$.  This means
that, still in the absence of boundary punctures, trivializations of $\E_i$ correspond to  choices of the sign of the  zero-mode on $S_i$.  Hence once
we trivialize all the $\E_i$, the right hand side of (\ref{odo}) is trivialized and $\h\Omega$ acquires a well-defined sign.  

Thus once theory $\T$ is included and the boundaries are equipped with trivializations of their spin bundles, the problem 
with the orientation of the moduli space is solved.   However, without some further ingredients, all correlation functions would vanish.  Indeed, 
summing over the signs of the trivializations of the $\varepsilon_i$ will imply summing over the sign of $\h\Omega$. 

 Moreover, what we have said does
not make sense for boundaries with NS spin structure, since their spin bundles cannot be globally trivialized.

The additional ingredient that has to be considered is a boundary puncture.  One postulates that locally, away from punctures, $\E_i$
is trivialized, but that this trivialization changes sign in crossing a boundary puncture.

With this rule, it is possible to incorporate NS as well as Ramond boundaries.   A simple example of a boundary component with NS spin
structure is the boundary of a disc (fig. \ref{disc}).  Its spin structure is of NS or antiperiodic 
type, and cannot be trivialized globally.  It can be trivialized on
the complement of one point, but then the trivialization changes sign in crossing that point.   In the theory of \cite{PST,Tes,BT,STa}, that
point would be interpreted as a boundary puncture.  So an NS boundary with one boundary puncture is possible in the theory, but
an NS boundary with no boundary punctures is not.  More generally, the number of punctures on a given NS boundary can be any positive odd
number, since the spin structure of an NS boundary can have a trivialization that jumps in sign any odd number of times in going around
the boundary circle.  

As an example, a disc with $n$ boundary punctures and $m$ bulk ones has a moduli space of dimension $2m+n-3$.  The fact that $n$ is odd means
that this number is even. This is actually a  necessary condition for some of the correlation functions 
$\int_{\bM}\prod_i \psi_i^{d_i}$  (eqn. (\ref{proft})) to be nonzero, since the cohomology classes $\psi_i$ are all of even degree.  

The spin structure of a Ramond puncture is globally trivial, so it is possible to have a Ramond boundary with no boundary punctures.
Of course, this is the case we started with.  More generally, a Ramond boundary can have any even number of punctures.

On any given boundary component of either NS or Ramond type, there are two allowed classes of piecewise trivialization of the spin structure.
One can pick an arbitrary trivialization at a given starting point 
(not one of the punctures), and then the extension of this over the rest of the circle is uniquely determined by the
condition that the trivialization jumps in sign whenever a boundary puncture is crossed.

\begin{figure}
 \begin{center}
   \includegraphics[width=1.6in]{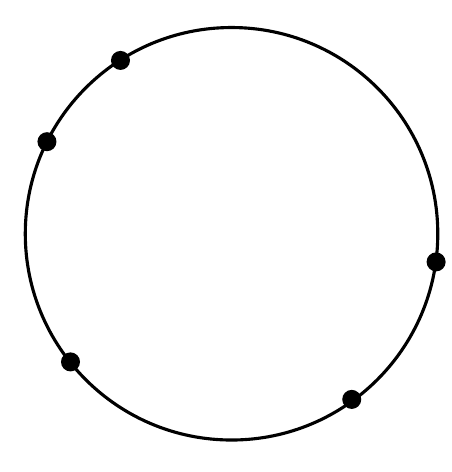}
 \end{center}  
\caption{\small A disc with five boundary punctures.  The spin bundle of the boundary circle $S$ is a real line bundle that is inevitably of NS type.
 This real line bundle is not trivial globally over $S$, but -- since the number
of boundary punctures is odd -- it can be trivialized
 on the complement of the boundary punctures in such a way that the trivialization changes sign whenever one
crosses a boundary puncture. \label{disc}}
\end{figure}

This description of boundaries and their punctures may seem bizarre at first, but we will see in section \ref{branes} that it is
not too difficult to give it a plausible physical interpretation.    But first, let us ask whether incorporating boundary punctures has
reintroduced any problem with the orientation of moduli space.  We will deal with this question by describing a consistent recipe 
\cite{PST,Tes,BT,STa} for dealing with the sign questions. We expect that this recipe could be deduced from the framework of section \ref{branes},
but we will not show this.

First let us consider the case of a boundary component $S$ with NS spin structure.  It has a circumference $b$ and it has an odd number $n$
of boundary punctures that have a natural cyclic order.  Let us pick an arbitrary starting point  $p\in S$ and relative to this label the punctures
in ascending order by angles $\alpha_1< \alpha_2<\cdots < \alpha_n$.  So $b$ and $\alpha_1,\dots,\alpha_n$ are the moduli that are associated
to $S$. To orient this parameter space, we can use the differential form 
\be\label{zony}\Upsilon= \d b \,\d\alpha_1\,\d\alpha_2\,\cdots \d\alpha_n. \ee
We note that $\Upsilon$ has a natural sign: because the number of $\alpha$'s is odd, moving a $\d\alpha$ from the end of 
the chain to the beginning does not affect the sign of $\Upsilon$.  Also, since $\Upsilon$ is of even degree, it commutes with similar factors
associated to other boundary components.   Therefore, an NS boundary component raises no problem in orienting the moduli space.

Now let $S$ have Ramond spin structure.  In this case, $n$ is even.   This has two consequences.  First, we get a sign change
if we move a $\d\alpha$ from the end of the chain to the beginning.  However, just as in the case $n=0$ that we started with,
the sign of $(-1)^\zeta$ depends on how one trivializes the spin structure of a Ramond boundary.  A consistent recipe is to define
the sign of $(-1)^\zeta$ using the trivialization that is in effect just to the right of  the starting point $p\in S$ relative to which we measured the $\alpha$'s.
Then moving one of the boundary punctures from the end of the chain to the beginning will reverse the sign of $\Upsilon$ while also
reversing the sign of $(-1)^\zeta$.  Also, because $n$ is even, $\Upsilon$ is of odd degree in the case of a Ramond boundary.
Therefore the $\Upsilon$ factors associated to different Ramond boundaries anticommute with each other.  Just as we discussed
for the case $n=0$, this compensates for the fact that $(-1)^\zeta$ is odd under exchanging any two Ramond boundaries.

\subsection{Branes}\label{branes}

\subsubsection{The $\zeta$-Instanton Equation And Compactness}

 In the present section, we will attempt to interpret the possibly strange-sounding  picture just described
in terms of the physics of branes.

For this, it will be helpful to use the second realization of theory $\T$ that was presented in section \ref{twor}.  This was based on topologically twisting a two-dimensional
theory with $(2,2)$ supersymmetry and a complex chiral superfield $\Phi$.   
The bottom component of $\Phi$ is a complex field $\phi$.   The theory also has a holomorphic superpotential, which in our application is $W(\Phi)=\frac{\i}{2}m^2\Phi^2$,
but we will write some formulas for a more general $W(\Phi)$.

The condition for a configuration of the $\phi$ field to be supersymmetric is the $\zeta$-instanton equation
\be\label{consym}\frac{\partial\phi}{\partial \bar z}+\frac{\partial \bar W}{\partial\bar\phi}=0.  \ee
This equation can be written $\bar\partial \phi +\d\bar z \partial_{\bar\phi}\bar W$, so it can be defined on a Riemann surface $\Sigma$ with
a distinguished, everywhere nonzero $(0,1)$-form $\d\bar z$.  (For example, such a form exists globally if $\Sigma$ is a Riemann surface of genus 1 or a domain in the complex plane.)  
If $W$ is quasihomogeneous,
as in our case, the equation is conformally-invariant  if $\phi$ is suitably interpreted.  The conformally-invariant version of the equation can be formulated
on any Riemann surface.
This conformally-invariant form of the equation is what one gets when one topologically twists the theory using the $R$-symmetry that exists for
quasi-homogeneous $W$.
   For example, in the case of a quadratic
$W$, after topological twisting, $\phi$ has to be interpreted as a section of a chiral spin bundle $\L\to \Sigma$, a square root of the canonical bundle $K\to \Sigma$.  In \cite{Wittenveryold}, a more
general case $W\sim\Phi^r$ was considered, and then in the topologically twisted version of the theory, $\phi$ 
is a section of an $r^{th}$ root of $K$ (this $r^{th}$ root may have singularities at specified points in $\Sigma$ where ``twist fields'' are inserted).  

Certain important properties hold whenever the $\zeta$-instanton equation can be defined, whether in a topologically-twisted version or simply
in a naive version in which $\phi$ is a complex field.  In particular, if $\Sigma$ has no boundary, then the $\zeta$-instanton equation has only
``trivial'' solutions.  This is proved in a standard way:  take the absolute value squared of the equation, integrate over $\Sigma$, and then integrate by parts,
to show that any solution satisfies
\be\label{tonco}0=\int_\Sigma |\d^2z|\left|\frac{\partial\phi}{\partial \bar z}+\frac{\partial \bar W}{\partial\bar\phi}\right|^2 = 
\int_\Sigma|\d^2z| \left(|\d \phi|^2 +\left|\frac{\partial W}{\partial\phi}\right|^2 + \left(\partial_z W+\partial_{\bar z}\bar W\right) \right).\ee
If $\Sigma$ has no boundary, we can drop the total derivatives $\partial_z W$ and $\partial_{\bar z}\bar W$, and we learn that on a closed surface $\Sigma$,
any solution has $\d\phi=0$ and $\partial W/\partial \phi=0$; in other words, $\phi$ must be constant and this constant must be a critical point of $W$. 
For a large class of $W$'s, this implies that, on a surface $\Sigma$ without boundary, the space of solutions of the $\zeta$-instanton equation is compact
(and in fact ``trivial'').  
  This compactness is an important ingredient in the well-definedness of the
twisted topological field theory constructions related to the $\zeta$-instanton equation.

\subsubsection{Boundary Condition In The $\zeta$-Instanton Equation}\label{bondo}

If $\Sigma$ has a boundary, then we have to pick a boundary condition on the $\zeta$-instanton equation.  
Let us first ignore the twisting and treat $\phi$ as an ordinary complex scalar field.  If we also set $W$ to 0, the equation for $\phi$ becomes the Cauchy-Riemann
equation saying that $\phi$ is holomorphic.  The topological $\sigma$-model associated to counting solutions of this equations is then an ordinary $A$-model.
Though the topological field theory associated to theory $\T$ is not an ordinary $A$-model
 -- because of the superpotential and because $\phi$ is twisted to have spin 1/2 -- it
will be useful to first discuss this more familiar case.

   A boundary condition for the Cauchy-Riemann equations that is sensible (elliptic) at least locally can be obtained by picking an arbitrary curve $\ell\in\C$
and asking that the boundary values of $\phi$ should lie in $\ell$. Adding a superpotential to get the $\zeta$-instanton equation does not affect this statement,
which only depends on the ``leading part'' of the equation (the terms with the maximum number of derivatives).    Here we may loosely call $\ell$ a brane, although to be more precise, it is the support of a brane.
As we will discuss later, there can be more than one brane with support $\ell$.  More generally, as is usual in brane physics, we may impose such a boundary
condition in a piecewise way.  For this, we pick several branes $\ell_\alpha$, we decompose the boundary $\partial\Sigma$ as a union of intervals $I_\alpha$ that
meet only at their endpoints, and for each $\alpha$,  we require that $I_\alpha$ should map to $\ell_\alpha$.  (A common endpoint of $I_\alpha$ and $I_\beta$
must then map to an intersection point of $\ell_\alpha$ and $\ell_\beta$.)

What sort of $\ell$ should we use?  At first sight, it may seem that the $A$-model is most obviously well-defined if $\ell$ is compact.  Actually, a compact closed curve
in $\C$ is a boundary, and with such a choice of $\ell$, the $A$-model with target $\C$ is actually anomalous, as explained from a physical point of view in
\cite{GMW}, section 13.5. This anomaly is an ultraviolet effect that is related to a boundary contribution to the fermion number anomaly on a Riemann
surface.  More intuitively, if $\ell$ is a closed curve in the plane, that it can be shrunk to a point and is not interesting topologically. Thus we should consider noncompact $\ell$, for example a straight line in $\C$.

With such a choice, we avoid the ultraviolet issues mentioned in the last paragraph, but the noncompactness of $\ell$ raises potential
infrared problems.  The space of solutions of the Cauchy-Riemann equation $\bar\partial\phi=0$, with boundary values in the noncompact space $\ell$,
is in general not compact, and this poses difficulties in defining the $A$-model with target $\C$.  

There are a number of approaches to resolving these difficulties, depending on what one wants.  One approach leads mathematically to the ``wrapped
Fukaya category.''  For our purposes, we want to use the superpotential $W$ to
prevent $\phi$ from becoming large.  This corresponds mathematically to the Fukaya-Seidel category \cite{FS}; for a physical interpretation, see \cite{GMW}, especially
sections 11.2.6 and 11.3.   To see the idea, let us return to the identity (\ref{tonco}), but now allow for the possibility that $\Sigma$ has a boundary.  For instance,
we can take $\Sigma$ to be the upper half $z$-plane.  Setting $z=x_1+\i x_2$, 
the identity becomes
\be\label{ronco}0= 
\int_\Sigma|\d^2z| \left(|\d \phi|^2 +\left|\frac{\partial W}{\partial\phi}\right|^2 \right)+2\int_{\partial\Sigma}\d x_1  {\mathrm{Im}}\,W.\ee
Now it becomes clear what sort of brane we should consider.  We should choose $\ell$ so that $\mathrm{Im}\,W\to \infty$ at $\infty$ along $\ell$.  
Then the boundary term in the identity will ensure that $\phi$ cannot become large along $\partial\Sigma$, and given this,  the bulk terms in the identity ensure
that $\phi$ cannot become large anywhere.  That is an essential technical step toward being able to define the $A$-model.

Let us implement this in our case that $W(\phi)=\frac{\i}{2}m\phi^2$, with $m>0$ and $\phi=\phi_1+\i\phi_2$.  We have 
$\mathrm{Im}\,W(\phi)=\frac{m}{2} (\phi_1^2-\phi_2^2)$.
Thus near infinity in the complex $\phi$ plane, there are two regions with $\mathrm{Im}\,W\to +\infty$:  this happens near the positive real $\phi$ axis
and also near the negative axis.   

A noncompact 1-manifold $\ell$ is topologically a copy of the real line, with two ends.   To ensure that $\mathrm{Im}\,W\to \infty$ at $\infty$ along $\ell$, we should
pick $\ell$ so that each of its ends is in one of the good regions near the positive or negative $\phi$ axis.  Beyond this, the precise choice of $\ell$ does not matter, 
because of the fact that the $A$-model is invariant under Hamiltonian symplectomorphisms of $\C$.  All that really matters is whether $\phi$ tends toward $+\infty$
or $-\infty$ at each of the two ends of $\ell$.  Moreover, if $\phi$ tends to infinity in the same direction at each end of $\ell$, it is ``topologically trivial'' in the sense
that it can be pulled off to infinity in the $\phi$-plane while preserving the fact that $\mathrm{Im}\,W\to \infty$ at $\infty$ along $\ell$.  So the only interesting case is
that $\phi$ tends to $-\infty$ at one end of $\ell$ and to $+\infty$ at the other.  Further details do not matter.  Therefore, we may as well simply take\footnote{This
$\ell$ can be described as a Lefschetz thimble for the superpotential $W$ associated to its unique critical point at $\phi=0$.  In general, in the Fukaya-Seidel
category, the most basic objects are such Lefschetz thimbles.}  $\ell$ to be the
real $\phi$ axis.    In other words, the boundary condition on $\phi$ is that it is real along $\partial\Sigma$, or in other words if $\phi=\phi_1+\i\phi_2$, then $\phi_2=0$
at $x_2=0$.

This has an interesting interpretation in the topologically twisted model that we are really interested in.  We recall that in this model, $\phi$ is a section of the
chiral spin bundle $\L$ of $\Sigma$.  The fiber of $\L$ at a point in $\Sigma$ is a complex vector space of dimension 1.   This is actually the same as a real
vector space of rank 2.  Thus, we can alternatively view the complex line bundle $\L\to\Sigma$ as a rank 2 real vector bundle $\S\to \Sigma$.  The resulting
$\S$ is simply the real, nonchiral spin bundle of $\Sigma$.  Thus, it is possible to view the real and imaginary parts of $\phi$ as a two-component real spinor field
over $\Sigma$.  In fact, we have already made much the same statement in eqn. (\ref{zilk}), where we asserted that the $\zeta$-instanton equation for
$\phi$ is equivalent to the massive Dirac equation for 
$\h\phi=\begin{pmatrix}\phi_1\cr -\phi_2\end{pmatrix}$.

Now recall that in section \ref{bound}, we defined a rank 1 real spin bundle $\E\to\partial\Sigma$ by saying that a section of $\E$ is a section $\h\phi$ of the rank 2 spin
bundle $\S$ of $\Sigma$ (restricted to $\partial\Sigma$) that satisfies $\gamma_\parallel\h\phi=\h\phi$.   (The opposite sign in this relation, $\gamma_\parallel
\h\phi=-\phi$, defines another equivalent real spin bundle of $\partial\Sigma$.)  For $\Sigma$ the upper half plane, the tangential gamma matrix is $\gamma_\parallel
=\gamma_1$, and the representation that we have used of the gamma matrices (eqn. (\ref{hd})) is such that $\gamma_1\h\phi=\h\phi$ is equivalent to $\phi_2=0$.

Thus, we can state the boundary condition that we have found in a way that makes sense in general for the twisted topological field theory under study.  
In bulk, that is away from $\partial \Sigma$, $\phi$ is a section of the chiral spin bundle $\L\to \Sigma$.   The boundary condition satisfied by $\phi$ is
that along $\partial\Sigma$, it is a section of the real spin bundle $\E\to\partial\Sigma$.    The merit of this boundary condition is the same as it is in the ordinary
$A$-model, which we used as motivation:  it ensures that the surface terms in eqn. (\ref{tonco}) vanish, and therefore that the only solution of the $\zeta$-instanton
equation on a Riemann surface $\Sigma$ with boundary is $\phi=0$.  

We can gain some more insight by comparison to the ordinary $A$-model.  To construct a brane with support $\ell$, we need to pick an orientation of $\ell$.
There are two possible orientations, so there are two possible branes, which we will call $\B'$ and $\B''$.  Neither one is distinguished relative to the other.

In the ordinary $A$-model, we could at our discretion introduce $\B'$ or $\B''$ or both.  The twisted model that is related to theory $\T$, in which $\phi$ is
a chiral spinor rather than a complex-valued field, is different in this respect.  The reason it is different is that $\B'$ and $\B''$ represent choices of orientation
of the real spin bundle $\E\to \partial\Sigma$, but in general this real spin bundle is unorientable.  Thus, if one goes all the way around a component of
$\partial\Sigma$ with NS spin structure, then $\B'$ and $\B''$ are exchanged.  Accordingly, in the model relevant to theory $\T$, if we introduce one of these branes,
we have to also introduce the other.  

Once we introduce branes $\B'$ and $\B''$, we are very close to the picture developed in the mathematical literature \cite{PST,Tes,BT,STa}.  The boundary
of $\Sigma$ is decomposed as a union of intervals $I_\alpha$ that have only endpoints in common, and  each interval is labeled by $\B'$ or $\B''$.  This labeling 
here means simply a chosen orientation of $\E\to\partial\Sigma$.  Since $\E$ is a real vector bundle of rank 1, a choice of orientation of $\E$ is (up to homotopy)
the same as a trivialization of $\E$, the language used in section \ref{anomcan}.

There is really just one more puzzle.  In the theory developed in \cite{PST,Tes,BT,STa}, whenever one crosses a boundary puncture, 
the orientation of $\E$
jumps.  Why is this true?

A quick answer is the following.  In general, for any brane $\B$,
$(\B,\B)$ strings in the $A$-model correspond to local operators that can be inserted on the boundary of the string in a region
of the boundary that is labeled by brane $\B$.  Our model is only locally equivalent to an $A$-model, but this is good enough to
discuss local operators.  In the case of the branes $\B'$ and $\B''$, as $\ell$ is contractible, the only interesting local $(\B',\B')$ or
$(\B'',\B'')$ operator is the identity operator.  However, in topological string theory, what we add to the action along the boundary of
the string worldsheet is really a descendant of a given local operator.  In the case of a boundary local operator $\O$, what we
want is the 1-form operator $\mathcal V$ that can be deduced from $\O$ via the descent procedure.  If $\O$ is the identity operator,
then $\V=0$. (Recall that $\V$ is characterized by $\{Q,\V\}=\d \O$, where $Q$ is the BRST operator of the theory; if $\O$ is the identity
operator, then $\d\O=0$ so $\V=0$.)  Therefore we cannot get anything interesting from $(\B',\B')$ or $(\B'',\B'')$ strings.  

The analogy with the standard $A$-model indicates that the space of $(\B',\B'')$ or $(\B'',\B')$ strings is also 1-dimensional (see 
sections \ref{bsg} and \ref{quantizing}), but now
a $(\B',\B'')$ or $(\B'',\B')$ string corresponds to a local operator that causes a jumping in the brane that labels the boundary, and
this is certainly not the identity operator.  Thus the gravitational descendant will not vanish.  

Another crucial detail concerns the statistics of the operators.  The identity operator is bosonic, so its 1-form descendant, if not zero,
would be fermionic.  A fermionic boundary puncture operator is not what we need for the theory of \cite{PST,Tes,BT,STa}, in which the coupling
parameters and correlation functions are all bosonic.  The analogy with the standard $A$-model indicates (section \ref{bsg}) that the
$(\B',\B'')$ and $(\B'',\B')$ local operators are fermionic, so that their 1-form descendants are bosonic.

There is also an important detail on which the analogy to the standard $A$-model is a little misleading, because it is only valid locally.
In an $A$-model with branes $\B'$ and $\B''$, the $(\B',\B'')$ and $(\B'',\B')$ local operators would be independent operators, and we would
potentially include them (or their 1-form descendants) with independent coupling parameters.  In the present context, there is not really any
way to say which is which of $\B'$ and $\B''$; one can only say that they differ by the orientation of the real spin bundle.\footnote{For example,
$\B'$ and $\B''$ are exchanged in going all the way around a circle with NS spin structure. Perhaps more fundamentally,
 orienting the real spin bundle
of one boundary of $\Sigma$ does not in general tell us how to choose such an orientation for  other boundaries. So we can say locally
how $\B'$ and $\B''$ differ but there is no global notion of which is which.}  So there is really only one type of
boundary puncture, which one can think of as $(\B',\B'')$ or $(\B'',\B')$, and correspondingly there is only one boundary coupling.

It follows, incidentally, that even if the identity $(\B',\B')$ or $(\B'',\B'')$ operator had a nontrivial 1-form gravitational descendant, it could not play a role.
We would have to identify these two operators, so we would have a single such operator with a fermionic coupling constant $\upupsilon$.  As the correlation
functions of topological gravity are bosonic, they could not depend on a single fermionic variable $\upupsilon$.

\subsubsection{Orientations and Statistics}\label{bsg}

Consider a brane $\B'$ in an arbitrary $A$-model with some target space $X$.  The support of $\B'$ is a Lagrangian submanifold $L\subset X$.
Take $\B'$ to have trivial Chan-Paton bundle.\footnote{For example, $L$ might be topologically trivial (as it is in our application, with
$L=\ell$).
We will ignore various subtleties related to the $K$-theory interpretation of branes; these
are not relevant for our purposes.}  If we consider $N$ copies of brane $\B'$, we get  an effective $U(N)$ gauge theory along $L$.

Another $M$ copies of brane $\B'$ would similarly support by themselves a $U(M)$ gauge theory.  If we combine $N$ copies of $\B'$
with $M$ more copies, we get a $U(N+M)$ gauge theory.

Now consider another brane $\B''$ that differs from $\B'$ only by reversing the orientation of $L$.  $M$ copies of $\B''$ would support
a $U(M)$ gauge theory.  However, if one combines $M$ copies of $\B''$ with $N$ copies of $\B'$, one does not get a gauge group $U(N+M)$.
Instead, one gets the supergroup $U(N|M)$ \cite{vafa-antibrane}.

\def\A{{\mathcal A}}
We will give a simple example to explain why this must be the case.  For a familiar setting, take $X$ to be a Calabi-Yau three-fold.  The effective
gauge theory for $N$ copies of a brane is actually a $U(N)$ gauge theory.  Let us denote the gauge field as $A$.  The theory  also has a 1-form field $\phi$ in the adjoint representation,
which describes fluctuations in the position of the brane.  The effective action 
is a multiple of the Chern-Simons three-form for the complex connection $\A=A+\i \phi$:
\be\label{zold}I= \frac{1}{g_{\mathrm{st}}}\int_L \mathrm{CS}(\A). \ee
Here $\mathrm{CS}(\A)=\Tr\, \left(\A\wedge \d \A+\frac{2}{3}\A\wedge \A\wedge \A\right) $ is the Chern-Simons three-form and $g_{\mathrm{st}}$ is the
string coupling constant.    There is no problem, given $L$ purely as a bare three-manifold, so define the three-form $\mathrm{CS}(\A)$.  But
to integrate a three-form over $L$ requires an orientation of $L$.  There is no natural choice, but a choice is part of
the definition of a brane with support $L$.   That is one way to understand the fact that in order to define a brane 
$\B'$ or $\B''$ with support $L$, one needs to endow $L$ with an orientation; and there are in fact two $A$-branes $\B'$ and $\B''$ with the
same support $L$ that differ
only by which orientation is chosen.  The sign of the effective action $I$ is opposite for $\B'$ relative to $\B''$.

Now if we bring together $N$ branes supporting a $U(N)$ Chern-Simons theory to $M$ more branes supporting a $U(M)$ Chern-Simons
theory with the same sign of the action, the two Chern-Simons theories can merge into a $U(N+M)$ Chern-Simons theory.   (The  expectation
value of the field $\phi$ can describe the breaking of $U(N+M)$ down to
$U(N)\times U(M)$.)  However, if the $U(M)$ and $U(N)$ Chern-Simons actions have opposite signs, they cannot possibly combine to
a $U(N+M)$ Chern-Simons theory.  Instead, they can combine to a $U(N|M)$ supergroup Chern-Simons theory.    We recall that the
supertrace of an $N|M$-dimensional matrix is defined, in an obvious notation, as
\be\label{murro}\mathrm{Str}\begin{pmatrix}U &V\cr W&X\end{pmatrix}=\mathrm{Tr}\,U-\mathrm{Tr}\,X, \ee
where the relative minus sign is just what we need so that the supertrace of a Chern-Simons three-form of $U(N|M)$ leads to opposite
signs for the $U(N)$ and $U(M)$ parts of the action.  

A consequence of going from $U(N+M)$ to $U(N|M)$ is that the statistics of the off-diagonal blocks $V$ and $W$ is reversed.
At the end of section \ref{bondo}, that is what we needed so that the $(\B',\B'')$ strings are fermionic, and have bosonic 1-form descendants.

The situation just described does not usually arise in physical string theory, because there one usually is interested in branes
that satisfy a stability condition involving the phase of the holomorphic volume form of the Calabi-Yau manifold, restricted to the brane.
For a given Lagrangian submanifold, this condition is satisfied at most for one orientation.

\subsubsection{Quantizing the String}\label{quantizing}

In the standard $A$-model, the space of local operators of type $(\B_1,\B_2)$, for any branes $\B_1$ and $\B_2$ that may or may not be
the same, is the same as the space of physical states found by quantization on an infinite strip with boundary conditions set by $\B_1$ at one
end and by $\B_2$ at the other end.  Here we will explain the analog of this for the model under consideration here, which is only locally
equivalent to a standard $A$-model.   

We will work on the strip $0\leq x_2\leq a$ in the $x_1x_2$ plane, for some $a$, and will treat $x_1$ as Euclidean ``time.''   In eqn. (\ref{ronco}), there is now
a boundary contribution at $x_2=a$, as well as the one at $x_2=0$ that was discussed previously.  The two
contributions have opposite signs, 
and to achieve compactness the boundary condition at $x_2=a$
should ensure that $\mathrm{Im}\,W\to -\infty$ at infinity.  Thus we take the boundary condition at $x_2=a$ to be $\phi_1=0$, while at $x_2=0$
it is $\phi_2=0$, as before.\footnote{\label{isre} This difference in boundary condition is related to something that will be explained in section \ref{bd}:
at $x_2=0$, $\sqrt{\d z}$ is real, while at $x_2=a$, $\sqrt{-\d z}$ is real.}

To find the space of physical states with these boundary conditions, the first step is to find the space of classical ground states.
With $x_1$ viewed as ``time,'' these are the $x_1$-independent solutions of the $\zeta$-instanton equation that satisfy the boundary
conditions at the two ends.   For solutions that depend only on $x_2$, the $\zeta$-instanton equation reduces to 
$\frac{\d\phi}{\d x_2}+m\bar\phi=0$.  The only solution of this linear first-order equation with $\phi_2=0$ at $x_2=0$ and $\phi_1=0$ at $x_2=a$
is $\phi=0$.  Moreover, this solution is nondegenerate, meaning that when we linearize around it, the linearized equation has trivial kernel.
(In the present case, this statement is trivial since the $\zeta$-instanton equation is already linear.) A nondegenerate classical solution
corresponds upon quantization to a single state.

If there were multiple classical vacua, we would have to consider possible tunneling effects to identity the quantum states that
really are supersymmetric ground states.
With only one classical vacuum, this step is trivial.  So in our problem, there is just one supersymmetric ground state.

One might be slightly puzzled that we seem to have used different boundary conditions and thus different branes at $x_2=a$ relative to $x_2=0$.
However, if we conformally map the strip to the upper half plane $x_2\geq 0$, mapping $x_2=-\infty$ in the strip to the origin $x_1=x_2=0$ in the boundarey of the
upper half plane, then this difference disappears.    What we have done, on both boundaries, is to require that $\phi$ should restrict on $\partial\Sigma$
to a section of the real spin bundle $\E\to\partial\Sigma$.

The space of supersymmetric ground states that we just obtained corresponds
to the space of local operators of type $(\B',\B')$, $(\B'',\B'')$, or $(\B',\B'')$ that can be inserted at $x_1=x_2$.  Since we did not
have to orient the spin bundles of the boundaries of the strip in order to determine that there is a 1-dimensional space of physical
states on the strip, the spaces of local operators of type $(\B',\B')$, $(\B'',\B'')$, or $(\B',\B'')$ are the same if understood just as vector spaces.
But these operators have different statistics, as explained in section \ref{bsg}.

\subsection{Boundary Degenerations}\label{bd}

\begin{figure}
 \begin{center}
   \includegraphics[width=6in]{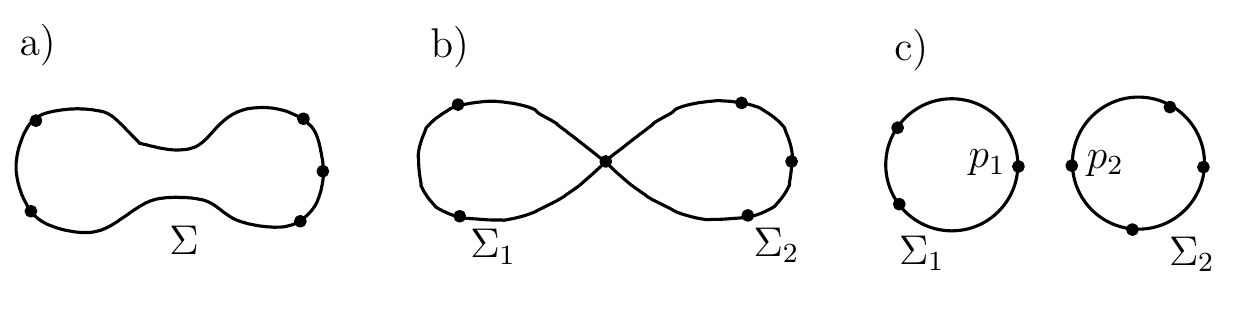}
 \end{center}  
\caption{\small (a)  A disc $\Sigma$ with $n$ boundary punctures that develops a narrow neck. (b) The neck collapses and $\Sigma$ degenerates
to the union of two discs $\Sigma_1$ and $\Sigma_2$ glued at a point.   (c)  The picture of part (b) can be recovered by gluing $p_1\in\Sigma_1$
to $p_2\in\Sigma_2$.  The original boundary punctures of $\Sigma$ are divided in some way between $\Sigma_1$ and $\Sigma_2$.  \label{deg}}
\end{figure}

So far we have concentrated on  questions concerning the orientation of the moduli space.  However, as explained in section \ref{prelims}, in trying to define
topological gravity on Riemann surfaces with boundary,
there is a second serious problem, which is that the moduli space of Riemann surfaces with boundary, with its Deligne-Mumford compactification,
itself has a boundary.  Because of this, intersection numbers such as the correlation functions 
$\int_{\bM}\prod_i \psi_i^{d_i}$ of topological
gravity (eqn. (\ref{proft})) are {\it a priori} not well-defined from a topological point of view.  We will explain schematically how this difficulty has been overcome, going just far enough to describe the simplest concrete computations.  For  full explanations, see \cite{PST,Tes,BT,STa}.

First let us give a simple example to illustrate the problem.  A disc $\Sigma$ with $n$ boundary punctures (and no bulk punctures)
has a moduli space $\bM$ of real dimension $n-3$.
The disc can degenerate in real codimension 1 by forming a narrow neck (fig. \ref{deg}(a)), which then pinches off (fig. \ref{deg}(b)) 
to make a singular Riemann surface $\Sigma$ that can be obtained by gluing together two discs $\Sigma_1$ and $\Sigma_2$ (fig. \ref{deg}(c)).
This occurs in real codimension 1, and thus fig. \ref{deg}(b) describes a component of $\partial\bM$, the boundary of $\bM$.  As a check, let us confirm that the configuration in fig. \ref{deg}(b) has precisely $n-4$ real moduli, so that it is of real codimension 1 in $\bM$.
$\Sigma_1$ and $\Sigma_2$ inherit the boundary punctures of $\Sigma$, say $n_1$ for $\Sigma_1$ and $n_2$ for $\Sigma_2$ with $n_1+n_2=n$.
In addition, $\Sigma_1$ and $\Sigma_2$ have one more boundary puncture $p_1$ or $p_2$ 
where the gluing occurs.  So in all, $\Sigma_1$ and $\Sigma_2$ have
respectively $n_1+1$ and $n_2+1$ boundary punctures, and moduli spaces of dimension $n_1-2$ and $n_2-2$.  The singular configuration in
fig. \ref{deg}(b) thus has a total of $(n_1-2)+(n_2-2)=n-4$ real moduli, as claimed.

Thus, we have confirmed the assertion that moduli spaces of Riemann surfaces with boundary are themselves manifolds (or orbifolds) with boundary. This presents a problem for defining intersection numbers.

Now let us reexamine this assuming that $\Sigma$ is endowed with a spin bundle $\S$ and that the induced real spin bundle $\E$ of
$\partial \Sigma$ is piecewise trivialized along $\partial\Sigma$, as described in section \ref{anomcan}.  We immediately run into  something interesting.
If $\Sigma$ is a disc, the spin bundle $\E\to\partial\Sigma$ is always of NS type, and the number $n$ of boundary punctures on a disc 
will have to be odd.
But when $\Sigma$ degenerates to the union
of two branches $\Sigma_1$ and $\Sigma_2$, with $n_1+1$ punctures on one side and $n_2+1$ on the other side, inevitably either $n_1+1$ or $n_2+1$
is even.  But in the theory that we are describing here, a disc is always supposed to have an odd number of boundary punctures.
 What this means in practice is that either $p_1$ or $p_2$ does not really behave as a boundary puncture in the
sense of this theory: the piecewise trivializations of the real spin bundles 
$\E_1\to \Sigma_1$ and $\E_2\to \Sigma_2$ jump in crossing either $p_1$ or in crossing $p_2$, but not both. This is explained more explicitly
shortly.  As a result, the cohomology classes
$\psi_i$ whose products we want to integrate to get the correlation functions  have the property that when restricted to $\partial\bM$,
they are pullbacks from a quotient space in which either $p_1$ or $p_2$ is forgotten.  Effectively, then, $\partial\bM$ behaves as if it is of real codimension 2
and the intersection numbers are well-defined.  

Now let us explain these assertions in more detail.  First we introduce a useful language.  In the following, $\Sigma$ will be a Riemann
surface, possibly with boundary.  We write $K$ for the complex canonical bundle of $\Sigma$ and $\S$ for its chiral spin bundle.   So $K$ is a complex
line bundle over $\Sigma$, and $\S$ is a complex line bundle over $\Sigma$ with a linear  map $w:\S\otimes \S\to K$ that establishes an isomorphism 
between $\S\otimes \S$ and $K$.

Along $\partial\Sigma$, it is meaningful to say that a one-form is real, and thus $K$, restricted to $\partial\Sigma$, has a real subbundle.  Moreover,
the Riemann surface $\Sigma$ is oriented and this induces an orientation of $\partial\Sigma$.  As a result, it is meaningful to say that a section of $K$,
when restricted to $\partial\Sigma$, is real and positive.  For example, if $\Sigma$ is the upper half of the complex $z$-plane, so that $\partial\Sigma$
is the real $z$ axis,  then the complex 1-form $\d z$ is real and positive when restricted to $\partial\Sigma$.  But if $\Sigma$ is the lower half of the
$z$-plane, then its boundary is the real $z$ axis now with the opposite orientation, and so in this case, $-\d z$ is real and positive along $\partial\Sigma$. 
 
This gives a convenient framework in which to describe the real spin bundle $\E$ of $\partial\Sigma$.  We say that a local section $\psi$ of
$\S\to \Sigma$ is real along $\partial\Sigma$ if the 1-form $w(\psi\otimes\psi)$ is real and positive when restricted to $\partial\Sigma$.  In this
case, we say that the restriction of $\psi$ to $\partial\Sigma$ is a section of $\E$.  This serves to define $\E$. For example, if $\Sigma$ is the upper half of the complex $z$-plane,
then a section $\psi$ of $\S$ with the property that $w(\psi\otimes \psi) =\d z$ is real along $\partial\Sigma$, and its restriction to $\partial\Sigma$
provides a section of $\E$.  We describe this more informally by writing $\psi=\sqrt{\d z}$.  Note that since $(-\psi)\otimes (-\psi)=\psi\otimes \psi$,
in this situation we also have $w((-\psi)\otimes (-\psi))=\d z$.  So just like the square root of a number, a square root of $\d z$ is only uniquely determined
up to sign.  If $\Sigma$ is the lower half of the complex $z$ plane, then a section $\psi$ of $\S$ that satisfies $w(\psi\otimes\psi)=-\d z$
is real and is a section of $\E$.  We describe this informally by writing $\psi=\pm \sqrt{-\d z}$ or $\psi=\pm \i\sqrt{\d z}$.

\begin{figure}
 \begin{center}
   \includegraphics[width=6in]{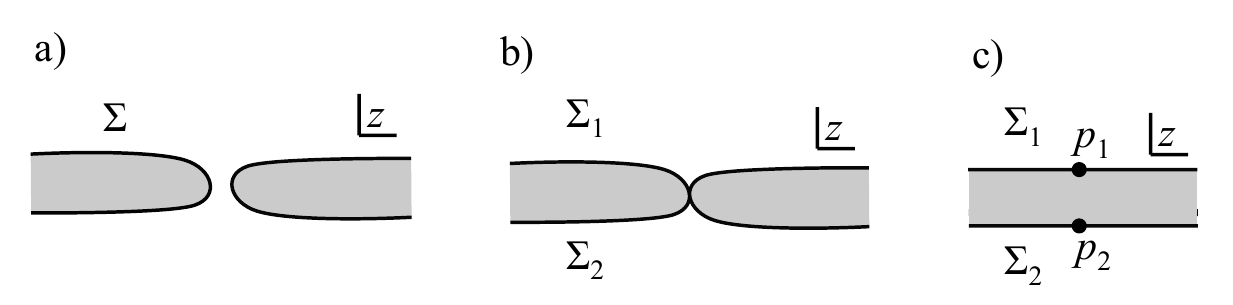}
 \end{center}  
\caption{\small (a) The complement of the shaded region of the complex $z$-plane is a Riemann surface $\Sigma$ with boundary.  It consists
of an upper and lower half plane  connected through a narrow neck.  (b) In real codimension 1, the neck collapses and
$\Sigma$ degenerates to a pair of branches $\Sigma_1$ and $\Sigma_2$ glued together along a double point. (c)  In this picture, the two branches
have been separated.  Now  $\Sigma_1$ and $\Sigma_2$
are upper and lower half-planes, respectively, with distinguished boundary punctures $p_1$ and $p_2$. Gluing $p_1$ to $p_2$ will return us to the singular configuration in (b).  \label{bdeg}}
\end{figure}

A trivialization of the real spin bundle $\E \to \partial \Sigma$ is given by any nonzero section of $\E$. For example, if $\Sigma$ is the upper
half $z$ plane, then $\E\to \partial\Sigma$ can be trivialized by $\psi=\pm\sqrt{\d z}$, and if $\Sigma$ is the lower half $z$ plane, then 
$\E\to\partial\Sigma$ can be trivialized by $\psi=\pm\i\sqrt{\d z}$.

With this in place, we can return to our problem.   In fig. \ref{bdeg}, we show the same open-string degeneration as in fig. \ref{deg}, but now we zoom in
on the important region where the degeneration occurs and do not specify what the Riemann surface $\Sigma$ looks like outside this region.  The open-string degeneration is drawn in
the figure ignoring spin structures and their trivializations.    In figure \ref{cdeg}, we repeat fig. \ref{bdeg}(a), but now providing information
about the trivializations of spin structures. 

 First of all, as there are no boundary punctures in this picture,\footnote{The Deligne-Mumford compactification is
defined in such a way that a degeneration never occurs at the location of an already existing puncture.  Hence in picture \ref{bdeg}(a), which shows
the part of $\Sigma$ in which an open-string degeneration occurs, we can assume that there are no boundary punctures.} the real spin bundle
of $\partial\Sigma$ is supposed to be trivialized everywhere in the picture.   The trivializations are easy to describe in the regions -- the upper and
lower left and right in the figure -- in which $\partial\Sigma$ is parallel to the real $z$ axis.    We will use the fact that as $\Sigma$ is a region in the
complex $z$ plane, the complex 1-form $\d z$ is defined everywhere on $\Sigma$; similarly it is possible to make a global choice of sign
of $\psi=\sqrt{\d z}$, though such a $\psi$ will not be everywhere real on $\partial\Sigma$.   The overall sign of what we mean by $\sqrt{\d z}$
will not be important in what follows.

\begin{figure}
 \begin{center}
   \includegraphics[width=2.5in]{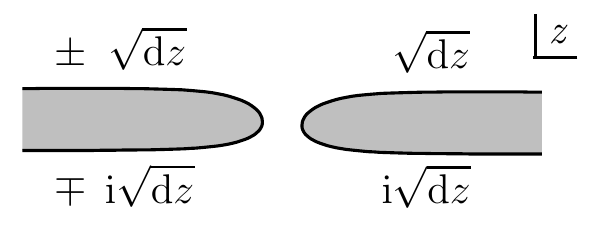}
 \end{center}  
\caption{\small Here we repeat fig. \ref{bdeg}(a), but now providing information on the trivialization of the spin bundle of $\partial\Sigma$. 
On the upper and lower left and right of the figure, $\partial\Sigma$ is parallel to the real $z$ axis and so the spin structure is trivialized
by a choice of $\pm\sqrt{z}$ (the upper regions) or $\pm \i\sqrt{\d z}$ (the lower regions). \label{cdeg}}
\end{figure}

 We begin on the upper right  of the picture
with $\E$ trivialized by $\psi=\sqrt{\d z}$.  (It would add nothing essentially new to use $-\sqrt{\d z}$ in the starting point, as the overall
sign of $\sqrt{\d z}$ is anyway arbitrary.)    Now on the upper left of the picture, we pick a trivialization $\pm \sqrt{\d z}$.  This sign is meaningful,
given that we used the trivialization $+\sqrt {\d z}$ on the upper right.  Now we continue through the narrow neck into the lower part of the picture.
As we do this, the boundary of $\partial\Sigma$ bends counterclockwise by an angle $\pi$ on the right of the figure and by an angle $-\pi$
on the left.  As a result, a section of $\S\to \partial\Sigma$ has to acquire a phase in order to remain real.   The trivialization of $\E$ that is defined
as $\sqrt{\d z}$ on the upper right will evolve to $\i\sqrt{\d z}$ on the lower right, and the trivialization of $\E$ that is defined as $\pm\sqrt{\d z}$ 
on the upper left will evolve to $\mp\i\sqrt{\d z}$ on the lower left.  

We see that with one choice of sign on the left part of the picture, the trivializations agree on the upper left and upper right of the figure
but not on the lower left and lower right; with the other choice of sign, matters are reversed. So when $\Sigma$ degenerates to the union
of two branches $\Sigma_1$ and $\Sigma_2$ that are to be joined by gluing a point $p_1\in\partial\Sigma_1$ to a point $p_2\in\partial\Sigma_2$, as in
fig. \ref{bdeg}(c),
the trivialization of the spin structure
of the boundary jumps in crossing $p_1$ but not in crossing $p_2$ or in crossing $p_2$ but not in crossing $p_1$.   In the construction studied in
\cite{PST,Tes,BT,STa}, precisely one of $p_1$ and $p_2$ plays no role and can be forgotten.  This is the basic reason that the boundary of $\bM$
behaves as if it is of real codimension two and the correlation functions are well-defined.  We provide more detail momentarily.

\subsection{Computations of Disc Amplitudes}\label{compdisc}

\def\tgst{\widetilde{\mathrm g}_{\mathrm{st}}}
Several concrete methods to compute in this framework have been deduced  \cite{PST,Tes,BT,STa}.   Here we will just describe
the simplest computations of disc amplitudes.

First let us discuss the proper normalization of a disc amplitude.  We write $\gst$ for the string coupling constant in topological gravity
of closed Riemann surfaces with its usual normalization, and $\tgst$ for the string coupling constant in the present theory.

In the standard approach, genus $g$ amplitudes are weighted by a factor of $\gst^{2g-2}$.    With theory $\T$ included,
this is replaced by $\tgst^{2g-2} 2^{g-1}$, where $2^{g-1}$ is the partition function of theory $\T$ (eqn. (\ref{onmo})).   The
relation between the two is thus 
\be\label{muflo} \tgst=\frac{\gst}{\sqrt 2}. \ee

A disc has Euler characteristic 1, so a disc amplitude is weighted by $1/\tgst=\sqrt 2/\gst$.  The partition function of theory $\T$ on
a disc is $1/2$ (as a disc has only one spin structure).  However, for any given set of boundary punctures, there are two possible
piecewise trivializations of the spin structure of the boundary, with the requisite jumps across boundary punctures.  These two
choices will contribute equally in the simple computations we will discuss, so we can take them into account by including a factor of 2.

The factors discussed so far combine to $2\cdot \frac{1}{2} \sqrt 2/\gst=\sqrt 2/\gst$.  In addition, in \cite{PST} it was found convenient
to include a factor of $1/\sqrt 2$ for every boundary puncture.   Thus,  let $\Sigma$ be a disc
 with $m$   boundary punctures and
$n$  bulk punctures labeled by integers $d_1,\dots,d_n$; let $\overline \M$ be the compactified moduli space of conformal structures
on $\Sigma$.    Then refining eqn. (\ref{proft}), the general disc ampitude is
\be\label{mondo}\langle \tau_{d_1}\tau_{d_2}\dots \tau_{d_n}\sigma^m\rangle_D =\frac{2^{(1-m)/2}}{\gst} \int_{\overline \M}
\psi_1^{d_1}\psi_2^{d_2}\cdots \psi_n^{d_n}.\ee   This formula agrees with eqn. (18) in \cite{PST}.
We have included factors of $\gst$ in this explanation, because that helps determine the factors of 2 that are needed to ensure that the
theory is consistent with the standard normalization in the case that a surface $\Sigma$ has no boundary.  However, in mathematical
treatments, $\gst$ is often set to 1, and we will do so in the rest of this section.
(No topological information is lost, since a given correlation function receives contributions only
from surfaces with a given Euler characteristic, and this determines the power of $\gst$.)  

In interpreting eqn. (\ref{mondo}), we consider the boundary punctures to be inequivalent and labeled, and we sum over all
possible cyclic orderings.  For example, let us compute $\la\sigma\sigma\sigma\ra$, which receives a contribution only from a disc
with three boundary punctures labeled 1,2,3.  There are two cyclic orderings (namely 123 and 132), and for each cyclic ordering, $\overline\M$
is just a point, with $\int_{\overline\M}1=1$.  So after settting $\gst=1$, eqn. (\ref{mondo}) with $n=0$, $m=3$, 
and including a factor of 2 from the sum over cyclic
orderings, gives
\be\label{purr}\langle\sigma^3\rangle = 1. \ee
Getting this formula was the motivation to include a factor $1/\sqrt 2$ for each boundary puncture.
Another simple formula is
\be\label{nurr} \langle \tau_0\sigma\rangle =1. \ee
This is again easy because the moduli space is a point.
With boundary punctures only, eqn. (\ref{purr}) is the only nonzero amplitude, for dimensional reasons, and similarly (\ref{nurr})
is the only additional nonzero disc amplitude with insertions of $\sigma$ and $\tau_0$ only.

\begin{figure}
 \begin{center}
   \includegraphics[width=4in]{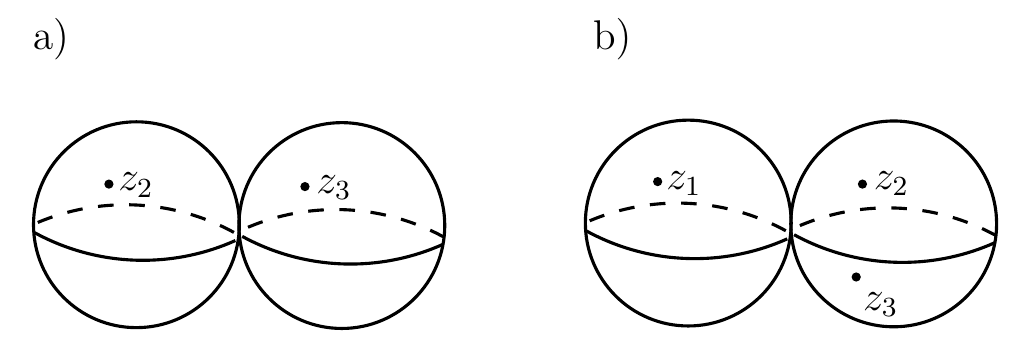}
 \end{center}  
\caption{\small (a) If the two-sphere $\Sigma$ degenerates to two branches with punctures $z_2$ and $z_3$ on opposite sides, then the
1-form $\rho=\d z/(z-z_2)(z-z_3)$ has poles on each branch, so in
particular it is nonzero on each branch.  (When $\Sigma$ degenerates, $\rho$ also acquires poles at the double point
that the two branches have in common, with equal and opposite residues on the two sides.) $\lambda$ also remains
nonzero.  (b) If instead $z_2$ and $z_3$
are on the same branch, then all poles of $\rho$ are on that branch and in fact $\rho=0$ on the other branch.  Since $\lambda$
is defined by setting $z=z_1$ in $\rho$, $\lambda$ vanishes if, as sketched here,
 $z_1$ is on the branch on which $\rho$ is identically zero.  \label{depiction}}
\end{figure}

The simplest method to compute arbitrary disc amplitudes is given by the recursion relations in Theorem 1.5 of \cite{PST},
and indeed the first of these relations is sufficient.   To explain it, first
 we recall the genus 0 recursion relations of \cite{Witten}.  It is convenient to define
\be\label{moxo}\ll \tau_{d_1}\tau_{d_2}\cdots \tau_{d_s}\rr=\left\la \tau_{d_1}\tau_{d_2}\cdots \tau_{d_s}\exp\left(\sum_{n=0}^\infty t_n\tau_n
\right)
\right\ra. \ee   Thus $\ll \tau_{d_1}\tau_{d_2}\cdots \tau_{d_s}\rr$ is an amplitude with specified insertions as shown, with all possible
additional insertions weighted by powers of the $t_n$.  We also write $\ll \tau_{d_1}\tau_{d_2}\cdots \tau_{d_s}\rr_0$ for the
genus 0 contribution to $\ll \tau_{d_1}\tau_{d_2}\cdots \tau_{d_s}\rr$.
Then one has the genus 0  recursion relation
\be\label{nuxx}\ll \tau_{d_1}\tau_{d_2}\tau_{d_3}\rr_0=\ll\tau_{d_1-1}\tau_0\rr_0 \ll \tau_0\tau_{d_2}\tau_{d_3}\rr_0.\ee
The proof goes roughly as follows.  For a smooth genus 0 surface $\Sigma$, we take  the complex $z$-plane plus a point at infinity. We
denote the specified punctures as $z_1,z_2,z_3$.
We will construct a convenient section $\lambda$ of the line bundle $\L_1\to \overline \M$ whose fiber is the cotangent bundle to $\Sigma$
at  $z_1$.
Let $\rho$ be the 1-form
\be\label{ruxx} \rho=(z_2-z_3)\frac{\d z}{(z-z_2)(z-z_3)}. \ee
It has poles at $z=z_2,z_3$, with residues 1 and $-1$, and elsewhere is regular and nonzero. These properties characterize $\rho$ uniquely, so $\rho$
does not depend on the coordinates used in writing the formula. Upon setting $z=z_1$ in $\rho$, we get a holomorphic 
section $\lambda$
of $\L_1\to\overline \M$; the divisor $\sf D$ of the zeroes of this section represents $c_1(\L_1)$.  But $\lambda$ never vanishes when
$\Sigma$ is smooth, because $\rho$ has no zeroes on the finite $z$-plane or at $z=\infty$.    If $\Sigma$ degenerates to
two components with $z_2$ and $z_3$ on opposite sides (fig. \ref{depiction}(a)), $\lambda$ is still everywhere nonzero.
But if $z_2$ and $z_3$ are contained in the same component (fig. \ref{depiction}(b)), then $\lambda$ vanishes on the other component.  Finally, then,
$\rho$ vanishes precisely if, as in the figure, $z_1$ is contained in the opposite component from the one containing $z_2$ and $z_3$.
Moreover, this is a simple zero (because $\rho$ has a simple zero at $z_2=z_3$).  So in $\tau_{d_1}=c_1(\L_1)^{d_1}$, we can replace one factor
of $c_1(\L_1)$ with a restriction to the divisor $\sf D$ that is depicted in fig. \ref{depiction}(b).   After making this substitution,
we are left with an insertion of $\tau_{d_1-1}$ on one branch and insertions of $\tau_{d_2}$ and $\tau_{d_3}$ on the other;
in addition, a new puncture corresponding to an insertion of $\tau_0$ appears on each branch, where the two branches meet.  All
this leads to the right hand side of eqn. (\ref{nuxx}).    It is not difficult to see that this recursion relation uniquely determines all genus zero amplitudes,
modulo the statement that the only nonzero amplitude with insertions of $\tau_0$ only is $\la\tau_0^3\ra_0=1$.

The disc recursion relation that we aim to describe can be formulated and proved in almost the same way.  Similarly to the previous case,
we define \be\label{ponzzo} \ll \tau_{d_1}\tau_{d_2}\cdots \tau_{d_s}\sigma^m\rr =\left\la\tau_{d_1}\tau_{d_2}\cdots \tau_{d_s}\sigma^m
\exp\left(\sum_{n=0}^\infty t_n\tau_n+ \v\sigma\right)\right\ra,\ee
and write $\ll  \tau_{d_1}\tau_{d_2}\cdots \tau_{d_s}\sigma^m\rr_D$ for the disc contribution.   The desired recursion relation is
\be\label{wonzo}\ll \tau_n\sigma\rr_D=\ll \tau_{n-1}\tau_0\rr_0\ll \tau_0\sigma\rr_D+\ll\tau_{n-1}\rr_D\ll\sigma^2\rr_D. \ee 
Given a knowledge of 
eqns. (\ref{purr}) and (\ref{nurr}) and vanishing of $\la \tau_0^n\sigma^m\ra_D$ for other values of $n,m$, 
it is not difficult to see that eqn. (\ref{wonzo}) determines all disc amplitudes in terms of the genus 0 amplitudes.  These in turn
can be determined, for example, from (\ref{nuxx}).

The proof of eqn. (\ref{wonzo}) is rather similar to the proof of the genus zero recursion relation (\ref{nuxx}).   However,
we will have to explain more fully what is meant in saying that one of the punctures in an open-string
degeneration should be forgotten.

Roughly speaking, we are going to again compute $c_1(\L_1)$, for one of the bulk punctures,
 from the zeroes of a convenient section $\lambda$ of $\L_1$.  However,
here because $\overline\M$ has a boundary, we have to discuss how to relate $c_1(\L_1)$ to the zeroes of a section.  

As discussed in section \ref{bd}, 
the boundary $\partial\overline\M$ of $\overline \M$ has a forgetful map in which precisely one of the extra boundary  punctures that appears
at an open-string degeneration is forgotten.  Let us write $\overline{\mathcal N}$ for the remaining moduli space when this puncture
is forgotten, so that the forgetful map is $\pi:\partial\overline\M\to \overline{\mathcal N}$.

Simplifying a little,\footnote{The general recipe has two further complications.  First, in general one is allowed to compute
using a multisection rather than a section.  This is important because the conditions on a section that we are about to state are difficult
to satisfy.  Second, the general procedure allows one to define $\prod_{i=1}^n c_1(\L_i)^{d_i}$, without defining the individual
$c_1(\L_i)$, by picking a multisection $s$ of $E=\oplus_{i=1}^n \L_i^{\oplus d_i}$.  This multisection should obey conditions
analogous to the ones that we will state momentarily.}  
the  recipe \cite{PST} is that $c_1(\L_1)$ can be represented by the zeros of any section $s$ of
$\L_1$ that is nonvanishing everywhere along $\partial\overline\M$, and whose restriction to $\partial\overline\M$ is a pullback
from $\overline{\mathcal N}$.  Alternatively, one can still calculate $c_1(\L_1)$ using any section $s$ of $\L_1$ that is everywhere nonzero
along the boundary, even if its restriction to the boundary is not a pullback.  But in this case,  $c_1(\L_1)$ is represented by a sum
of two contributions, one involving in the usual way the zeroes of $s$, and the second measuring the failure of the restriction of $s$
to be a pullback.

Setting $z=x+\i y$, we take a smooth disc $D$ to be the closed upper half-plane $y\geq 0$ plus a point at infinity.   On the left
hand side of eqn. (\ref{wonzo}), we see a distinguished bulk puncture that we place at $z_1=x_1+\i y_1$, $y_1>0$, and a distinguished
boundary puncture that we place at $x_0$.  
In the present case, there is a convenient section $\lambda$ of $\L_1$ that is everywhere nonzero along the boundary, but whose
restriction to the boundary is not a pullback.  
To construct it, rather as before, we set
\be\label{poho}\rho=(\bar z_1-x_0)\frac{\d z}{(z-\overline z_1)(z-x_0)}.\ee
This 1-form is regular and nonzero throughout $D$, except at the boundary point $x_0$.  Evaluating $\rho$ at  $z=z_1$, we get a section $\lambda$
of $\L_1$ that is regular and nonzero as long as $D$ is smooth.

\begin{figure}
 \begin{center}
   \includegraphics[width=4.0in]{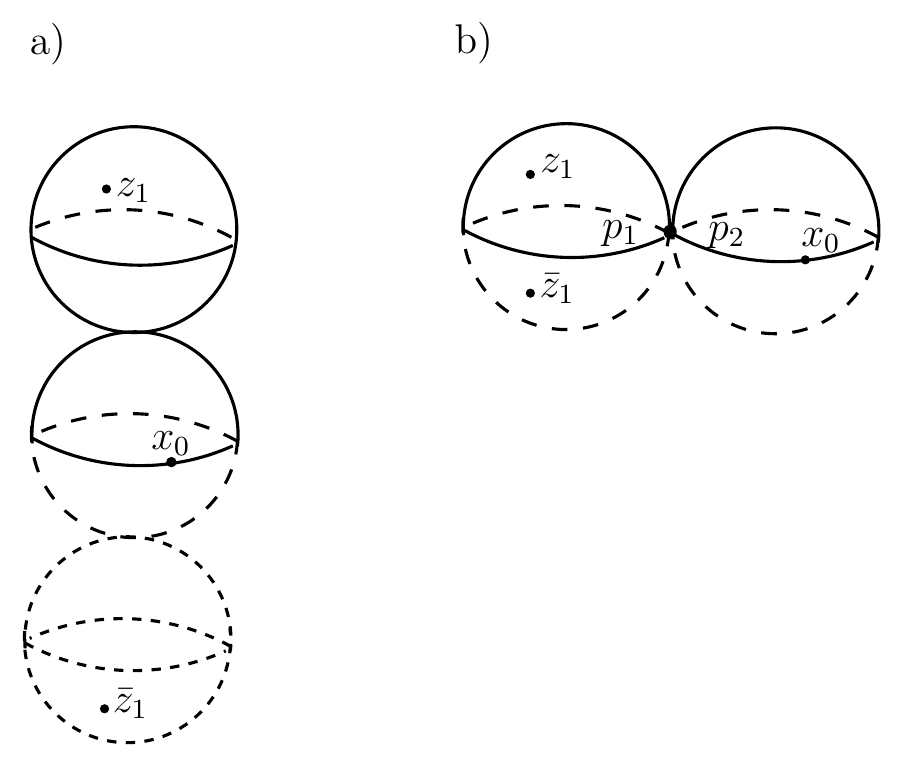}
 \end{center}  
\caption{\small (a) A disc $D$ splits up into the union of a disc and a sphere (upper half of the drawing).  
If the bulk puncture $z_1$ is contained in the sphere,
then the section $\lambda$ vanishes.  To see this, take the closed oriented double cover, obtained here by adding additional
components (lower half of the drawing, sketched with dotted lines).  It is a union of three spheres connected at double points.  The differential $\rho$ has poles
only on the bottom two components and vanishes identically on the top component.  So, setting $z=z_1$ to define $\lambda$, we learn
that, with $z_1$ being in the top component, $\lambda$ vanishes.  (b)  The same disc $D$ splits into
a union of two discs, again comprising the upper half of the drawing.
  The interesting case is that $z_1$ and $x_0$ are on opposite sides, as shown.  The oriented double cover
(the full drawing including the bottom half) 
is a union of two spheres.  $\rho$ has poles at $x_0$ and $\bar z_1$ and so is nonzero on both branches; hence $\lambda\not=0$ along
this divisor.  On the branch
containing $z_1$, $\rho$ has an additional pole at the point labeled $p_1$ where the two branches meet.  Therefore $\lambda$ depends on
$p_1$, and, if $p_1$ is the boundary puncture that is forgotten by the forgetful map $\pi:\partial\overline\M\to\overline{ \mathcal N}$,
then along this component of the boundary, $\lambda$ is not a pullback.
\label{splitup}}
\end{figure}

At a closed-string degeneration, where $D$ splits up into the union of a two-sphere and a disc (fig. \ref{splitup}(a)), $\lambda$
has a simple zero if and only if $z_1$ is on the two-sphere component.  This is responsible for the first term on the right hand side of the recursion relation (\ref{wonzo}).  
At an open-string degeneration, where $D$ splits up into the union of two discs (fig. \ref{splitup}(b)), $\lambda$ remains everywhere
nonzero.  However, in case the boundary puncture that is supposed to be forgotten is in the same component as $z_1$, $\lambda$
restricted to $\partial\overline{\M}$ is not a pullback from $\overline{\mathcal N}$.  The second term on the right hand
side of eqn. (\ref{wonzo}) corrects for this failure.  See fig. \ref{splitup} for an explanation of the statements about the behavior 
of $\lambda$ at degenerations.

\section{Interpretation Via Matrix Models}\label{intmat}

\subsection{The Loop Equations}\label{loopeqs}

Let us now briefly recapitulate the representation of topological gravity in terms of random matrix models. The simplest models are single matrix models of the form
\be
\label{matrixintegral}
Z =\frac{1 }{ {\rm vol}(U(N))} \int \d\Phi \cdot \exp\left(-\frac{1}{ \gst} \Tr\, W(\Phi)\right)
\ee
Here $\Phi$ is a hermitian $N \times N$ matrix integrated with the Euclidean measure for each matrix element, $W(x)$ is a complex polynomial, say of degree $d+1$, and $\gst$ is the string coupling constant. Since we divide by the volume of the ``gauge group'' $U(N)$, this integral should be considered the zero-dimensional analogue of a gauge theory---we integrate over matrices $\Phi$ modulo gauge transformations
\be
\Phi \to U \cdot \Phi \cdot U^{-1}.
\ee
In general, if $\mathrm{Re}\,W$ is not bounded below, one needs to complexify the matrix $\Phi$ and pick a suitable integration contour in the space of complex matrices to make the integral well-defined.   For a formal expansion in powers of $\gst$ and even for the formal
expansion in powers of $1/N$ that we will make shortly, this is not necessary and we can  consider (\ref{matrixintegral}) as a formal expression. 

\begin{figure}
 \begin{center}
   \includegraphics[width=3.5in]{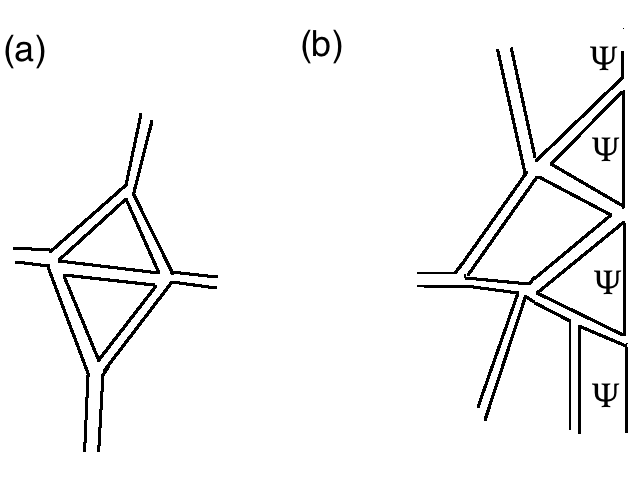}
 \end{center}  
\caption{\small (a)  Feynman diagrams of the matrix model are naturally ribbon graphs.  The two sides of a ribbon represent
the flow of the two ``indices'' of an $N\times N$ matrix $M^i{}_j$, $i,j=1,\dots, N$. The edges of the ribbons form closed loops.  Gluing
a disc to each such loop, the union of the ribbons and the discs is a two-manifold $\Sigma$ without boundary on which the given
Feynman diagram can be drawn.   (The edges of the ribbon are oriented -- not shown
here -- because the two indices transform according to inequivalent, dual representations of $U(N)$.  As a result, $\Sigma$
has a natural orientation.  (A similar model with symmetry group $O(N)$ or $Sp(N)$ leads to unoriented two-manifolds.) 
(b)  New variables $\Psi,$  $\bar\Psi$ transforming in
the $N$-dimensional representation of $U(N)$ and its dual are added to the matrix model.  
Because $\Psi^i$ and $\bar\Psi_j$, $i,j=1,\dots,N$ carry only a single 
``index'' -- rather than the two indices of the matrix $M^i{}_j$ -- their propagator is naturally represented by a single line rather than
the double line of the matrix propagator.   These single lines provide boundaries of the surface $\Sigma$, so now we get a ribbon
graph on $\Sigma$ with $\Psi$ propagating on the boundary of $\Sigma$, as shown.  For the model described in the text,
the $\Psi$ propagtor is $1/z$ and this gives a factor $1/z^L$ where $L$ is the length of the boundary.
 \label{IndexLoops}}
\end{figure}

In a perturbative expansion near a critical point of $W(\Phi)$,  the  Feynman diagrams become so-called ``fat" or ribbon graphs 
that can be conveniently represented (see fig.\ \ref{IndexLoops}(a)) by a double line \cite{thooft}.  These are graphs, in general with $\ell$ loops, that can
be naturally drawn on some oriented two-manifold of genus $g$.  The contribution of such a graph to the expansion of the matrix
integral is weighted by a factor 
\be
 (\gst N )^\ell \gst^{2g-2}.
\ee
The large $N$ or 't Hooft limit is obtained by taking the rank $N$ of the matrix to infinity and simultaneously the coupling $\gst$ to zero, 
keeping fixed the combination 
\be
\mu = \gst N.
\ee
 In the limit,  all graphs with a fixed genus and an arbitrary number of holes contribute in the same order, 
 so  the matrix integral has an asymptotic expansion of the form
\be
Z \sim \exp\left(\sum_{g \geq 0}  {\gst}^{2g-2} {\cal F}_g\right),
\ee
where ${\cal F}_g$ is the contribution of ribbon graphs of genus $g$. In general, the matrix integral
 depends on the coefficients of the potential $W$ and the particular critical point around which the expansion is made.  We describe
 the critical points at the end of this section.

Matrix integrals are governed by Virasoro constraints that are associated to the vector fields $L_n\sim -\Tr\, \Phi^{n+1} \frac{\partial}{\partial\Phi}$.
Though these constraints can be deduced directly from that representation of $L_n$, a fuller understanding with details that we will
need below can be obtained by diagonalizing the matrix as $\Phi = U \Lambda U^{-1}$, with $U$ unitary and 
$\Lambda=\mathrm{diag}(\lambda_1,\lambda_2,\cdots,\lambda_N)$.  The integral over $U$ cancels the factor of $1/\mathrm{vol}(U(N))$
in the definition of the matrix integral, and the integral becomes 
\be
\label{evint}
Z = \int \d^N\!\lambda \prod_{I<J} \left(\lambda_I - \lambda_J\right)^2 \exp\left(- \sum_I \frac{1}{ \gst} W(\lambda_I)\right).
\ee
If $x_i,\,i=1,\dots,d$ are the  critical points of the polynomial $W(x)$, then the critical points of the matrix function $\Tr\,W(\Phi)$ are found by setting each $\lambda_I$
equal to one of the $x_i$.   A critical point is labeled 
  by the number $N_i$ of eigenvalues with $\lambda_I=x_i$. (Note that the eigenvalues $\lambda_I$
are only defined up to permutation.) The large $N$ limit is taken is such a way that the ``filling fractions''
\be
\mu_i = \gst N_i,\qquad i=1,\ldots,d,
\ee
are all kept finite. These parameters characterize the saddle-points, and 
together with the coefficients of the polynomial $W(x)$  play the role of moduli of the matrix model.   (In our application, because
it only involves a local portion of the spectral curve, we will not really see these parameters.)

To derive the Virasoro constraints on the matrix integral, one can start with 
\be
\label{levint}
0 = \int \d^N\!\lambda\sum_K \frac{\partial}{\partial\lambda_K}\left( \frac{1}{x-\lambda_K} \prod_{I<J} \left(\lambda_I - \lambda_J\right)^2 \exp\left(- \sum_I \frac{1}{ \gst} W(\lambda_I)\right)\right).
\ee
This implies the identity 
\be\label{brevint} \left\langle \left(\sum_K\frac{1}{x-\lambda_K}\right)^2-\frac{1}{\gst}\sum_K \frac{W'(\lambda_K)}{x-\lambda_K}\right\rangle=0,\ee
where the symbol $\langle \cdots \rangle$ is defined by
\be\label{morz} \langle A\rangle= \frac{1}{Z} \int \d^N\!\lambda \,A\,\prod_{I<J} \left(\lambda_I - \lambda_J\right)^2 \exp\left(- \sum_I \frac{1}{ \gst} W(\lambda_I)\right).\ee
In eqn.\ (\ref{brevint}) we see the matrix resolvent $\Tr\,(x-\Phi)^{-1}=\sum_K \/(x-\lambda_K)^{-1}$, but as we will see a slightly more
convenient variable is
\be\label{worz} J(x)= \frac{1}{2}W'(x)-{\gst} \Tr
\frac{1}{x-\Phi}= \frac{1}{2}W'(x)-{\gst} \sum_K
\frac{1}{x-\lambda_K}. \ee
The identity (\ref{brevint}) is equivalent to 
\be\label{norz} \langle J(x)^2 \rangle = \left\langle \frac{1}{4}W'(x)^2-\gst\sum_K \frac{W'(x)-W'(\lambda_K)}{x-\lambda_K}\right\rangle , \ee
and we note that if $W$ is a polynomial, then 
\be\label{zorz} f(x) = -\gst\left\langle \sum_K \frac{W'(x)-W'(\lambda_K)}{x-\lambda_K}\right\rangle \ee
is a polynomial in $x$, as is
\be\label{porz}  P(x)= \frac{1}{4}W'(x)^2+f(x) .\ee
    If $W$ is a general function $W=\sum_{n\geq 0} \u_nx^n$ regular at $x=0$, then $P(x)$ is no longer
a polynomial but is regular at $x=0$.

 When we insert the expression $J(x)$ inside the matrix integral  (\ref{evint}), where we now consider a general function
\be
W(x) = \sum_{n\geq 0} \u_n x^n,
\ee
it can be written as a differential operator
\be
\label{diffop}
J(x) = \frac{1}{2}\sum_{n\geq 0} \left\{ (n+1) \u_{n+1} x^{n} + 2 \gst^2
 \frac{\partial }{ \partial \u_n} x^{-n-1} \right\}.
\ee
Comparing to standard formulas in conformal field theory, we are led to set
\be\label{iffo}J(x)=\frac{\gst}{\sqrt 2}\partial\varphi(x) \ee
where $\varphi(x)$ is a chiral boson in a $c=1$ conformal field theory
with canonical two-point function $\partial\varphi(x)\partial\varphi(y)\sim 1/(x-y)^2$.
Thus 
\be\label{loxx} \partial\varphi(x)={\sqrt 2}\left(\frac{W'(x)}{2\gst} -\sum_K \frac{1}{x-\lambda_K}\right), \ee and formally
\be\label{wiffo} \varphi(x) = \sqrt{2} \left(\frac{W(x)}{2\gst} - \sum_K\log (x-\lambda_K)\right)=\sqrt 2\left(\frac{W(x)}{2\gst}- \log \det (x-\Phi)\right).\ee
The corresponding stress tensor is 
\be\label{strest}T(x) =\frac{1}{2}(\partial\varphi)^2 = \frac{1}{\gst^2}J(x)^2. \ee 
Making the standard mode expansion 
\be\label{modex} T(x)= \sum_{k\in\Z}\frac{L_k}{x^{k+2}},\ee
the equation (\ref{norz}) becomes a set of differential equations for the partition function,
\be\label{odex} \sum_{k\in\Z}\frac{L_k}{x^{k+2}}Z=\gst^2 P(x)Z. \ee
Since $P(x)$ is regular at $x=0$, it contributes only to
the terms in eqn.\ (\ref{odex}) with $k\leq -2$ and those terms serve to determine\footnote{If $W$ is a polynomial
of degree $d+1$, then $P(x)$ is a polynomial of degree $2d$.  The condition on $W$ means that $\u_n=0$ for $n>d+1$, which leads
to $L_n=0$ for $n<-2d-2$.    The terms in eqn.\ (\ref{odex}) with $-2\geq k\geq -2d-2$ determine $P$ and the ones with $k<-2d-2$
are trivial identities.  If we consider a general function $W=\sum_{n\geq 0} \u_n x^n$ that is regular at $x=0$, then $P$ has a similar
power series expansion around $x=0$, and all of the constraints (\ref{odex}) with $n\leq -2$ serve to determine $P$.}
$P(x)$.   However,
for $k\geq -1$, $P(x)$ does not contribute to eqn.\ (\ref{odex}) and we get differential equations satisfied by $Z$:
\be\label{podex} L_n Z=0, ~~~n\geq -1. \ee
In this range of $n$, the $L_n$ are 
\begin{eqnarray}
\label{vir}
L_{-1} & = & \sum_{k \geq 1} k \u_k \frac{\partial}{ \partial \u_{k-1}}, \\
L_n &  =  & \sum_k k \u_k \frac{\partial}{ \partial \u_{k+n}} + \gst^2 \sum_{i+j=n} \frac{\partial^2 }{ \partial \u_i \partial \u_j}, \qquad n \geq 0.
\end{eqnarray}

\def\CC{{\mathcal C}}
So far, all of this is true for any $N$; we have not made any large $N$ approximation.  For any function $h$, the quantity
$\langle \gst \Tr \, h(\Phi)\rangle$ has a limit for large $N$, and for any two functions $h_1,h_2$, one has a large $N$ factorization
\be\label{mubb}\langle \gst^2\Tr\,h_1(\Phi)\,\Tr\,h_2(\Phi)\rangle \overset{N\to\infty}{\longrightarrow} \langle \gst\Tr\, h_1(\Phi)\rangle
\langle \gst\Tr\, h_2(\Phi)\rangle. \ee 
These properties can be demonstrated by an elementary study of the matrix integral. 
 In particular, both $\langle J\rangle$ and $f(x)$ have large $N$ limits, and in the large $N$ limit 
\be\label{nucu}\langle J(x)^2\rangle = \langle J(x)\rangle^2. \ee
We define
\be\label{ucu} y=\langle J(x)\rangle_0, \ee
where the subscript denotes the the large $N$ limit.
Eqn.\ (\ref{norz}) becomes for large $N$ a hyperelliptic equation for $y$
\be\label{curvex}y^2=P(x)=\frac{1}{4}W'(x)^2+f(x) \ee
and defines what is known as the spectral curve $\CC$.  In eqn. (\ref{curvex}), $y$, $W$, and $f$ all depend on the ``coupling parameters''
$\u_i$, though this is not shown explicitly. 
Remarkably, the spectral curve fully captures the solution of the matrix model. That is, all the perturbative functions ${\cal F}_g$ can be completely calculated using the geometric data of the spectral curve \cite{eynard-orantin}.

Suppose that $W$ is a polynomial of degree $d+1$, thus with $d$ critical points $p_1,\dots,p_d$.
Concretely, when one takes the large $N$ limit of the matrix integral, the first step is to pick a critical point of the matrix
function $\Tr\,W(\Phi)=\sum_{K=1}^N W(\lambda_K)$ about which to expand. The critcal points of
this matrix function are found simply by setting each of the $\lambda_K$ equal to one of the $p_j$.  Up to a permutation of the $\lambda$'s,
the critical points are classified by the number $N_i$ of eigenvalues that equal $p_i$.  The $N_i$ are subject to one
constraint
\be\label{ponzo}\sum_{i=1}^{d+1}N_i=N. \ee
A large $N$ limit is obtained in general by taking $N\to\infty$ keeping fixed
\be\label{rono}\mu_i=\gst N_i .\ee
In the large $N$ limit, the $\mu_i$ behave as continuous variables constrained only by
\be\label{ono}\sum_i \mu_i = \mu. \ee
Thus if $W$ is of degree $d+1$, there are $d$ ``moduli'' $\mu_i$ that appear in constructing the large $N$ limit.
We note  from eqn. (\ref{zorz}) 
that $f(x)$ is a polynomial of degree $d-1$ in $x$ and so has $d$ coefficients.  These $d$ coefficients are functions
of the moduli $\mu_i$ of the matrix model.   So except for constraints coming from keeping the $N_i$ real and positive, a $d-1$-parameter
family of $f$'s can arise, even for fixed $W$, by varying the critical
point about which one expands the matrix model.

In the above derivation, for finite $N$, we discovered that the matrix integral is governed by an operator-valued conformal field
$\partial\varphi(x)$.  For finite $N$, this field depends on the parameters of the matrix model, namely the $\u_i$ and $N$, as well
as $x$.  In the large $N$ limit, the matrix integral can be defined by an expansion around a particular saddle point, and then
new parameters appear.   For the ``bare'' matrix integral, without trying to compute the expectation value of the resolvent,
the extra parameters are the $\mu_i$.  When one tries to compute the expectation value of the resolvent, there is an
additional binary choice, since $\langle J(x)\rangle$ is governed by a quadratic equation with two roots.  In the large $N$ limit,
and also in the more refined double scaling limit in which $N\to\infty$ with $\mu=\gst N$ fixed, the formalism with the conformal
field $\partial\varphi$ remains valid, but this field now depends on additional parameters -- the $\mu_i$ and the choice of sign of $\langle
J(x)\rangle$.  

In our application, the $\mu_i$ will not be very important, since we will consider only the local behavior near a particular branch
point.  However, the extension of the conformal formalism to include the choice of sign of $\langle J(x)\rangle$ is important.
It means that $\partial\phi$ should be interpreted as a conformal field on the spectral curve $\CC$, the double cover of the $x$-plane
that is defined by the hyperelliptic equation (\ref{curvex}).   
The hyperelliptic curve has an involution $y\to -y$ that exchanges the two choices of the sign of $\langle J(x)\rangle$.  Since $\partial
\varphi$ is defined as a multiple of $J(x)$ (eqn. (\ref{iffo})), $\partial\varphi$ is odd under the hyperelliptic involution.

\subsection{Double-Scaling Limits And Topological Gravity}\label{doublescaling}

Topological gravity and other models of two-dimensional gravity coupled to matter  are obtained by taking a suitable double-scaling limit of the generic matrix model. These scaling limits are best understood in terms of the underlying spectral curve. For the so-called $(2,2p-1)$ minimal model CFT coupled to gravity, the corresponding spectral curve takes the form
\be
y^2 \sim x^{2p-1}.
\ee
This limiting curve can be obtained by starting from the generic case $y^2 = P(x)$, where $P$ is a polynomial of degree $2p$, and then making $2p-1$ branch points coincide and sending the remaining one to infinity. In particular for topological gravity, which corresponds to the case $p=1$, we choose to write the underlying curve as
\be
\label{topgravcurve}
\frac 1 2 y^2 = x
\ee
This curve can be obtained, for example, from the simple Gaussian matrix model, with a quadratic polynomial $W(x)=x^2$.  In this
example, the polynomial $P$ is $P(x)=x^2-c$, with a constant $c$.  There are branch points at $x=\pm \sqrt c$.   After shifting $x$
by a constant and 
 ``zooming in" to a single branch point, one gets the curve of eqn. (\ref{topgravcurve}).

In the limit that the spectral curve $\CC$ is described by eqn. (\ref{topgravcurve}), the operator-valued conformal field $\partial\varphi$
takes a simple form.  Because it is odd under the hyperelliptic involution $y\to -y$, its expansion in powers of $x$ has only
half-integer powers. We will choose to  parametrize the expansion as
 \be
\label{delphi}
\frac 12 \gst  \partial\varphi(x) =  x^\frac{1}{ 2} - \sum_{n\geq 0} (n+\frac{1}{ 2})s_n x^{n-\frac{1}{ 2}} - \frac 1 4  \gst^2 \sum_{n\geq 0}
\frac{\partial }{ \partial s_n} x^{-n- \frac{3}{ 2}}
\ee
Here $\varphi$ is what would usually be called a twisted chiral boson on the complex $x$-plane, with a twist field at $x=0$
(and another at $x=\infty$).      The $s_n$ are functions of the parameters $u_n$ of an underlying matrix model; the precise relationship
depends upon exactly what matrix model one starts with before passing to the limit in which the spectral curve $\CC$ reduces to the
curve $y^2=2 x$.   This relationship is not very important for us.  

What is important is the relationship between the $s_n$ and the corresponding parameters $t_n$ of topological gravity -- the parameters
that were introduced in eqn. (\ref{zoft}).  This relationship turns out to be
\be
t_n = \frac {(2n+1)!!}{2^n}  s_n.
\ee
This statement is part of the relationship between the matrix model and intersection theory on $\M_{g,n}$, as proved in \cite{K} as well as\cite{M2,OP,KL}. (Note that the factor $2^n$, which is not entirely standard, is a consequence of our particular normalization of the spectral curve in eqn.\ (\ref{topgravcurve}).)

Inserting these expressions into the loop equations then gives the familiar Virasoro constraints
\be
L_{n} Z =0,\qquad n\geq -1,
\ee
where the operators $L_n$ are modes of the stress tensor $T=\frac{1}{2}(\partial\varphi)^2$, with $\partial\varphi$ now given by
eqn. (\ref{delphi}).  That is, we have
\begin{eqnarray}
L_{-1} & = & - \frac{\partial}{\partial s_0} + \sum_{k\geq 1}  (k+\frac{1}{2}) s_k \frac{\partial}{ \partial s_{k-1}} + \frac{1}{2 \gst^2} s_0^2 ,\\
L_{0}  & = & - \frac{\partial}{\partial s_1} +  \sum_{k\geq 0}  (k+\frac{1}{2}) s_k \frac{\partial}{ \partial s_{k}} + \frac{1}{16}, \\
L_n &  =  & - \frac{\partial}{\partial s_{n+1}} + \sum_{k\geq 0}  (k+\frac{1}{2}) s_k \frac{\partial}{ \partial s_{k+n}} + 
\frac{1}{8} \gst^2 \sum_{i+j=n-1} \frac{\partial^2 }{ \partial s_i \partial s_j}, \qquad n \geq 1.
\end{eqnarray}
Note that these equations fix the normalization of the partition function. In particular if we set all variables $s_n=0$ for $n>0$, the $L_{-1}$ constraint gives the genus zero contribution (using $s_0=t_0$)
\be
\frac{\partial}{\partial t_0} F_0 = \frac1 2 {t_0^2}
\ee
corresponding to three closed-string punctures on the sphere 
\be
\langle \tau_0^3 \rangle_0 =1.
\ee
Note that in that case, with only $t_0$ non-zero, the spectral curve becomes
\be
\label{topgravcurveshifted}
\frac 1 2 y^2=  x -  t_0.
\ee

Returning to a theme from section \ref{volint},
we are now also in a position to write the spectral curve that corresponds to the model computing the volumes of the moduli space of curves. As we have seen in equation (\ref{gudd}), in that case the values of the coupling constants are 
\be
t_n = \frac{(-1)^n \xi^{n-1}}{(n-1)!},\qquad n \geq 2,
\ee
which corresponds to
\be
s_n= \frac{ n (-1)^n 2^{2n} \xi^{n-1}}{(2n+1)!},\qquad n\geq 2.
\ee
Plugging this into (\ref{delphi}) we find
\be
y = \frac{\sin(2 \sqrt{\xi x})}{2\sqrt{2\xi}}
\ee
which is, up to normalization coventions, the known expression for the spectral curve \cite{Eynard}.

Perhaps we should add another word about eqn.\ (\ref{delphi}).  Because the modes of $\partial\varphi(x)$ proportional to $x^{n-1/2}$,
$n\geq 0$, commute, we can just declare them to be multiplication by commuting variables $s_n$.  In a derivation that starts with
a matrix model based on a function $W(x)=\sum u_nx^n$, the $s$'s would be complicated functions of the $u$'s; the precise functions would
depend on exactly how one zooms in on a critical point to get to the spectral curve $y^2= 2 x$.  Once the coefficients of $x^{n-1/2}$,
$n\geq 0$ are fixed as $s_n$, the coefficients of other terms in $\partial\varphi(x)$ are uniquely determined by the commutation
relations and operator product expansion satisfied by $\partial\varphi(x)$.  

\subsection{Branes And Open Strings}\label{branesop}

Before we consider open strings within topological gravity, let us first discuss the formulation of open strings in a general random matrix model.\footnote{Early
references include \cite{Kostov,Minahan,J1,ZY,IT,J2}.} Open strings are naturally included by adding vector degrees of freedom. Let $\Psi, \overline{\Psi}$ be a pair of conjugate $U(N)$ vectors. We can choose these to be bosonic or fermionic variables. The natural interaction with the matrix variable $\Phi$ takes the form
\be
\int d\Psi\, d\overline{\Psi} \cdot \exp \left\{ - z \overline{\Psi}^T \Psi  + \overline{\Psi}^T \cdot \Phi \cdot  \Psi \right\}
\ee
The effect of adding these additional variables is that now the ribbon graph is naturally drawn on a two-manifold $\Sigma$ with boundary
(fig. \ref{IndexLoops}(b)).  The propagator of the vector variables has a factor $1/z$, leading to a factor $1/z^L$, where $L$ is the
length of the boundary of $\Sigma$.  

The integral over  $\Psi$ and $\overline{\Psi}$ just gives a determinant
\be
\det(z-\Phi)^{\pm 1}
\ee
(apart from an irrelevant constant factor that could be absorbed in normalizing the measure).  
Here the sign in the exponent is $-1$ or $+1$ if  $\Psi, \overline{\Psi}$ are bosons or fermions.  In terms of the Feynman diagram
expansion, this sign means that for fermions, one will get an extra $-1$ for every component of the boundary of $\Sigma$.

Instead of including the variables $\Psi,$ $\bar\Psi$ in the model, it is equivalent to simply consider a matrix model with
an extra factor of $\det(z-\Phi)^{\pm 1}$ in the integrand.  However, it turns out that it is slightly more convenient to accompany
this factor with a prefactor $e^{\mp W(z)/2 \gst}$, which is a ``trivial'' modification in the sense that it does not depend on the matrix variables.
Thus we consider the modified matrix model based on the integral
\be\label{loboxx}
\frac{1 }{{\rm vol}(U(N))} \int d\Phi \cdot \exp\left(-\frac{1}{\gst} \Tr\, W(\Phi)\right) \det(z-\Phi)^{\pm 1}e^{\mp W(z)/2 \gst}
\ee
Loosely 
\be\label{thelf} V(z)=\det(z-\Phi)e^{-W(z)/2\gst},   ~~~~V^*(z) =\det(z-\Phi)^{-1}e^{W(z)/2\gst} \ee
are ``operators'' that create a brane or antibrane with the ``modulus'' $z$.  
We will see that $z$ has the interpretation of a value of $x$, which parametrizes the base of the hyperelliptic spectral curve\footnote{Thus, $z$ parametrizes
a pair of points on the spectral curve that are exchanged by the hyperelliptic involution $y\to -y$.  This is somewhat analogous to the fact that in section \ref{branes},
the brane had locally two components, which globally are exchanged by a sort of monodromy.} $y^2=P(x)$.
More generally, one could add several sets of vector degrees of freedom $\Psi_a, $ $\bar\Psi_a$,  $a=1,\dots, r$, each
with its own modulus $z_a$.     For definiteness, we will consider the case of insertion of just one factor of $V$:
\be\label{obo}Z_V(z)=\frac{1 }{{\rm vol}(U(N))} \int d\Phi \cdot \exp\left(-\frac{1}{\gst} \Tr\, W(\Phi)\right) \det(z-\Phi) e^{-W(z)/2 \gst}.
\ee

It is not difficult to derive the modification of the Virasoro equations that reflects the presence of a
brane.  Repeating
the derivation of eqn. (\ref{brevint}), we get
\be\label{revint} \left\langle \left(\sum_K\frac{1}{x-\lambda_K}\right)^2-\frac{1}{\gst}\sum_K \frac{W'(\lambda_K)}{x-\lambda_K}-\sum_K\frac{1}{(x-\lambda_K)(z-\lambda_K)}\right\rangle_{{ V(z)}}=0.\ee
Here $\langle A\rangle_{V(z)}$ is defined, by analogy with eqn. (\ref{morz}), as the expectation of $A$ in the matrix integral $Z_{V(z)}$.
However, it turns out that it is slightly more convenient to make the insertion of $V$ explicit and to write the equivalent  identity 
\be\label{wevint}\left\langle\left( \left(\sum_K\frac{1}{x-\lambda_K}\right)^2-\frac{1}{\gst}\sum_K \frac{W'(\lambda_K)}{x-\lambda_K}-\sum_K\frac{1}{(x-\lambda_K)(z-\lambda_K) } \right)  \cdot V(z)\right\rangle=0,\ee
where $\langle A\rangle$ is defined precisely as in eqn. (\ref{morz}), with the original matrix integral $Z$.  

We can write the integral (\ref{obo}) as 
\be\label{lobo}
Z_V =\frac{1 }{{\rm vol}(U(N))} \int d\Phi \cdot \exp\left(-\frac{1}{\gst} \Tr\, \left(W(\Phi)-\gst\log(z-\Phi)\right)\right) e^{- W(z)/2 \gst},
\ee
suggesting that in the definition of $J(x)$, we should just shift $W(\Phi)\to W(\Phi)-\gst \log(z-\Phi)$ and hence $W'(x)\to W'(x)+\gst/(z-x)$.
So we define
\be\label{robo} J(x) =   \frac{1}{2}W'(x)+\frac{\gst}{2(z-x)}-{\gst} \sum_K
\frac{1}{x-\lambda_K} \ee
and again
\be\label{wobo} T(x)=\frac{J(x)^2}{\gst^2}. \ee
Because we multiplied the partition function with the factor $e^{-W(z)/2\gst}$, which introduces an extra explicit dependence on the coefficients $u_n$, the formula for $J(x)$ as a differential operator is still given by equation (\ref{diffop}).
We get the identity
\be\label{tofog}
\left\langle T(x) V(z) \right\rangle = \left(P(x) +\frac {1}{4}\frac {1}{(x-z)^2} + \frac{1}{x-z}\frac{\partial }{\partial z}\right) \left\langle V(z) \right\rangle
\ee
where now 
\be\label{ubbx}P(x) = \frac{1}{4}W'(x)^2+f(x) - \frac{1}{2} \gst \frac{W'(z)-W'(x)}{z-x}.\ee
and the definition of $f(x)$ becomes
\be\label{tzorz} f(x) = -\gst   \frac 1 {\left\langle V(z) \right\rangle}  \left\langle \sum_K \frac{W'(x)-W'(\lambda_K)}{x-\lambda_K}\cdot V(z) \right\rangle. \ee
$P(x)$ has the same essential properties as before: it is a polynomial of degree $2d$ if $W(x)$ is a polynomial of degree $d+1$,
and if $W(x)$
has a general expansion $\sum_{n\geq 0}\u_n x^n$, then $P(x)$ is regular at $x=0$.  Moreover, $P(x)$ is regular at $x=z$.

Eqn.\  (\ref{tofog}) has a nice interpretation.    We can interpret $V(z)$ as an insertion on the spectral curve (which generically is
locally parametrized by $x$) of a primary field of
conformal dimension $h=1/4$.  On the right hand side of eqn. (\ref{tofog}), we see the expected singular contributions to the
$T(x)V(z)$ operator product expansion,
\be\label{expterms}
T(x) \cdot V(z) \sim \frac{h }{(x-z)^2} V(z) +\frac {1}{ x-z} \partial_z V(z) + \ldots
\ee
 as well as regular terms that are contained in $P(x)$.  
Indeed, comparing eqn. (\ref{wiffo}) to the definition (\ref{thelf}), we see that 
\noindent
we can identify $V$ and $V^*$ in terms of the conformal field $\varphi$ as 
\be
V(z) = e^{- \varphi(z)/\sqrt{2}} ,\qquad V^*(z) = e^{\varphi(z)/\sqrt{2}}
\ee 
These are indeed standard expressions for conformal primaries of dimension $1/4$.  In the large $N$ limit the scalar $\varphi$ and therefore also the vertex operator $V$ can be expressed in terms of the spectral curve data
\be
\varphi(z) \sim \frac {\sqrt 2} \gst \left( \int^z y(x)\d x + {\cal O}(\gst)\right)
\ee
Since there are two roots in the hyperelliptic spectral curve, there are two saddle-points that dominate the expectation value of $V(z)$ 
\be
\label{wkb}
\left\langle V(z) \right\rangle \sim \left\{ A e^{-\frac 1 \gst \int^z \!  y(x)dx} + B e^{\frac 1 \gst \int^z \! y(x)dx} \right\} \left(1 + {\cal O}(\gst)\right)
\ee
for some coefficients $A,B$ given by the one-loop correction. These two contributions, that only appear in string perturbation theory, can be considered as the manifestation of the two branes ${\cal B}'$ and ${\cal B}''$ as discussed in the A-model in section \ref{branesop}.  Note that they are interchanged by flipping the sign of $\gst$.

\def\c{{\sf c}}\def\o{{\sf o}}
Just as before, the terms in eqn.\ (\ref{tofog}) involving negative powers of $x$ give Virasoro constraints
\be
L_n Z_V=0,\qquad n\geq -1, 
\ee
while the terms involving nonnegative powers determine $P(x)$ or are trivial identities.    However, there are additional 
terms in the Virasoro generators.  We write the Virasoro generators as $L_n=L_n^\c+L_n^o$, where superscripts $\c$ and $\o$
represent ``closed-string'' and ``open-string'' contributions.   $L_n^\c$  comes from $T(x)$ on the left hand
side of eqn. (\ref{tofog}) and is given by the same formula (\ref{vir}) as before.
To find $L_n^\o$, we move the singular terms in eqn (\ref{tofog}) to the left hand side of the equation and expand in powers of $1/x$:
\be\label{polgo}-\frac{1}{x-z}\frac{\partial}{\partial z}-\frac{1}{4(x-z)^2}=-\sum_{k=0}^\infty \frac{1}{x^{k+1}}\left(z^k\frac{\partial}{\partial z}
+\frac{1}{4}k z^{k-1}\right). \ee
Thus 
\be\label{olgo}L_k^\o = -z^{k+1}\frac{\partial}{\partial z}-\frac{1}{4}(k+1)z^k.\ee
Consequently
\be\label{tvir}
L_n=L_n^\c+L_n^\o = \sum_k k u_k \frac{\partial}{ \partial u_{k+n}} + \gst^2 \sum_{i+j=n} \frac{\partial^2 }{ \partial u_i \partial u_j}
 -z^{n+1}\frac{\partial}{\partial z}-\frac{1}{4}(n+1)z^n.
\ee

On top of these Virarsoro constraints, there is another useful relation that should be added. Recall that with the introduction of the brane modulus $z$, the partition function depends on one more variable, and we expect to find an accompanying relation to determine the matrix model. This extra relation  can be considered as the analogue of the BPZ equation for degerate fields. It is obtained as the limit of the expression $T(x) V(z)$ when we take $x$ to $z$. 
The equation can be derived by observing that \cite{ACDKV}
\be
\frac {\partial^2} {\partial z^2} Z_V = \left\langle \left( \sum_K \frac 1 {(z-\lambda_K)}\right)^2 -\sum_K \frac 1{(z-\lambda_K)^2} - \frac 1 {2 \gst} W''(z) \right\rangle_{V(z)} .
\ee
In the right-hand side we recognize part of the loop equation (\ref{wevint}) in the case $x=z$. Combining the two equations we obtain a second-order differential equation in $z$
\be
\label{BPZ}
\left( \frac {\partial^2} {\partial z^2} - Q(z) \right) Z_V = 0
\ee
where now
\be
\label{defQ}
Q(z)= \lim_{x\to z} P(x)= \frac{1}{4}W'(z)^2 -  \frac 1 2 \gst W''(z) + g(z),
\ee
with
\be
g(z) = \lim_{x\to z} f(x) = -\gst  \frac 1 {\left\langle V(z) \right\rangle}  \left\langle \sum_K \frac{W'(z)-W'(\lambda_K)}{z-\lambda_K}\cdot V(z) \right\rangle. 
\ee
Note that $g(z)$ can in general be a complicated function of $z$, not necessarily polynomial. Together, the Virasoro constraints combined with equation (\ref{BPZ}) determine the behavior of the open string partition function as a function of the couplings $t_n$ and $z$.

Let us now consider these equations in the double-scaling limit, where the spectral curve takes the form 
\be
\frac 1 2 y^2 =  x - t_0.
\ee
In the absence of any further deformations---that is, without any other closed string insertions than the bulk puncture $t_0$---the open string partition function $Z_V(z)$ is very simple to compute. We obtain this case by taking the limit of the Gaussian model $W(x) = ax^2$, for which we find
\be
Q(x) = a^2 x^2 - c, \qquad c=\gst (2N+1).
\ee
and zoom again in on one of the branch points. In this limit the function $Q(z)$ becomes simply $Q= 2(z - t_0)$, and consequently equation (\ref{BPZ}) becomes the Airy equation
\be
\left(\frac 1 2 \gst^2 \frac {\partial^2} {\partial z^2} - z + t_0 \right) Z_V = 0.
\ee
The solution is the Airy function
\be
\label{airy}
Z_V(z) = \int dv \ e^{\frac 1 {\gst} (-vz + v^3/6+  v t_0)}
\ee
In this case one can also directly take the double-scaling limit of the exact expression for $Z_V(z)$ in the Gaussian model, where it given by the $N$-th eigenfunction of the harmonic oscillator, see {\it e.g.} the discussion in \cite{MMSS}. 

We know claim that the brane partition function, as computed in the double-scaled matrix model, is related to the topological gravity partition function 
by a Laplace transform
\be
Z_{\mathrm{top}}(v) = \int dz \ e^{\frac 1 \gst vz} Z_V(v).
\ee
Something similar has been encountered in the B-model. It has been claimed in \cite{AV, ADKMV} for example, that there is an important subtlety if one introduce branes on a spectral curve. One can insert branes  at a fixed value $x=z$ or at a fixed value of $y=v$. These two brane insertions are exchanged by a Laplace (or Fourier\footnote{Note that all these functional transforms are here considered as operations on formal power series.}) transform.

We have to compare this answer with the calculation in topological gravity where one computes insertions of the bulk puncture operator $\tau_0$ and the boundary puncture operator $\sigma$ 
\be
Z_{\mathrm{top}}(v) = \exp \sum_{\chi=-n} \gst^n \langle e^{t_0\tau_0 + v\sigma}\rangle
\ee
as a sum over surfaces with Euler number $-n$. In the absence of other operators, as discussed in section \ref{compdisc},  only two non-vanishing contributions are
expected: the disc with three insertions of $\sigma$, or with one insertion of $\sigma$ and one of $\tau_0$:
\be
\langle \sigma^3 \rangle_D = 1,\qquad \langle \tau_0 \sigma \rangle_D = 1.
\ee
So the correct answer should be
\be
\label{topdisc}
Z_{\mathrm{top}}(v) = e^{\frac 1 \gst( v^3/6 + v t_0)},
\ee
which is consistent  with the matrix model calculation (\ref{airy}).

One can now include arbitrary closed string perturbations and use this identification for the full partition functions. This becomes clear by considering the combined Virasoro constraints. If one takes into account the above Laplace transformation, these now take the form
\begin{eqnarray}
L_n^\c Z_{\mathrm{top}}(v) &  = & \int dz \ e^{\frac 1 \gst vz} \left[\frac{1}{4} (n+1)z^n + z^{n+1} \frac{\partial}{\partial z}\right] Z_V(z) \\
& = & \gst^n
\left[-\frac{3}{4} (n+1)\left(\frac{\partial}{ \partial v}\right)^n  - v \left(\frac{\partial }{\partial v}\right)^{n+1} \right] Z_{\mathrm{top}}(v).
\end{eqnarray}
This is indeed the expression given in \cite{PST}. This completes the identification of the double-scaled matrix model with the open-closed topological string partition function.

%The fact that brane insertion is given by the vertex operator $V(x)= e^{-\varphi(x)/\sqrt 2}$ can be further substantiated. As we discussed, in the double-scaling limit relevant to topological gravity, the field $\partial\varphi(x)$ has the mode expansion given in (\ref{delphi}). With the normalization $t_n = (2n+1)!! s_n$ and the relation $\partial/\partial t_n = \tau_n = \psi^n$ we use this expansion to express the field $\varphi(x)$ directly in terms of the tautological classes as
%\be \sqrt 2 \varphi(x) = \frac 2 {3 \gst} x^{\frac 2 3} + \frac 1 \gst \sum_{n \geq 0} \frac 1 {(2n+1)!!} t_n x^{n+\frac 1 2} +  \gst \int db\ %e^{b x^{\frac 1 2}} e^{\frac 1 2 b^2 \psi} \ee
%Here the operator $\exp \frac 1 2 b^2 \psi$ can be seen as creating a loop on the world-sheet of length $b$.  Using this expression %we can write the vertex operator
%\be V(x)= e^{-\varphi(x)/\sqrt 2} = \exp \int db\ e^{b x^{\frac 1 2}} e^{\frac 1 2 b^2 \psi} \ee

We thank D. Freed, R. Penner,
 and J. Solomon for comments on the manuscript. Research of EW supported in part by NSF Grant PHY-1606531.

\bibliographystyle{unsrt}

\begin{thebibliography}{99}

\bibitem{M1}
M. Mirzakhani, ``Simple Geodesics  and Weil-Petersson Volumes
Of Moduli Spaces of Bordered Riemann Surfaces,'' Invent. Math. {\bf 167}
(2007) 179-222.

\bibitem{M2}M. Mirzakhani, ``Weil-Petersson Volumes And Intersection Theory On The Moduli Space Of Curves,''
Journal of the American Mathematical Society {\bf 20} (2007) 1-23.



\bibitem{PST}
R. Pandharipande, J. P. Solomon, and R. J. Tessler, ``Intersection Theory On Moduli of Disks, Open KdV, and Virasoro,'' arXiv:1409.2191.



\bibitem{Tes}
R. Tessler, ``The Combinatorial Formula For Open Gravitational Descendants,''  arXiv:1507.04951.

\bibitem{BT}
A. Buryak and R. J. Tessler, ``Matrix Models And A Proof
Of The Open Analog Of Witten's Conjecture,'' arXiv:1501.07888.

\bibitem{STa}
J. P. Solomon and R. J. Tessler, to appear.

\bibitem{Wein}
D. Weingarten, ``Euclidean Quantum Gravity On A Lattice,'' Nucl. Phys. {\bf B210} [FS6] (1982) 229.

\bibitem{Kaz} V. Kazakov, ``Bilocal Regularization Of Models Of Random Surfaces,'' Phys. Lett. {\bf 150B} (1985)  282.

\bibitem{David}
F. David, ``Randomly Triangulated Surfaces In Two Dimensions,'' Nucl.
Phys. {\bf B257} [FS14] (1985) 45.

\bibitem{Amb} 
J. Ambjorn, B. Durhuus, and J. Frohlich, Nucl. Phys. {\bf B257} [FS14] (1985) 433.

\bibitem{KKM}  V. Kazakov, I. Kostov, and A. Migdal, ``Critical Properties Of Randomly Triangulated Planar
Random Surfaces,'' 
Phys. Lett. {\bf 157B}  (1985) 295.

\bibitem{BK} E. Brezin and V.  Kazakov, ``Exactly Solvable Field Theories Of Closed Strings,'' Phys. Lett. {\bf B236} (1990) 144.

\bibitem{DS} M.  Douglas and S. Shenker,  ``Strings In Less Than One Dimension,'' Nucl. Phys. {\bf B335} 635 (1990).

\bibitem{GM} D. J. Gross and A.  Migdal,  ``Nonperturbative Two-Dimensional Quantum Gravity,'' Phys. Rev. Lett. {\bf 64} (1990) 127.

\bibitem{FGZ}
P. Di Francesco, P. H. Ginsparg, and J. Zinn-Justin, ``2-D Gravity and Random Matrices,''
Phys. Rept. {\bf 254} (1995) 1-133.




\bibitem{Wittenone}
E. Witten, ``On The Structure Of The Topological Phase Of Two-Dimensional Gravity,''
Nucl. Phys. {\bf B340} (1990) 281-332.

\bibitem{Witten}
E. Witten, ``Two-Dimensional Gravity And Intersection Theory On Moduli Space,''
Surveys Diff. Geom. {\bf 1} (1991) 243-310.


\bibitem{D}
M. Douglas, ``Strings In Less Than One Dimension
and the Generalized KdV Equations,''
Phys. Lett. {\bf 238B} (1990) 176-80.

\bibitem{DVV} Robbert Dijkgraaf, H. L. Verlinde, and E. P. Verlinde,
``Loop Equations and Virasoro Constraints in Nonperturbative 2-D Quantum Gravity,''
Nucl. Phys. {\bf B348} (1991) 435-56.

\bibitem{K}M. Kontsevich, ``Intersection Theory On The Moduli Space Of Curves And The Matrix Airy Function,'' Commun. Math. Phys. {\bf 147} (1992)
1-23.

\bibitem{OP} A. Okounkov and R. Pandharipande, 
``Gromov-Witten Theory, Hurwitz Numbers, and Matrix Models,'' Proc. Symp. Pure Math. {\bf 80.1} (2009) 325-489.
math.AG/0101147.

\bibitem{KL} M. E. Kazarian and S. K. Lando, ``An Algebro-Geometric Proof of Witten's Conjecture,'' J. Am. Math. Soc. {\bf 20} (2007) 1079-89.

\bibitem{Wittengauge}
E. Witten, ``On Quantum Gauge Theories In Two Dimensions,'' Commun. Math. Phys. {\bf 141} (1991) 153-209. 

\bibitem{Kostov}
I. K. Kostov,  ``Exactly Solvable Field Theory Of $D=0$ Closed And Open Strings,''
Phys. Lett. {\bf B238} (1990) 181-86.

\bibitem{Minahan}
J. A. Minahan, ``Matrix Models With Boundary Terms
and The Generalized Painlev\'{e} II Equation,'' Phys. Lett. {\bf B268} (1991) 29-34.

\bibitem{J1} S. Dalley, C. V. Johnson, T. R. Morris, and A. Watterstam, Mod. Phys. Lett. {\bf A7} (1992) 2753-62, peh-th/9206060.

\bibitem{ZY}
Z. Yang, ``Dynamical Loops in $D=1$ Random Matrix Models,'' Phys. Lett. {\bf B257} (1991) 40-44.

\bibitem{IT}
Y. Itoh and Y. Tanii, ``Schwinger-Dyson Equations Of Matrix Models For Open And Closed Strings,''
Phys. Lett. {\bf B289} (1992) 335-41, hep-th/9202080.

\bibitem{J2}C. V. Johnson, ``On Integrable $c<1$ Open String Theory,'' Nucl. Phys. {\bf B414} (1994) 239-66, hep-th/9301112.




\bibitem{BH}
E. Br\'{e}zin and S. Hikami,  ``Random Matrix, Singularities, and Open/Closed Intersection Numbers,'' Commun. Math. Phys. {\bf 02} (2013) 035.

\bibitem{BH2}
E. Br\'{e}zin and S. Hikami, {\it Random Matrix Theory With An External Source} (Springer Briefs in Mathematical Physics, vol. 19). 

\bibitem{Alex}
A. Alexandrov, ``Open Intersection Numbers and Free Fields,'' arXiv:1606.06712.

\bibitem{Ok}
A. Okounkov, ``Random Trees and Moduli of Curves,'' arXiv:math/0309075.

\bibitem{Wolpert1}
S. Wolpert, ``Chern Forms And The Riemann Tensor For The Moduli Space Of Curves,'' Invent. Math. {\bf 85} (1986) 119-45.

\bibitem{Wolpert2}
S. Wolpert, ``On The Homology Of The Moduli Space Of Stable Curves,'' Ann. Math. {\bf 118} (1983) 491-523.

\bibitem{Z}
P. Zograf, ``On The Large Genus Asymptotics Of Weil-Petersson Volumes,'' arXiv:0812.0544.

\bibitem{Penner}
R. Penner, ``Weil-Petersson Volumes,'' J. Diff. Geom. {\bf 35} (1992) 599-608.


\bibitem{Goldman}
W. Goldman, ``The Symplectic Nature Of Fundamental Groups Of Surfaces,''
Adv. Math. {\bf 54} (1984) 200-25.

\bibitem{ABott}
M. F. Atiyah and R. Bott, ``The Yang-Mills Equations Over Riemann Surfaces,'' Phil. Trans. Roy. Soc. Longon {\bf A308} (1982) 523.




\bibitem{Wittengaugetwo}
E. Witten, ``Two-Dimensional Gauge Theory Revisited,'' J. Geom. Phys. {\bf 9} (1992) 303-68.



\bibitem{McS}
G. McShane, ``Simple Geodesics and A Series Constant Over Teichmuller Space,'' Invent. Math. {\bf 132}
(1998) 607-32.

\bibitem{Eynard}
B. Eynard, ``Recursion Between Mumford Volumes of Moduli Spaces,''
Ann. Henri Poincar\'e (2011) 12: 1431.

\bibitem{KMZ}
R. Kauffman, Yu. Manin, and D. Zagier, ``Higher Weil-Petersson Volumes of Moduli Spaces Of Stable $n$-Pointed Curves,'' 
arXiv:alg-geom/9604001.


\bibitem{MZ}
Yu. Manin and P. Zograf, ``Invertible Cohomological Field Theories And Weil-Petersson Volumes,'' arXiv:math/9902051.




\bibitem{ABT}
A. Alexandrov, A. Buryak, and R. Tessler, ``Refined Open Intersection Numbers and the Kontsevich-Penner Matrix Model,''
arXiv:1702.02319. 


\bibitem{AS}
M. F. Atiyah and I. M. Singer, ``The Index Of Elliptic Operators, V,'' Ann. Math. {\bf 93} (1971) 139-49.

\bibitem{Aspin}
M. F. Atiyah, ``Riemann Surfaces and Spin Structures,'' Ann. Scientifique de l'\'{E}.N.S. {\bf 4} (1971) 47-62.


\bibitem{WittenFP}
E. Witten, ``Fermion Path Integrals and Topological Phases,'' Rev. Mod. Phys. {\bf 88} (2016) 035001, arXiv:1508.04715.

\bibitem{Wittenveryold}
E. Witten, ``Algebraic Geometry Associated To Matrix Models Of Two-Dimensional Gravity,'' in {\it Topological Methods In Modern Mathematics},
I. R. Goldberg and A. V. Phillips, eds. (Publish or Perish, Inc., Houston TX, 1993).


\bibitem{Kitaev}
A. Kitaev, ``Unpaired Majorana Fermions In Quantum Wires,'' Usp. Fiz. Nauk. (Suppl.) {\bf 171} 9, arXiv:cond-mat/0010440.

\bibitem{KS}
A. Kapustin and N. Seiberg, ``Coupling a QFT to a TQFT and Duality,'' arXiv:1401.0740.

\bibitem{Wittentopsigma}
E. Witten, ``Topological Sigma Models,'' Commun. Math. Phys. {\bf 118} (1988) 411-449.

\bibitem{GMW}
D. Gaiotto, G. Moore, and E. Witten, ``Algebra Of The Infrared: String Field Theoretic
Structures In Massive ${\mathcal N}=(2,2)$ Field Theory In Two Dimensions,'' arXiv:1506.04087.

\bibitem{Freed}
D. Freed, ``Two Index Theorems In Odd Dimensions,''  Comm. An. Geom. {\bf 6} (1998) 317-29, dg-ga/9601005.

\bibitem{FS}
P. Seidel, ``Fukaya $A_\infty$ Structures Associated to Lefschetz Fibrations, I, II,'' arXiv:0912.3932, arXiv:1404.1352.

\bibitem{vafa-antibrane}
  C.~Vafa,
  ``Brane/anti-Brane Systems and $U(N|M)$ Supergroup,''
  hep-th/0101218.

\bibitem{thooft}
G. 't Hooft, ``A Planar Diagram Theory For Strong Interactions,''
Nucl. Phys. {\bf B72} (1974) 461-73.

\bibitem{eynard-orantin}
B. Eynard and N. Orantin, ``Invariants of Algebraic Curves and Topological Expansion,''
Comm. in Number Theory and Physics, Vol. 1 (2007), 347–552.


\bibitem{ACDKV} 
  M.~Aganagic, M.~C.~N.~Cheng, R.~Dijkgraaf, D.~Krefl and C.~Vafa,
  ``Quantum Geometry of Refined Topological Strings,''
  JHEP {\bf 1211}, 019 (2012).

\bibitem{MMSS}
J.~M.~Maldacena, G.~W.~Moore, N.~Seiberg and D.~Shih,
  ``Exact vs. Semiclassical Target Space of the Minimal String,''
  JHEP {\bf 0410}, 020 (2004).

\bibitem{AV}
M. Aganagic and C. Vafa, 
``Mirror Symmetry, D-Branes and Counting Holomorphic
Discs,'' arXiv:hep-th/0012041.

\bibitem{ADKMV} 
 M.~Aganagic, R.~Dijkgraaf, A.~Klemm, M.~Marino and C.~Vafa,
 ``Topological Strings and Integrable Hierarchies,''
  Commun.\ Math.\ Phys.\  {\bf 261}, 451 (2006).

\end{thebibliography}

\end{document}